\pdfoutput=1

\documentclass[11pt,twoside,a4paper,cmspaper,final,collab]{cms-tdr}
\def\svnVersion{403853}\def\cmsCernNoTag{CERN-EP/2016-227}\def\cmsCernDate{\today}\def\cmsMessage{Published in Physical Review D as \href{http://dx.doi.org/10.1103/PhysRevD.95.092001}{\doi{10.1103/PhysRevD.95.092001.}}}

\begin{document}\cmsNoteHeader{TOP-16-008}

\hyphenation{had-ron-i-za-tion}
\hyphenation{cal-or-i-me-ter}
\hyphenation{de-vices}
\RCS$Revision: 389283 $
\RCS$HeadURL: svn+ssh://svn.cern.ch/reps/tdr2/papers/TOP-16-008/trunk/TOP-16-008.tex $
\RCS$Id: TOP-16-008.tex 389283 2017-02-24 13:44:26Z hindrich $
\newlength\cmsFigWidth
\ifthenelse{\boolean{cms@external}}{\setlength\cmsFigWidth{0.85\columnwidth}}{\setlength\cmsFigWidth{0.4\textwidth}}
\ifthenelse{\boolean{cms@external}}{\providecommand{\cmsLeft}{top\xspace}}{\providecommand{\cmsLeft}{left\xspace}}
\ifthenelse{\boolean{cms@external}}{\providecommand{\cmsRight}{bottom\xspace}}{\providecommand{\cmsRight}{right\xspace}}
\ifthenelse{\boolean{cms@external}}{\providecommand{\cmsLLeft}{Top\xspace}}{\providecommand{\cmsLLeft}{Left\xspace}}
\ifthenelse{\boolean{cms@external}}{\providecommand{\cmsRRight}{Bottom\xspace}}{\providecommand{\cmsRRight}{Right\xspace}}
\ifthenelse{\boolean{cms@external}}{\providecommand{\NA}{}}{\providecommand{\NA}{{\ensuremath{\text{---}}}}}
\ifthenelse{\boolean{cms@external}}{\providecommand{\cmsTable}[1]{\relax#1}}{\providecommand{\cmsTable}[1]{\resizebox{\textwidth}{!}{#1}}}
\newcommand{\W}{\ensuremath{\PW}\xspace}
\newcommand{\FIG}[1]{Fig.~\ref{#1}\xspace}
\newcommand{\TAB}[1]{Table~\ref{#1}\xspace}
\newcommand{\tqh}{\ensuremath{\PQt_\mathrm{h}}\xspace}
\newcommand{\tql}{\ensuremath{\PQt_\ell}\xspace}
\newcommand{\Mtop}{\ensuremath{m_\PQt}\xspace}
\newcommand{\MW}{\ensuremath{m_{\PW}}\xspace}
\newcommand{\lpj}{$\ell$+jets\xspace}
\newcommand{\Dn}{\ensuremath{D_{\nu,\text{min}}}\xspace}
\newcolumntype{x}{D{,}{\text{--}}{4.5}}

\providecommand{\AMCATNLO}{\textsc{mg}5\_\text{a}\textsc{mc@nlo}\xspace}
\newcommand{\PYTHIAA}{\textsc{pythia}8\xspace}

\cmsNoteHeader{TOP-16-008}

\title{Measurement of differential cross sections for top quark pair production using the lepton+jets final state in proton-proton collisions at \texorpdfstring{13\TeV}{13 TeV}}

\date{\today}

\abstract{
Differential and double-differential cross sections for the production of top quark pairs in proton-proton collisions at 13\TeV are measured as a function of jet multiplicity and of kinematic variables of the top quarks and the top quark-antiquark system. This analysis is based on data collected by the CMS experiment at the LHC corresponding to an integrated luminosity of 2.3\fbinv. The measurements are performed in the lepton+jets decay channels with a single muon or electron in the final state. The differential cross sections are presented at particle level, within a phase space close to the experimental acceptance, and at parton level in the full phase space. The results are compared to several standard model predictions.
}

\hypersetup{%
pdfauthor={CMS Collaboration},%
pdftitle={Measurement of differential cross sections for top quark pair production using the lepton+jets final state in proton-proton collisions at 13 TeV},%
pdfsubject={CMS},%
pdfkeywords={CMS, physics, top, met}}

\maketitle

\section{Introduction}
Studying the differential production cross sections of top quark pairs (\ttbar) at high energies is a crucial ingredient in testing the standard model and searching for sources of new physics, which could alter the production rate. In particular, the differential \ttbar cross sections probe predictions of quantum chromodynamics (QCD) and facilitate the comparisons of the data with state-of-the-art calculations. In addition, some of the measured distributions, especially distributions of invariant mass and rapidity of the \ttbar system, can be used to improve our understanding of parton distribution functions (PDFs).

A measurement of the \ttbar differential and double-differential production cross sections as a function of jet multiplicity and of kinematic variables of the top quarks and the \ttbar system is presented. The measurement is based on proton-proton collision data at a center-of-mass energy of 13\TeV corresponding to an integrated luminosity of 2.3\fbinv~\cite{LUMI}. The data were recorded by the CMS experiment at the CERN LHC in 2015. This measurement makes use of the \ttbar decay into the \lpj ($\ell=\Pe,\mu$) final state, where, after the decay of each top quark into a bottom quark and a \W boson, one of the \W bosons decays hadronically and the other one leptonically. The $\tau$ lepton decay mode is not considered here as signal. The differential cross sections are presented as a function of the transverse momentum \pt and the absolute rapidity $\abs{y}$ of the hadronically (\tqh) and the leptonically (\tql) decaying top quarks; as a function of \pt, $\abs{y}$, and mass $M$ of the \ttbar system. The cross section is also measured as a function of the number of additional jets in the event. In addition, the differential cross sections as a function of $\pt(\tqh)$ and $\pt(\ttbar)$ are measured in bins of jet multiplicity and double-differential cross sections for the following combinations of variables are determined: $\abs{y(\tqh)}$ \vs $\pt(\tqh)$, $M(\ttbar)$ \vs $\abs{y(\ttbar)}$, and $\pt(\ttbar)$ \vs $M(\ttbar)$.

This measurement continues a series of differential \ttbar production cross section measurements in proton-proton collisions at the LHC. Previous measurements at 7\TeV~\cite{Chatrchyan:2012saa,Aad:2015eia} and 8\TeV~\cite{Khachatryan:2015oqa,Aad:2015mbv,Aad:2015hna,Khachatryan:2015fwh,Khachatryan:2149620} have been performed in various \ttbar decay channels.

The differential cross sections are presented in two different ways, at particle level and at parton level. For the particle-level measurement a proxy of the top quark is defined based on experimentally accessible quantities like jets, which consist of quasi-stable particles with a mean lifetime greater than 30\unit{ps}. These are described by theoretical calculations that, in contrast to pure matrix-element calculations, involve parton shower and hadronization models. These objects are required to match closely the experimental acceptance. A detailed definition is given in Section~\ref{PSTOP}. Such an approach has the advantage that it reduces theoretical uncertainties in the experimental results by avoiding theory-based extrapolations from the experimentally accessible portion of the phase space to the full range, and from jets to partons. However, such results cannot be compared to parton-level calculations.

For the measurement at parton level, the top quarks are defined directly before decaying into a bottom quark and a \W boson. For this analysis the parton-level \ttbar system is calculated at next-to-leading order (NLO) and combined with a simulation of the parton shower. No restriction of the phase space is applied for parton-level top quarks.

The experimental signature is the same for both measurements and consists of two jets coming from the hadronization of $\PQb$ quarks (b jets), two jets from a hadronically decaying \W boson, a transverse momentum imbalance associated with the neutrino, and a single isolated muon or electron.

This paper is organized as follows: In Section~\ref{SIM} we provide a description of the signal and background simulations, followed by the definition of the particle-level top quarks in Section~\ref{PSTOP}. After a short overview of the CMS detector and the particle reconstruction in Section~\ref{DET}, we describe the object and event selections in Sections~\ref{EVS} and \ref{EVTSEL}, respectively. Section~\ref{TTREC} contains a detailed description of the reconstruction of the \ttbar system. Details on the background estimation and the unfolding are presented in Sections~\ref{BKG} and \ref{UNFO}. After a discussion on systematic uncertainties in Section~\ref{UNC}, the results are finally presented in Section~\ref{RES}.

\section{Signal and background modeling}
\label{SIM}
The Monte Carlo programs \POWHEG~\cite{Nason:2004rx,Frixione:2007vw,Alioli:2010xd,Campbell:2014kua} (v2) and \MADGRAPH{}5\_a\MCATNLO~\cite{Alwall:2014hca} (v2.2.2) (\AMCATNLO) are used to simulate \ttbar events. They include NLO QCD matrix element calculations that are combined with the parton shower simulation of \PYTHIA~\cite{Sjostrand:2006za,Sjostrand:2007gs} (v8.205) (\PYTHIAA) using the tune CUETP8M1~\cite{Skands:2014pea}. In addition, \AMCATNLO is used to produce simulations of \ttbar events with additional partons. In one simulation all processes of up to three additional partons are calculated at leading order (LO) and combined with the \PYTHIAA parton shower simulation using the MLM~\cite{MLM} algorithm. In another simulation all processes of up to two additional partons are calculated at NLO and combined with the \PYTHIAA parton shower simulation using the FxFx~\cite{Frederix:2012ps} algorithm. The default parametrization of the PDF used in all simulations is NNPDF30\_nlo\_as\_0118~\cite{Ball:2014uwa}. A top quark mass $\Mtop=172.5$\GeV is used. When compared to the data, simulations are normalized to an inclusive \ttbar production cross section of $832^{+40}_{-46}$\unit{pb}~\cite{Czakon:2011xx}. This value is calculated with next-to-NLO (NNLO) precision including the resummation of next-to-next-to-leading-logarithmic (NNLL) soft gluon terms. Its given uncertainty is due to the choice of hadronization/factorization scales and PDF.

In all simulations, event weights are calculated that represent the usage of the uncertainty eigenvector sets of the PDF. There are also event weights available that represent the changes of factorization and renormalization scales by a factor of two or one half. These additional weights allow for the calculation of systematic uncertainties due to the PDF and the scale choices. For additional uncertainty estimations we use \POWHEG{}+\PYTHIAA simulations with top quark masses of 171.5 and 173.5\GeV, with parton shower scales varied up and down by a factor of two, and a simulation with \POWHEG combined with \HERWIGpp~\cite{Bahr:2008pv} (v2.7.1) using the tune EE5C~\cite{Seymour:2013qka}.

The main backgrounds are produced using the same techniques. The \AMCATNLO generator is used for the simulation of \W boson production in association with jets, $t$-channel single top quark production, and Drell--Yan (DY) production in association with jets. The \POWHEG generator is used for the simulation of single top quark associated production with a \W boson ($\PQt\PW$) and \PYTHIAA is used for multijet production. In all cases the parton shower and the hadronization are described by \PYTHIAA. The \W boson and DY backgrounds are normalized to their NNLO cross sections~\cite{Li:2012wna}. The single top quark processes are normalized to NLO calculations~\cite{Kant:2014oha,Kidonakis:2012rm}, and the multijet simulation is normalized to the LO calculation~\cite{Sjostrand:2007gs}.

The detector response is simulated using \GEANTfour{}~\cite{Allison:2006ve}. Afterwards, the same reconstruction algorithms that are applied to the data are used. Multiple proton-proton interactions per bunch crossing (pileup) are included in the simulation. To correct the simulation to be in agreement with the pileup conditions observed during the data taking, the average number of pileup events per bunch crossing is calculated for the measured instantaneous luminosity. The simulated events are weighted, depending on their number of pileup interactions, to reproduce the measured pileup distribution.

\section{Particle-level top quark definition}
\label{PSTOP}
The following list describes the definitions of objects constructed from quasi-stable particles, obtained from the predictions of \ttbar event generators before any detector simulation. These objects are further used to define the particle-level top quarks.
\begin{itemize}
\item Muons and electrons that do not have their origin in a decay of a hadron are selected and their momenta are corrected for the final-state radiation effects. The anti-\kt jet algorithm~\cite{Cacciari:2008gp, Cacciari:2011ma} with a distance parameter of 0.1 is used to cluster the leptons and photons not originating from hadron decays. Those photons that are clustered together with a selected lepton are assumed to have been radiated by the lepton and their momenta are added to the lepton momentum. However, the lepton is only selected if the original \pt is at least half of their corrected \pt.
\item All neutrinos that do not have their origin in a decay of a hadron are selected.
\item Jets are clustered by the anti-\kt jet algorithm with a distance parameter of 0.4. All quasi-stable particles are considered, excluding the selected neutrinos and leptons together with their radiated photons.
\item $\PQb$ jets at particle level are defined as those jets that contain a $\PQb$ hadron. As a result of the short lifetime of $\PQb$ hadrons, only their decay products should be considered for the jet clustering. However, to allow their association to a jet, the $\PQb$ hadrons are also included with their momenta scaled down to a negligible value. This preserves the information of their directions, but they have no impact on the jet clustering itself.
\end{itemize}

Based on the invariant masses $M$ of these objects, we construct a pair of particle-level top quarks in the \lpj final state. Events with exactly one muon or electron with $\pt > 30$\GeV and an absolute pseudorapidity $\abs{\eta} < 2.5$ are selected. We take the sum of the four-momenta of all selected neutrinos as the neutrino momentum $p_\nu$ from the leptonically decaying top quark and find the permutation of jets that minimizes the quantity
\ifthenelse{\boolean{cms@external}}{
\begin{multline}
K^2 = [M(p_\nu + p_{\ell} + p_{{\PQb}_\ell}) - \Mtop]^2
+ [M(p_{\mathrm{j}_1} + p_{\mathrm{j}_2}) - \MW]^2 \\+ [M(p_{\mathrm{j}_1} + p_{\mathrm{j}_2} + p_{{\PQb}_\mathrm{h}}) - \Mtop]^2,
\label{PSTOPE1}
\end{multline}
}{
\begin{equation}
K^2 = [M(p_\nu + p_{\ell} + p_{{\PQb}_\ell}) - \Mtop]^2 + [M(p_{\mathrm{j}_1} + p_{\mathrm{j}_2}) - \MW]^2 + [M(p_{\mathrm{j}_1} + p_{\mathrm{j}_2} + p_{{\PQb}_\mathrm{h}}) - \Mtop]^2,
\label{PSTOPE1}
\end{equation}
}
where $p_{\mathrm{j}_{1/2}}$ are the four-momenta of two light-flavor jet candidates, $p_{\PQb_{\ell/\mathrm{h}}}$ are the four-momenta of two \PQb-jet candidates, $p_\ell$ is the four-momentum of the lepton, and $\MW = 80.4$\GeV is the mass of the \W boson. All jets with $\pt > 25$\GeV and $\abs{\eta} < 2.5$ are considered. At least four jets are required, of which at least two must be $\PQb$ jets. If there are more than two $\PQb$ jets, we allow $\PQb$ jets as decay products of the proxy for the hadronically decaying \W boson. Due to a limited efficiency of the $\PQb$ jet identification at detector level this improves the agreement between the reconstructed top quarks and the particle-level top quarks. The remaining jets with the same kinematic selection are considered as additional jets at particle level.

\begin{figure}[tbhp]
\centering
\includegraphics[width=0.49\textwidth]{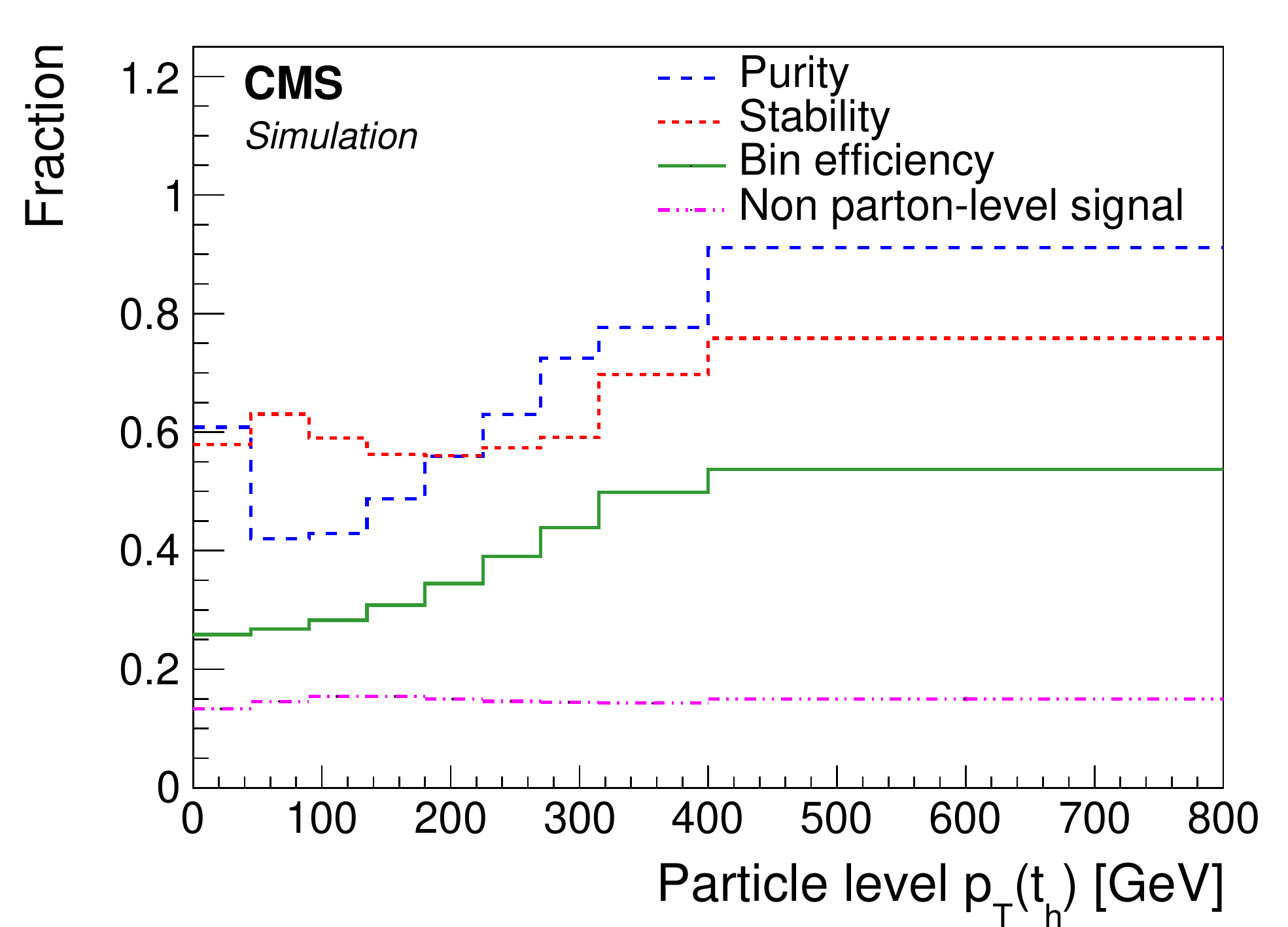}
\includegraphics[width=0.49\textwidth]{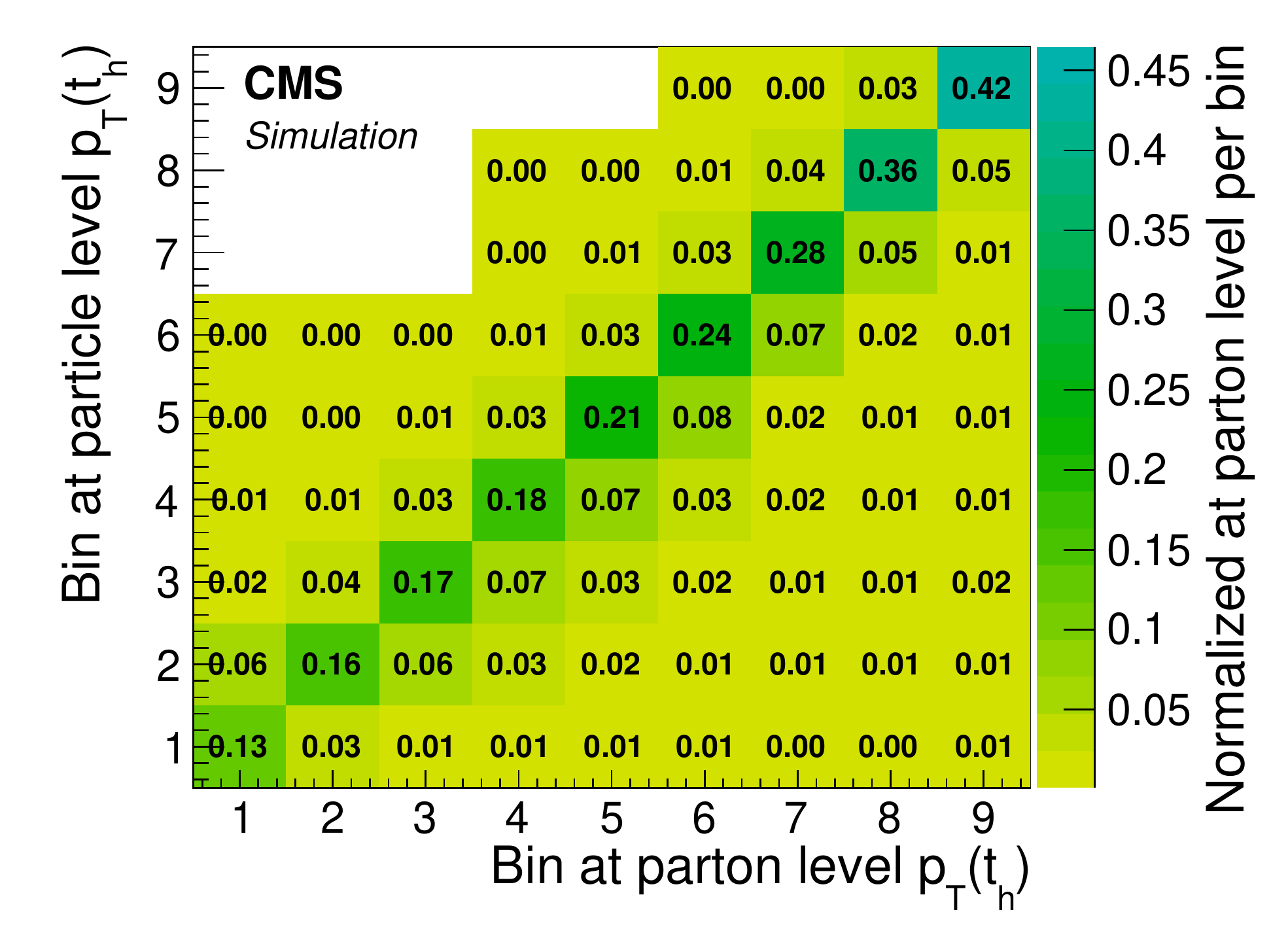}\\
\caption{Comparison between the $\pt(\tqh)$ at particle and parton level, extracted from the \POWHEG{}+\PYTHIAA simulation. \cmsLLeft: fraction of parton-level top quarks in the same bin at particle level (purity), fraction of particle-level top quarks in the same bin at parton level (stability), ratio of the number of particle- to parton-level top quarks, and fraction of events with a particle-level top quark pair that are not considered as signal events at parton level. \cmsRRight: bin migrations between particle and parton level. The \pt range of the bins can be taken from the \cmsLeft panel. Each column is normalized to the number of events per column at parton level in the full phase space.}
\label{PSTOPF1}
\end{figure}

It should be remarked that events with a hadronic and a leptonic particle-level top quark are not required to be \lpj events at the parton level. As an example, in \FIG{PSTOPF1} the relation between the $\pt(\tqh)$ values at particle and parton level is shown.

\section{The CMS detector}
\label{DET}
The central feature of the CMS detector is a superconducting solenoid of 6\unit{m} internal diameter, providing a magnetic field of 3.8\unit{T}. Within the solenoid volume are a silicon pixel and strip tracker, a lead tungstate crystal electromagnetic calorimeter (ECAL), and a brass and scintillator hadron calorimeter (HCAL), each composed of a barrel and two endcap sections. Forward calorimeters extend the $\eta$ coverage provided by the barrel and endcap detectors. Muons are measured in gas-ionization detectors embedded in the steel flux-return yoke outside the solenoid. A more detailed description of the CMS detector, together with a definition of the coordinate system and relevant kinematic variables, can be found in Ref.~\cite{Chatrchyan:2008zzk}.

The particle-flow (PF) event algorithm~\cite{CMS-PAS-PFT-09-001,CMS-PAS-PFT-10-001} reconstructs and identifies each individual particle with an optimized combination of information from the various elements of the CMS detector. The energy of photons is directly obtained from the ECAL measurement, corrected for zero-suppression effects. The energy of electrons is determined from a combination of the electron momentum at the primary interaction vertex as determined by the tracker, the energy of the corresponding ECAL cluster, and the energy sum of all bremsstrahlung photons spatially compatible with originating from the electron track. The energy of muons is obtained from the curvature of the corresponding track. The energy of charged hadrons is determined from a combination of their momentum measured in the tracker and the matching ECAL and HCAL energy deposits, corrected for zero-suppression effects and for the response function of the calorimeters to hadronic showers. Finally, the energy of neutral hadrons is obtained from the corresponding corrected ECAL and HCAL energy.

\section{Physics object reconstruction}
\label{EVS}
This analysis depends on the reconstruction and identification of muons, electrons, jets, and missing transverse momentum associated with a neutrino. Only leptons are selected that are compatible with originating from the primary vertex, defined as the vertex at the beam position with the highest sum of $\pt^2$ of the associated tracks. Leptons from \ttbar decays are typically isolated, i.e., separated in $\Delta R = \sqrt{(\Delta \phi)^2 + (\Delta \eta)^2}$ from other particles. A requirement on the lepton isolation is used to reject leptons produced in decays of hadrons.

The muon isolation variable is defined as the sum of the \pt of all tracks, except for the muon track, originating from the \ttbar interaction vertex within a cone of $\Delta R = 0.3$. It is required to be less than 5\% of the muon \pt. The muon reconstruction and selection~\cite{Chatrchyan:2012xi} efficiency is measured in the data using tag-and-probe techniques~\cite{TNPREF}. Depending on the \pt and $\eta$ of the muon it is 90--95\%.

For electrons the isolation variable is the sum of the \pt of neutral hadrons, charged hadrons, and photon PF candidates in a cone of $\Delta R = 0.3$ around the electron. Contributions of the electron to the isolation variable are suppressed excluding a small region around the electron. This isolation variable is required to be smaller than 7\% of the electron \pt. An event-by-event correction is applied that maintains a constant electron isolation efficiency with respect to the number of pileup interactions~\cite{Cacciari:2007}. The measured reconstruction and identification~\cite{Khachatryan:2015hwa} efficiency for electrons is 70--85\% with a \pt and $\eta$ dependence.

Jets are reconstructed from PF objects clustered using the anti-\kt jet algorithm with a distance parameter of 0.4 using the \textsc{FastJet} package~\cite{Cacciari:2011ma}. Charged particles originating from a vertex of a pileup interaction are excluded. The total energy of the jets is corrected for energy depositions from pileup. In addition, \pt- and $\eta$-dependent corrections are applied to correct for detector response effects~\cite{JET}. Those jets identified as isolated muons or electrons are removed from consideration.

For the identification of $\PQb$ jets the combined secondary vertex algorithm~\cite{BTV} is used. It provides a discriminant between light-flavor and $\PQb$ jets based on the combined information of secondary vertices and the impact parameter of tracks at the primary vertex. A jet is identified as $\PQb$ jet if the associated value of the discriminant exceeds a threshold criterion. Two threshold criteria are used: a tight threshold with an efficiency of about 70\% and a light-flavor jet rejection probability of 95\%, and a loose one with an efficiency of about 80\% and a rejection probability of 85\%.

The missing transverse momentum \ptvecmiss is calculated as the negative of the vectorial sum of transverse momenta of all PF candidates in the event. Jet energy corrections are also propagated to improve the measurement of \ptvecmiss.
\section{Event selection}
\label{EVTSEL}
Events are selected if they pass single-lepton triggers. These require $\pt > 22$\GeV for electrons and $\pt > 20$\GeV for muons, as well as various quality and isolation criteria.

To reduce the background contributions and optimize the \ttbar reconstruction additional, more stringent, requirements on the events are imposed. Events with exactly one muon or electron with $\pt > 30$\GeV and $\abs{\eta} < 2.1$ are selected. No additional muons or electrons with $\pt > 15$\GeV and $\abs{\eta} < 2.4$ are allowed. In addition to the lepton, at least four jets with $\pt > 30$\GeV and $\abs{\eta} < 2.4$ are required. At least two of these jets must be tagged as $\PQb$ jets. At least one jet has to fulfill the tight \PQb-jet identification criterion while for the second $\PQb$ jet only the loose criterion is required. At least one of the two jets with the highest value of the $\PQb$ tagging discriminant and at least one of the remaining jets are required to have $\pt > 35$\GeV.

We compare several kinematic distributions of the muon and electron channels to the simulation to verify that there are no unexpected differences. The ratios of the measured to the expected event yields in the two channels agree within the uncertainty in the lepton reconstruction and selection efficiencies. In the remaining steps of the analysis the two channels are combined by adding their distributions.
\section{Reconstruction of the top quark-antiquark system}
\label{TTREC}
The goal of the \ttbar reconstruction is the correct identification of reconstructed objects as parton- or particle-level top quark decay products. To test the performance of the reconstruction algorithm an assignment between detector level and particle- (parton-) level objects is needed. For the particle-level measurement this relationship is straightforward. Reconstructed leptons and jets can be matched spatially to corresponding objects at the particle level. For the parton-level measurement we need to define how to match the four initial quarks from a \ttbar decay with reconstructed jets. This is not free of ambiguities since a quark does generally not lead to a single jet. One quark might shower into several jets or multiple quarks might be clustered into one jet if they are not well separated. We introduce an unambiguous matching criterion that matches the reconstructed jet with the highest \pt within $\Delta R = 0.4$ to a quark from the \ttbar decay. If two quarks are matched with the same jet, the event has a merged topology and is considered as ``not reconstructible'' in the context of this analysis.

The same matching criterion is also used to assign particle-level jets to the \ttbar decay products at parton level. Those particle-level jets with $\pt > 25$\GeV and $\abs{\eta} < 2.5$, which are not assigned to one of the initial quarks, are considered as additional jets at parton level.

For the reconstruction of the top quark-antiquark system all possible permutations of jets that assign reconstructed jets to the decay products of the \ttbar system are tested and a likelihood that a certain permutation is correct is evaluated. Permutations are considered only if the two jets with the highest $\PQb$ tagging probabilities are the two \PQb-jet candidates. In addition, the \pt of at least one \PQb-jet candidate and at least one jet candidate from the \W boson decay have to be above 35\GeV. In each event the permutation with the highest probability is selected. The likelihoods are evaluated separately for the particle- and the parton-level measurements.

The first reconstruction step involves the determination of the neutrino four-momentum $p_\nu$. This is performed using the algorithm described in Ref.~\cite{Betchart:2013nba}. The idea is to find all possible solutions for the three components of the neutrino momentum using the two mass constraints $(p_\nu + p_\ell)^2 = m_{\PW}^2$ and $(p_\nu + p_\ell + p_{{\PQb}_\ell})^2 = m_\PQt^2$. Each equation describes an ellipsoid in the three-dimensional momentum space of the neutrino. The intersection of these two ellipsoids is usually an ellipse. We select $p_\nu$ as the point on the ellipse for which the distance $D_{\nu,\text{min}}$ between the ellipse projection onto the transverse plane and \ptvecmiss is minimal. This algorithm leads to a unique solution for the longitudinal neutrino momentum and an improved resolution for the transverse component. The minimum distance $D_{\nu,\mathrm{min}}$ can also be used to identify the correct ${\PQb}_\ell$. In the cases with an invariant mass of the lepton and the ${\PQb}_\ell$ candidate above $m_\PQt$ no solution can be found and we continue with the next permutation.

\begin{figure*}[tbhp]
\centering
\includegraphics[width=0.49\textwidth]{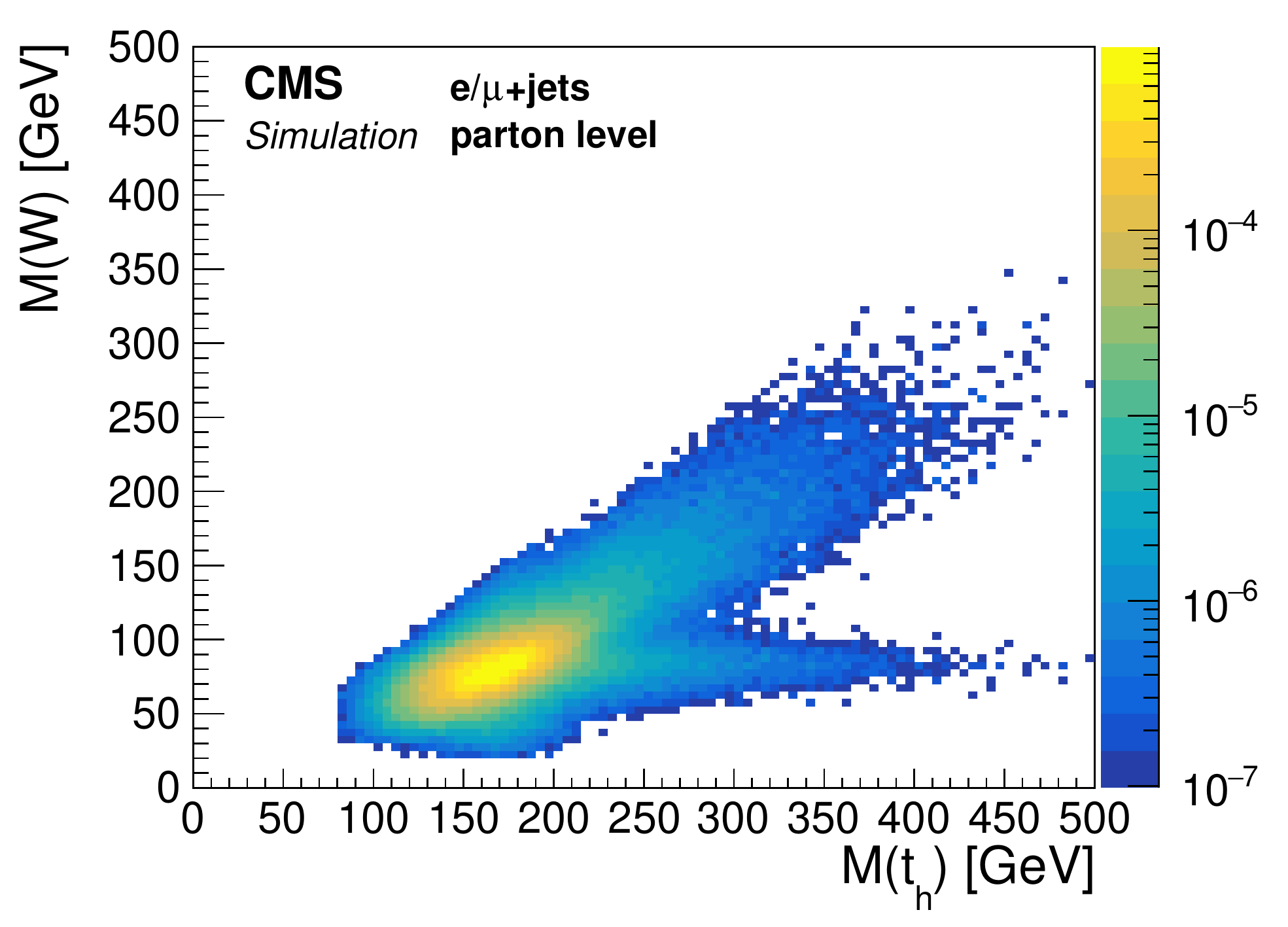}
\includegraphics[width=0.49\textwidth]{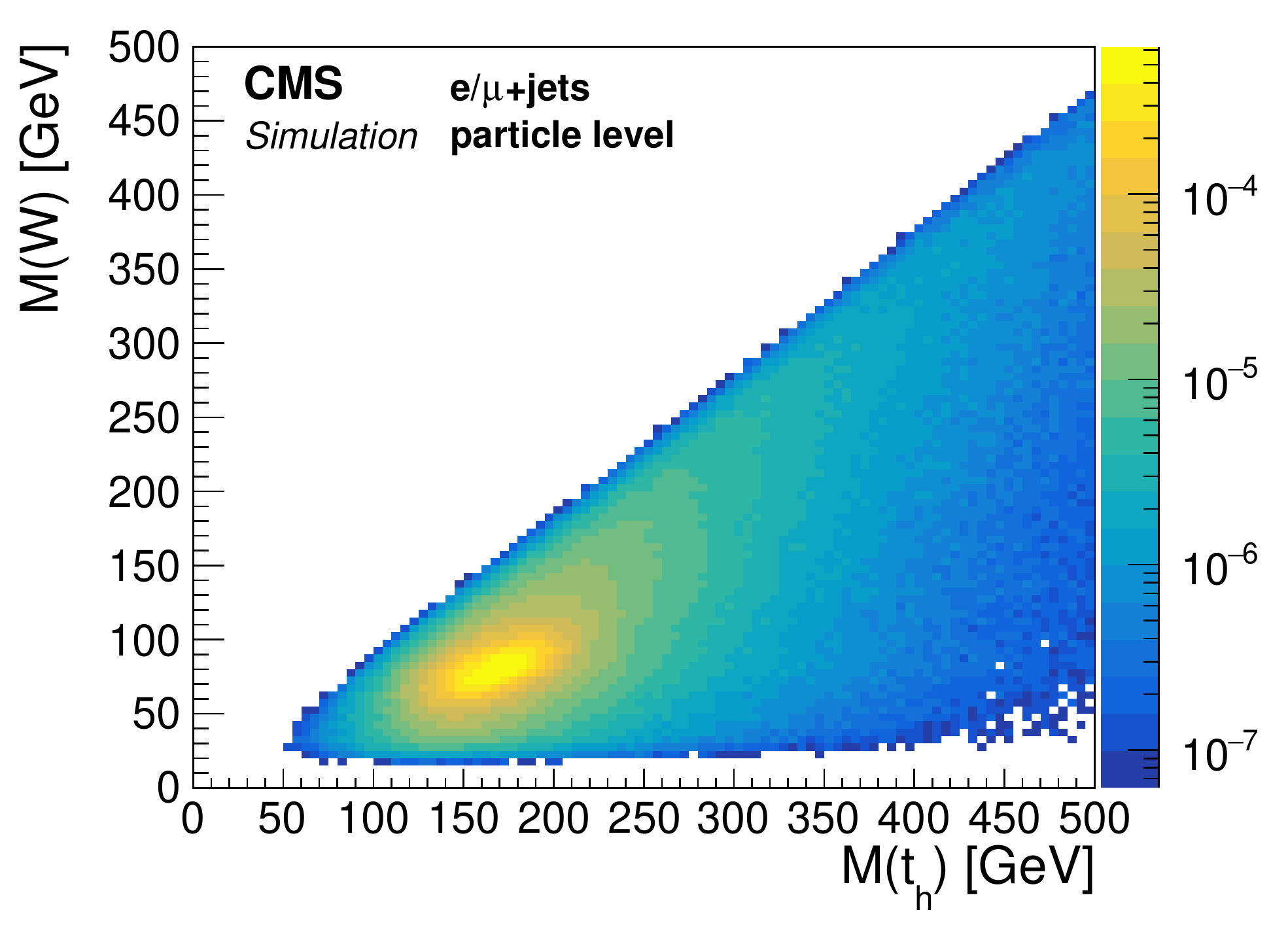}
\includegraphics[width=0.49\textwidth]{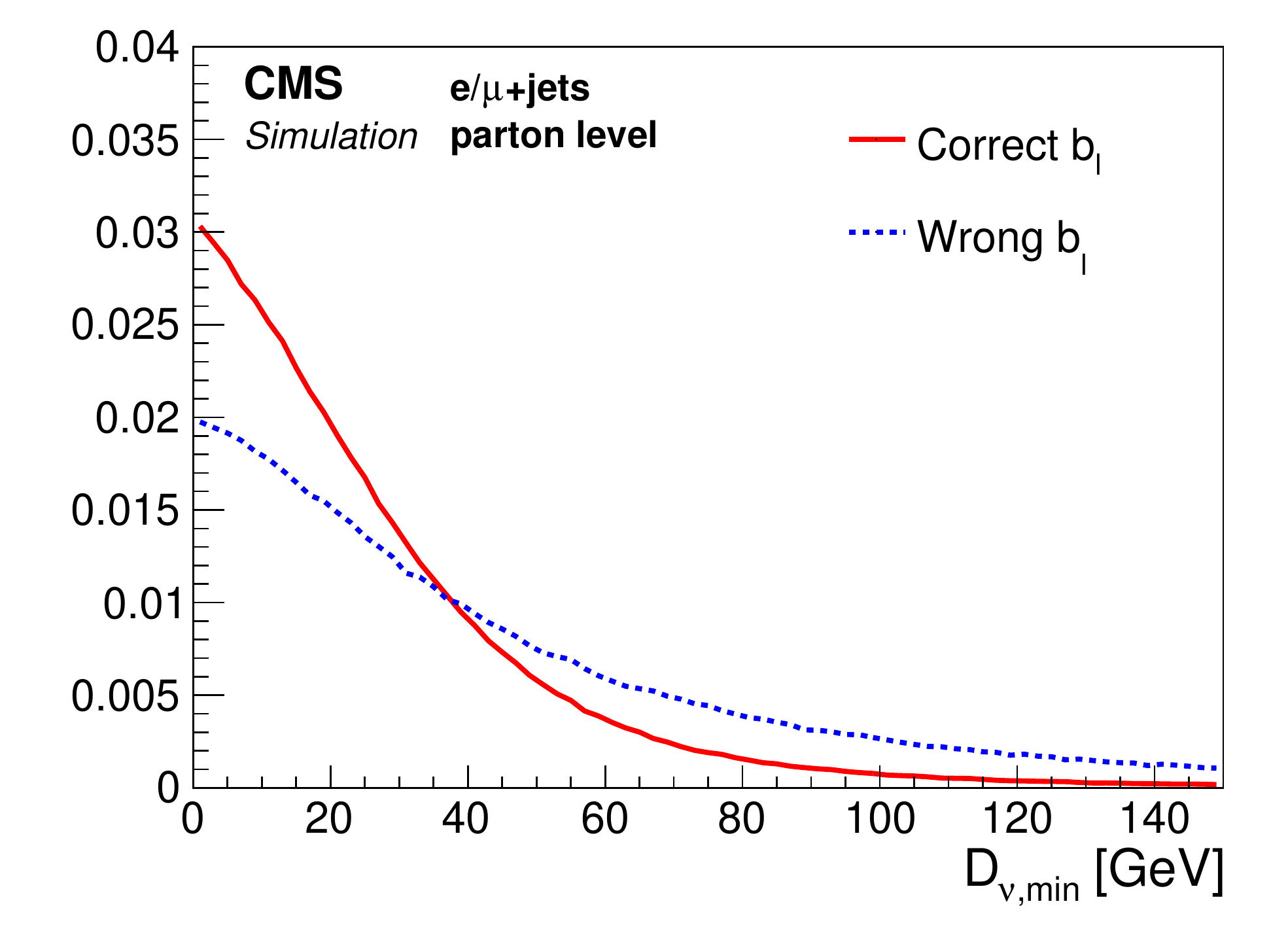}
\includegraphics[width=0.49\textwidth]{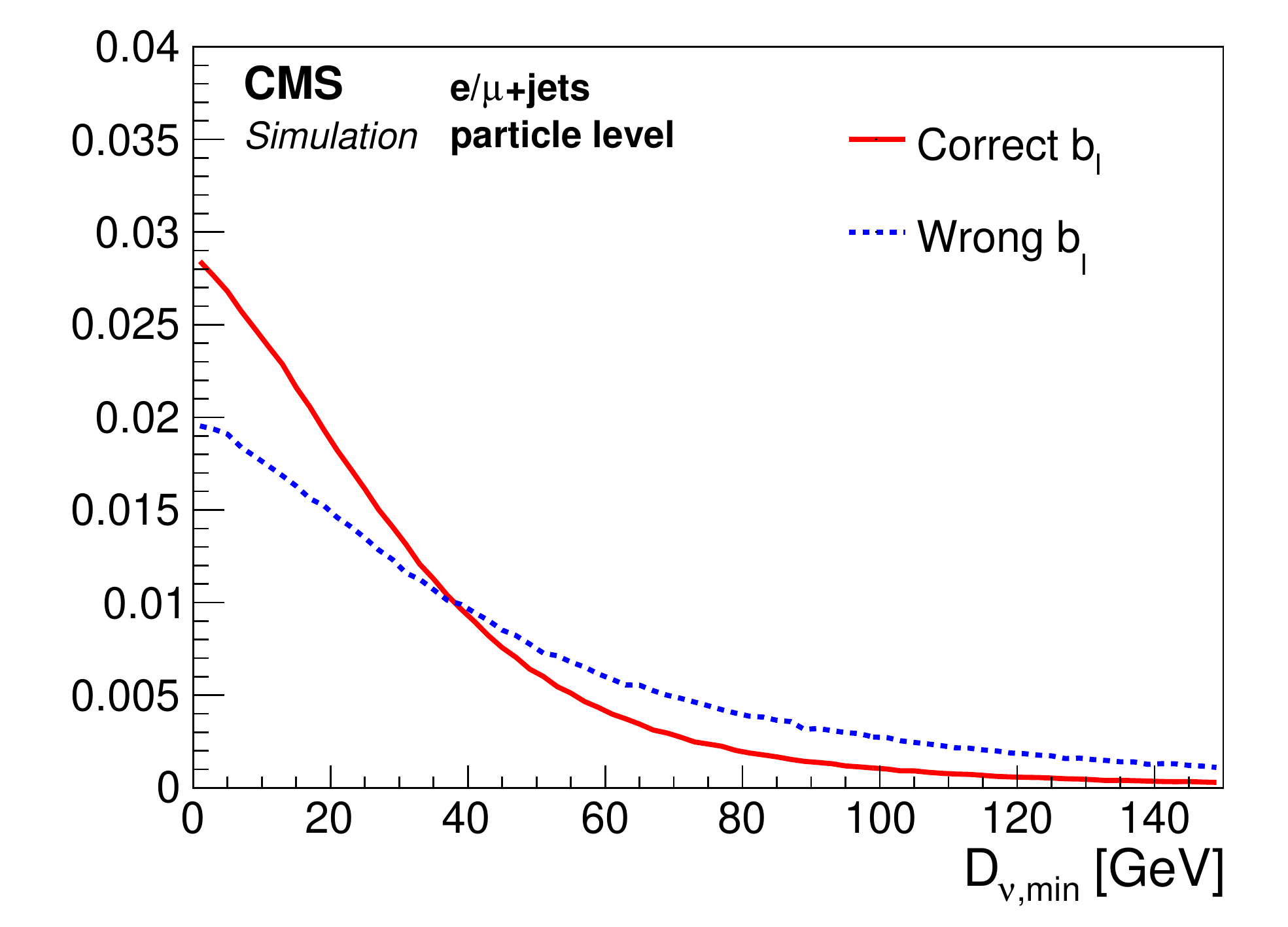}
\caption{Top: normalized two-dimensional mass distribution of the correct reconstructed hadronically decaying \W bosons $M(\PW)$ and the correct reconstructed top quarks $M(\tqh)$ for the parton- (left) and the particle- (right) level measurements. Bottom: normalized distributions of the distance \Dn for correctly and wrongly selected $\PQb$ jets from the leptonically decaying top quarks. The distributions are taken from the \POWHEG{}+\PYTHIAA \ttbar simulation.}
\label{TTRECF2}
\end{figure*}

The likelihood $\lambda$ is maximized to select the best permutation of jets. It uses constraints of the top quark and \W boson masses on the hadronic side and the $D_{\nu,\mathrm{min}}$ value from the neutrino reconstruction, and is defined through
\begin{equation}
-\log(\lambda) = -\log(P_m(m_2, m_3)) -\log(P_{\nu}(D_{\nu,\mathrm{min}})), \label{TTRECEQ1}
\end{equation}
where $P_m$ is the two-dimensional probability distribution of the invariant masses of correctly reconstructed \W bosons and top quarks. This probability is calculated for the invariant mass of the two jets $m_2$ tested as the \W boson decay products, and the invariant mass of the three jets $m_3$ tested as the decay products of the hadronically decaying top quark. The distributions for the correct jet assignments, taken from the \POWHEG{}+\PYTHIAA simulation and normalized to unity, are shown in \FIG{TTRECF2} for the particle- and parton-level measurements. Permutations with probabilities of less than 0.1\% of the highest value are rejected. This part of the likelihood is sensitive to the correct reconstruction of the hadronically decaying top quark, modulo a permutation of the two jets from the \W boson, but none of the measured kinematic variables will be affected by this ambiguity.

The probability $P_{\nu}$ describes the distribution of \Dn for a correctly selected ${\PQb}_\ell$. In \FIG{TTRECF2} the normalized distributions of \Dn for ${\PQb}_\ell$ and for other jets are shown. On average, the distance \Dn for correctly selected ${\PQb}_\ell$ is smaller and has a lower tail compared to the distance obtained for other jets. Permutations with values of $\Dn > 150$\GeV are rejected since they are very unlikely to originate from a correct ${\PQb}_\ell$ association. This part of the likelihood is sensitive to the correct reconstruction of the leptonically decaying top quark.

\begin{figure}[tbhp]
\centering
\includegraphics[width=0.49\textwidth]{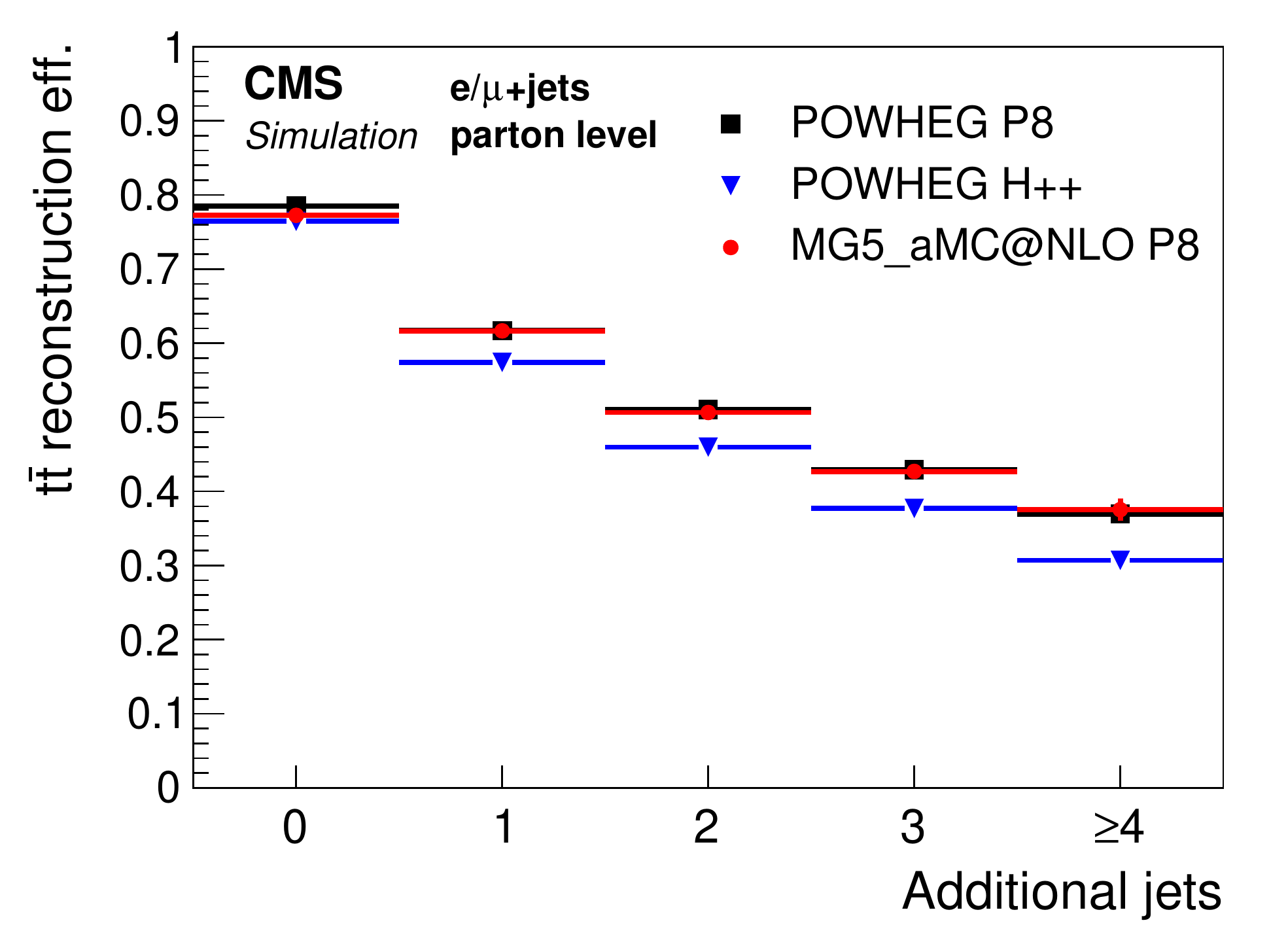}
\includegraphics[width=0.49\textwidth]{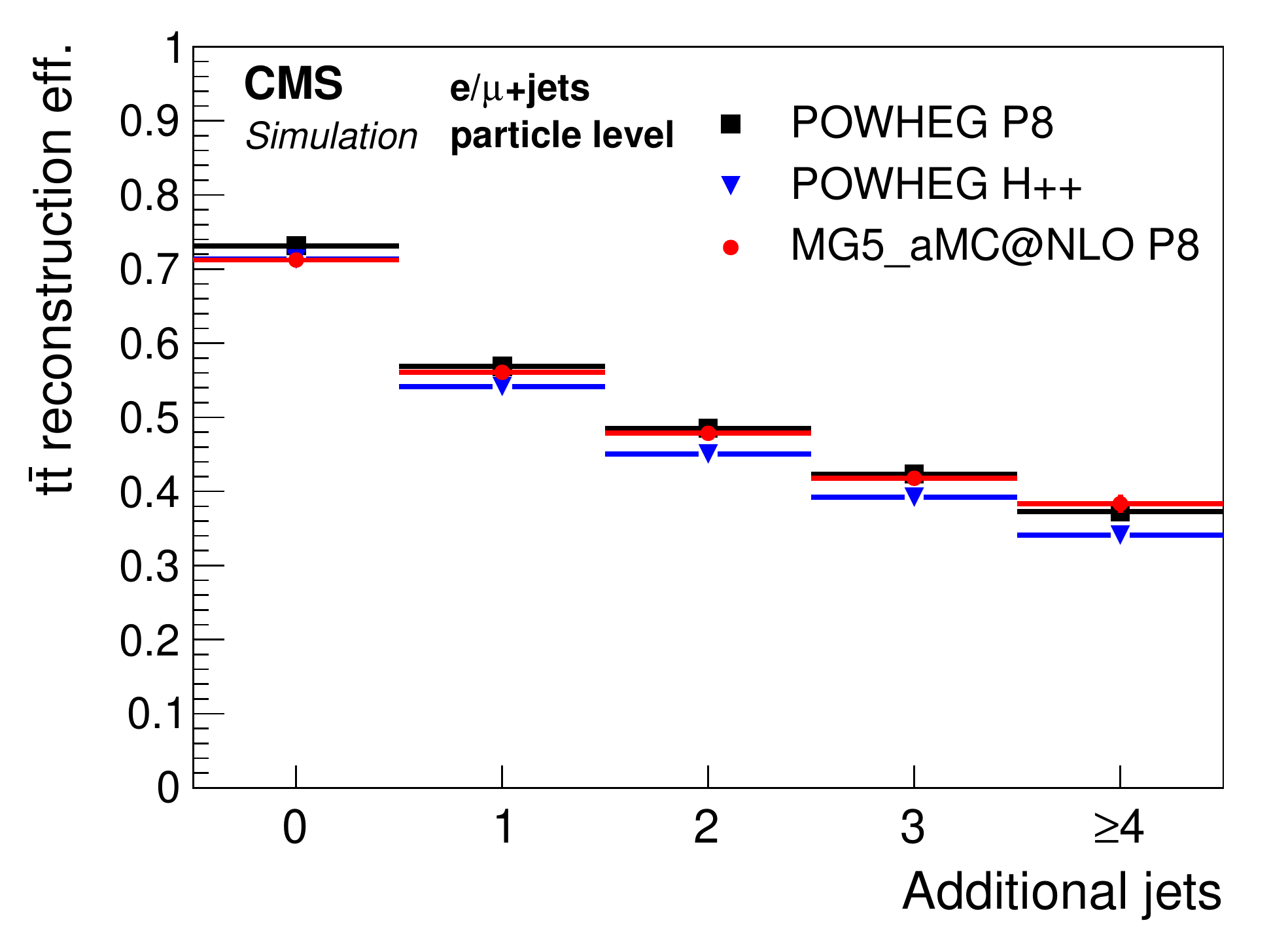}
\caption{Reconstruction efficiency of the \ttbar system as a function of the number of additional jets for the parton- (\cmsLeft) and particle- (\cmsRight) level measurements calculated based on the simulations with \POWHEG{}+\PYTHIAA (P8), \POWHEG{}+\HERWIGpp (H++), and \AMCATNLO+\PYTHIAA.}
\label{TTRECF3}
\end{figure}

\begin{figure}[tbhp]
\centering
\includegraphics[width=0.49\textwidth]{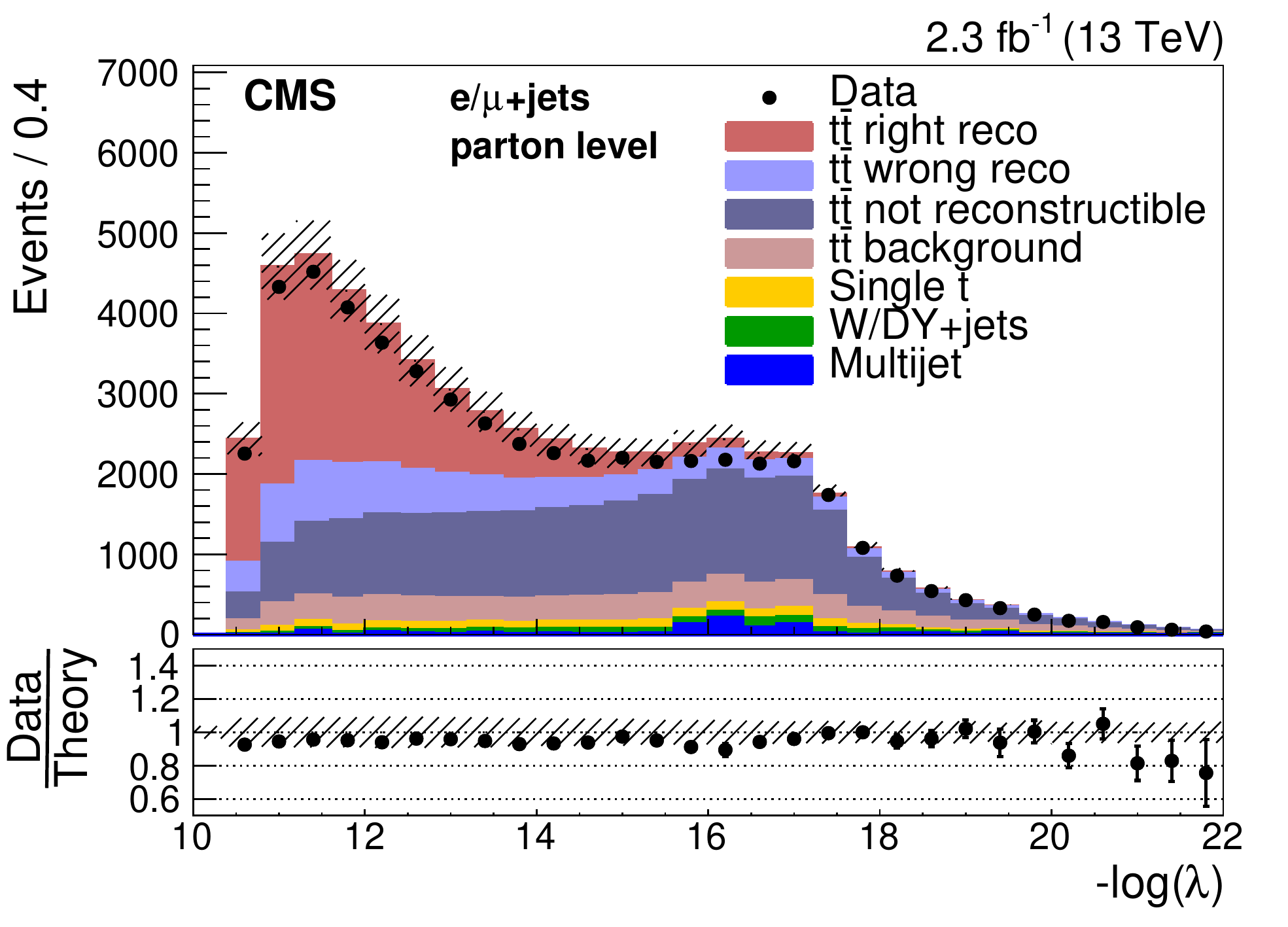}
\includegraphics[width=0.49\textwidth]{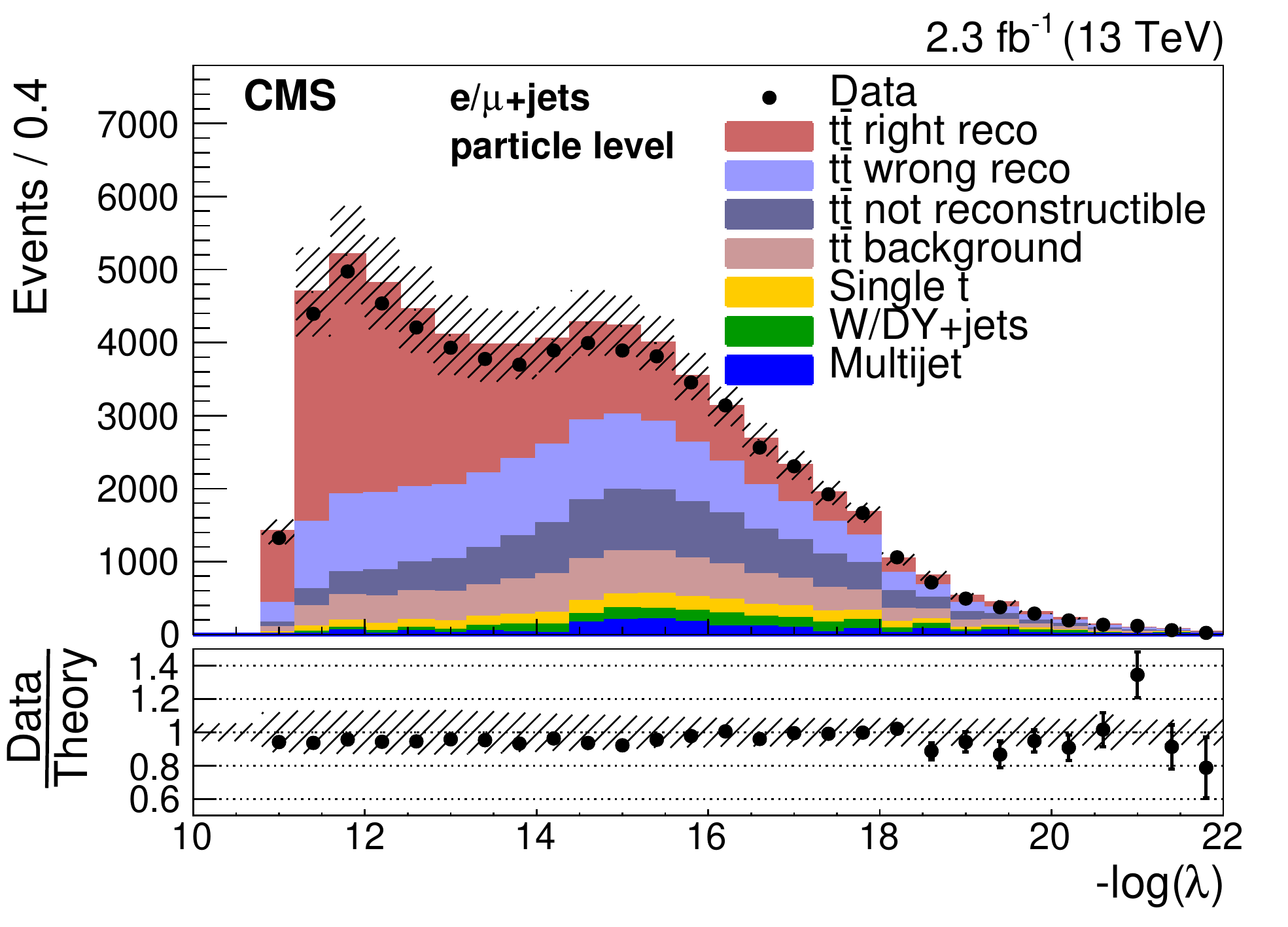}
\caption{Distribution of the negative log-likelihood for the selected best permutation in the parton- (\cmsLeft) and the particle- (\cmsRight) level measurements in data and simulations. The simulation of \POWHEG{}+\PYTHIAA is used to describe the \ttbar production. Experimental (cf. Section~\ref{UNC}) and statistical uncertainties (hatched area) are shown for the total simulated yield, which is normalized to the measured integrated luminosity. The ratios of data to the sum of the expected yields are provided at the bottom of each panel.}
\label{TTRECF4}
\end{figure}

The likelihood $\lambda$ combines the probabilities from the reconstruction of the hadronically and leptonically decaying top quarks and provides information on reconstructing the whole \ttbar system. The performance of the reconstruction algorithm is tested using the three \ttbar simulations generated with \POWHEG combined with \PYTHIAA or \HERWIGpp, and \AMCATNLO{}+\PYTHIAA where we use the input distributions $P_m$ and $P_{\nu}$ from \POWHEG{}+\PYTHIAA. The efficiency of the reconstruction algorithm is defined as the probability that the most likely permutation, as identified through the maximization of the likelihood $\lambda$, is the correct one, given that all decay products from the \ttbar decay are reconstructed and selected. These efficiencies as a function of the jet multiplicity are shown in \FIG{TTRECF3}. Since the number of permutations increases drastically with the number of jets, it is more likely to select a wrong permutation if there are additional jets. The small differences observed in different simulations are taken into account for the uncertainty estimations. We observe a lower reconstruction efficiency for the particle-level measurement. This is caused by the weaker mass constraints for a particle-level top quark, where, in contrast to the parton-level top quark, exact matches to the top quark and \W boson masses are not required. This can be seen in the mass distributions of \FIG{TTRECF2} and the likelihood distributions in \FIG{TTRECF4}. Here the signal simulation is divided into the following categories: correctly reconstructed \ttbar systems (\ttbar right reco), events where all decay products are available, but the algorithm failed to identify the correct permutation (\ttbar wrong reco), \lpj \ttbar events where at least one decay product is missing (\ttbar not reconstructible), and nonsignal \ttbar events (\ttbar background). However, the lower reconstruction efficiency of the particle-level top quark is compensated by the higher number of reconstructible events.

\begin{figure*}[tbhp]
\centering
\includegraphics[width=0.49\textwidth]{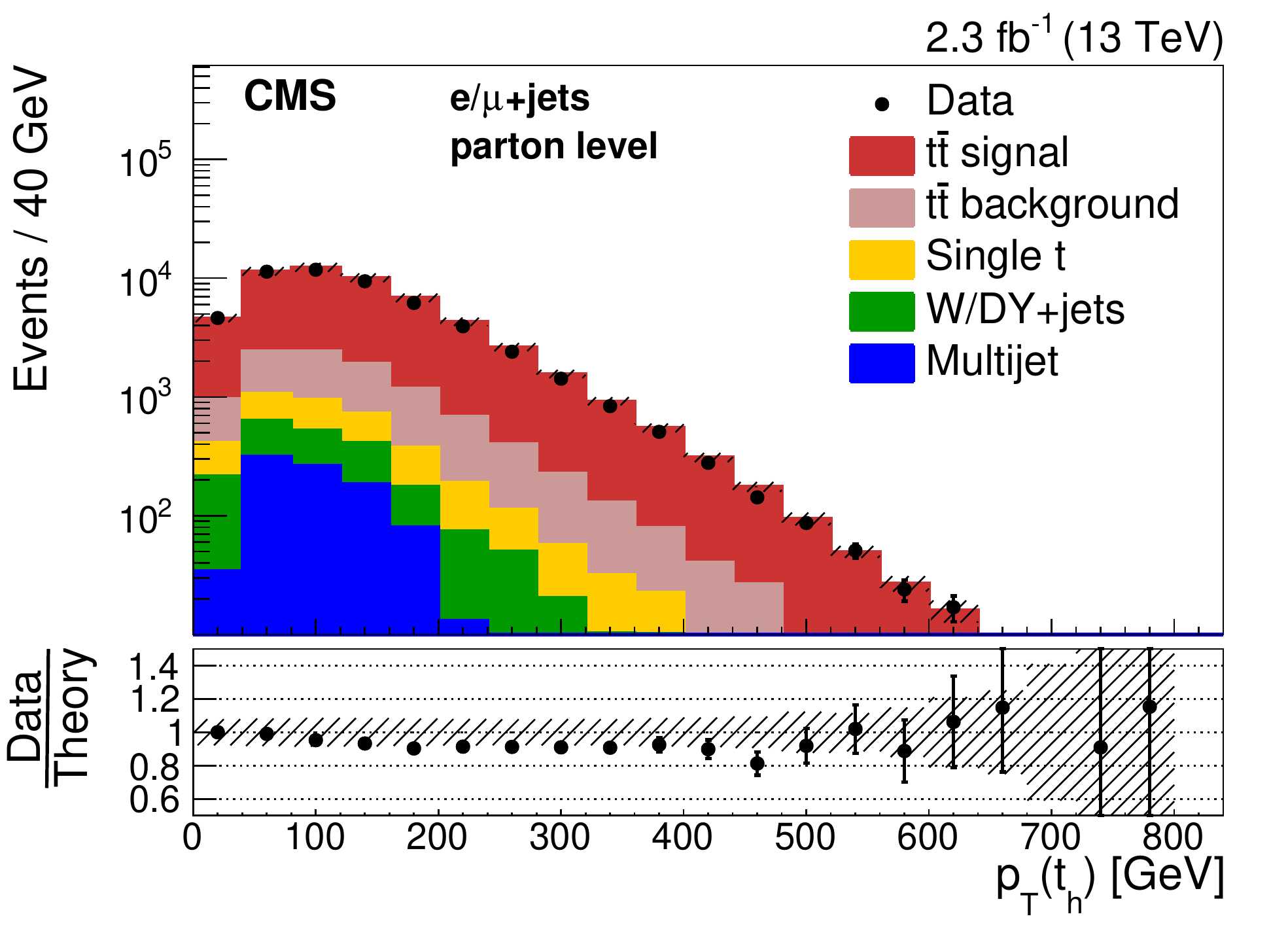}
\includegraphics[width=0.49\textwidth]{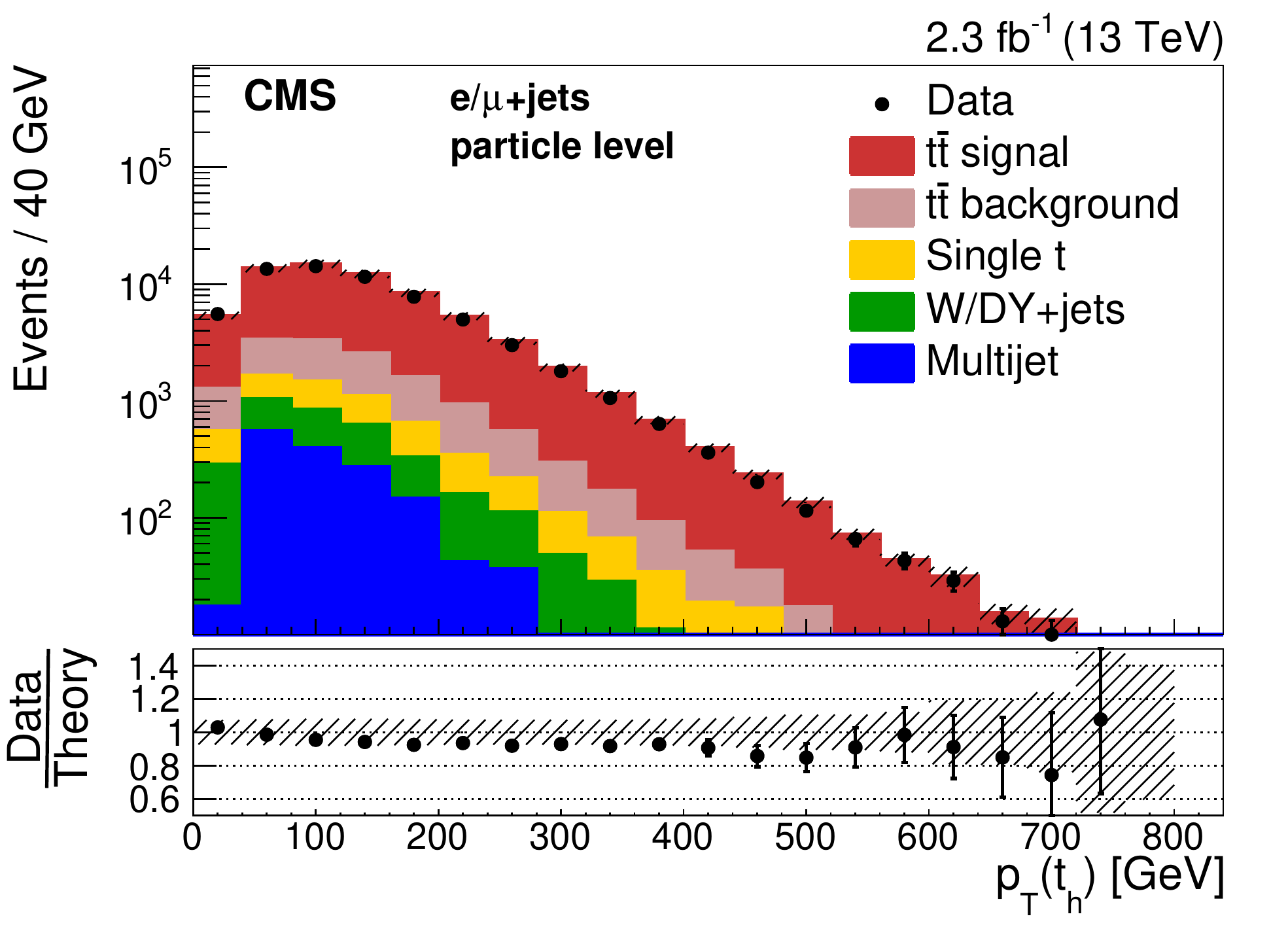}\\
\includegraphics[width=0.49\textwidth]{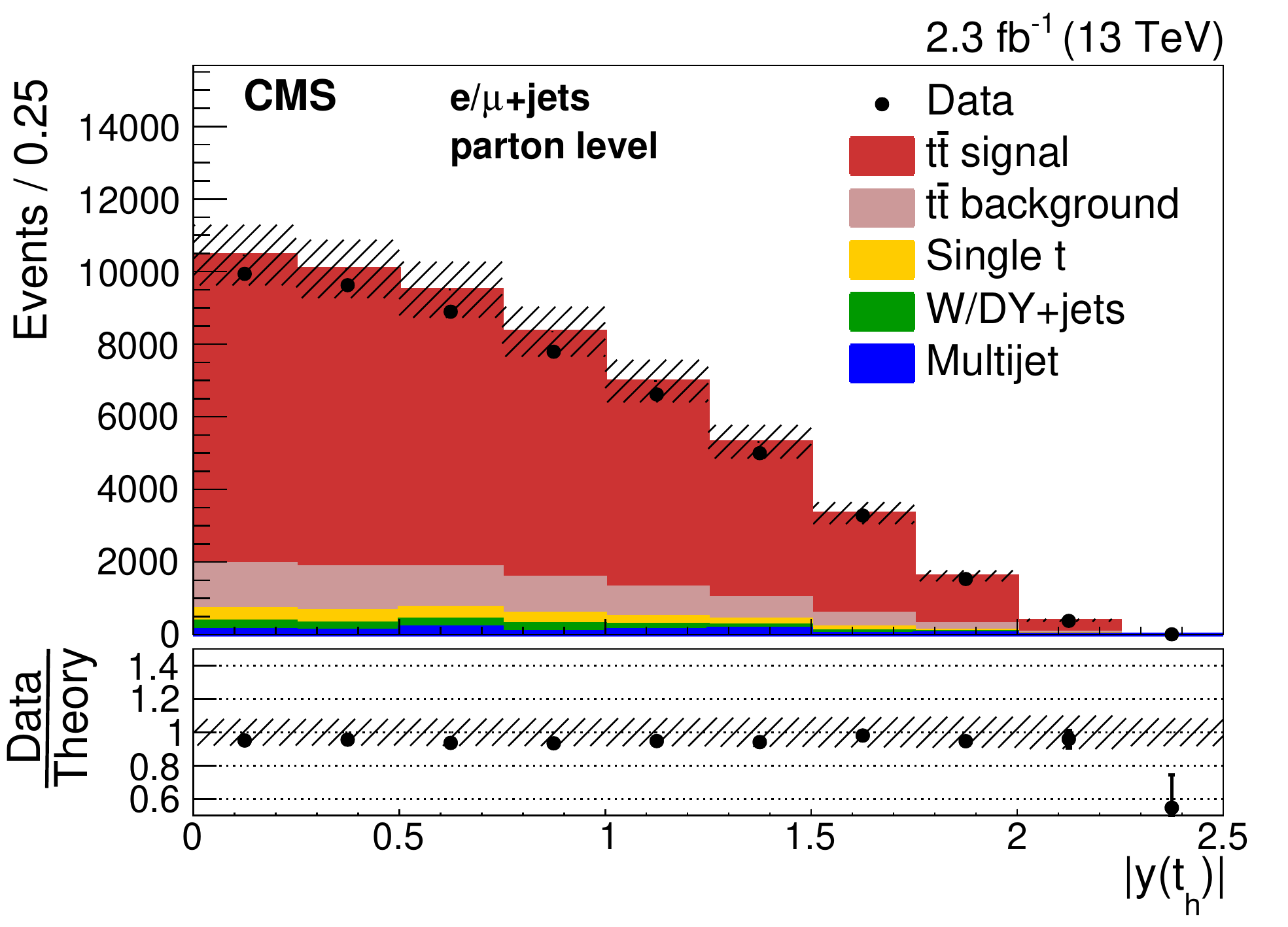}
\includegraphics[width=0.49\textwidth]{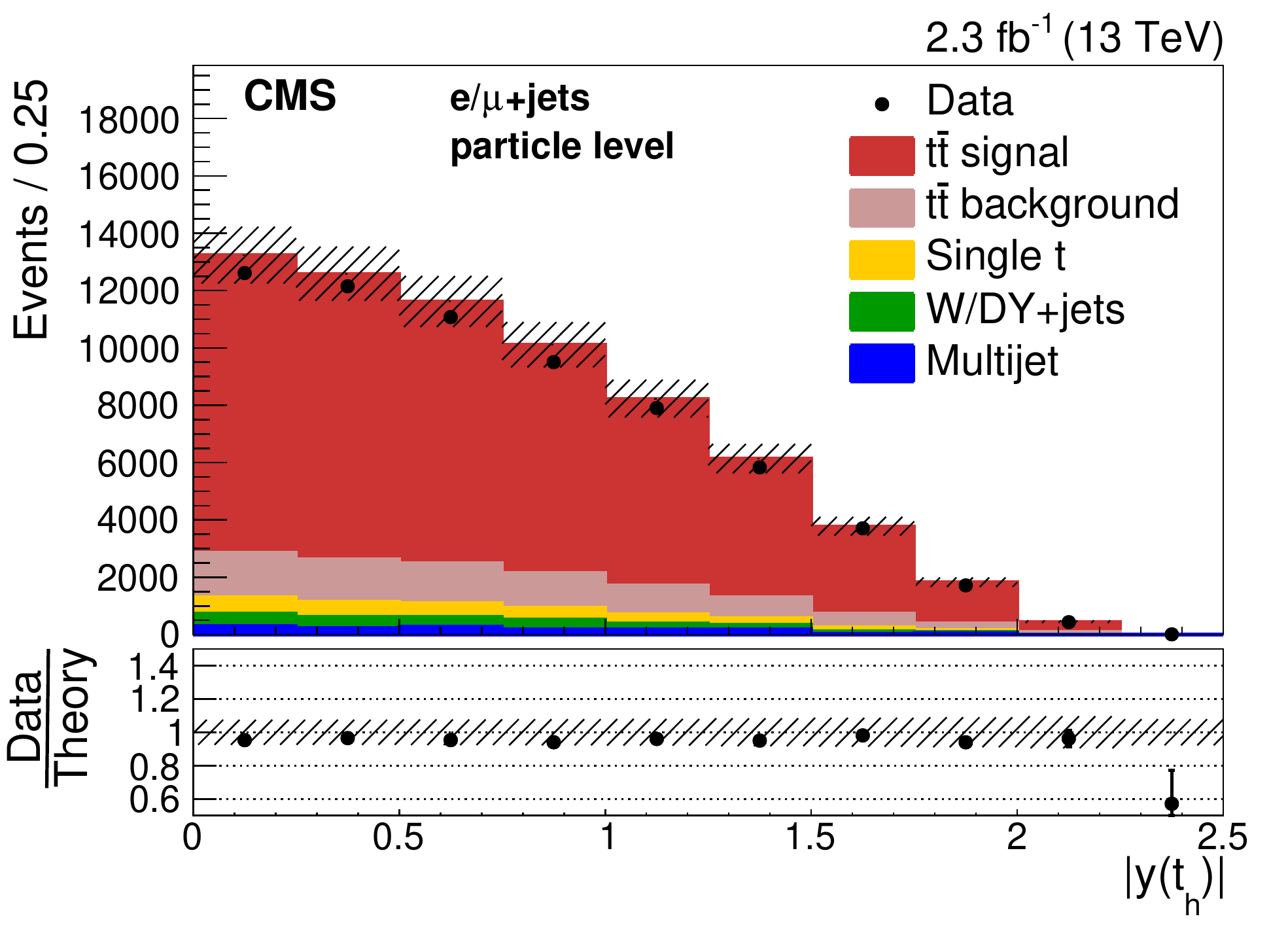}\\
\caption{Comparisons of the reconstructed $\pt(\tqh)$ (top) and $\abs{y(\tqh)}$ (bottom) in data and simulations for the parton (left) and the particle (right) level. The simulation of \POWHEG{}+\PYTHIAA is used to describe the \ttbar production. Experimental (cf. Section~\ref{UNC}) and statistical uncertainties (hatched area) are shown for the total simulated yield, which is normalized according to the measured integrated luminosity. The ratios of data to the expected yields are given at the bottom of each panel.
}
\label{TTRECF4a}
\end{figure*}

\begin{figure*}[tbhp]
\centering
\includegraphics[width=0.49\textwidth]{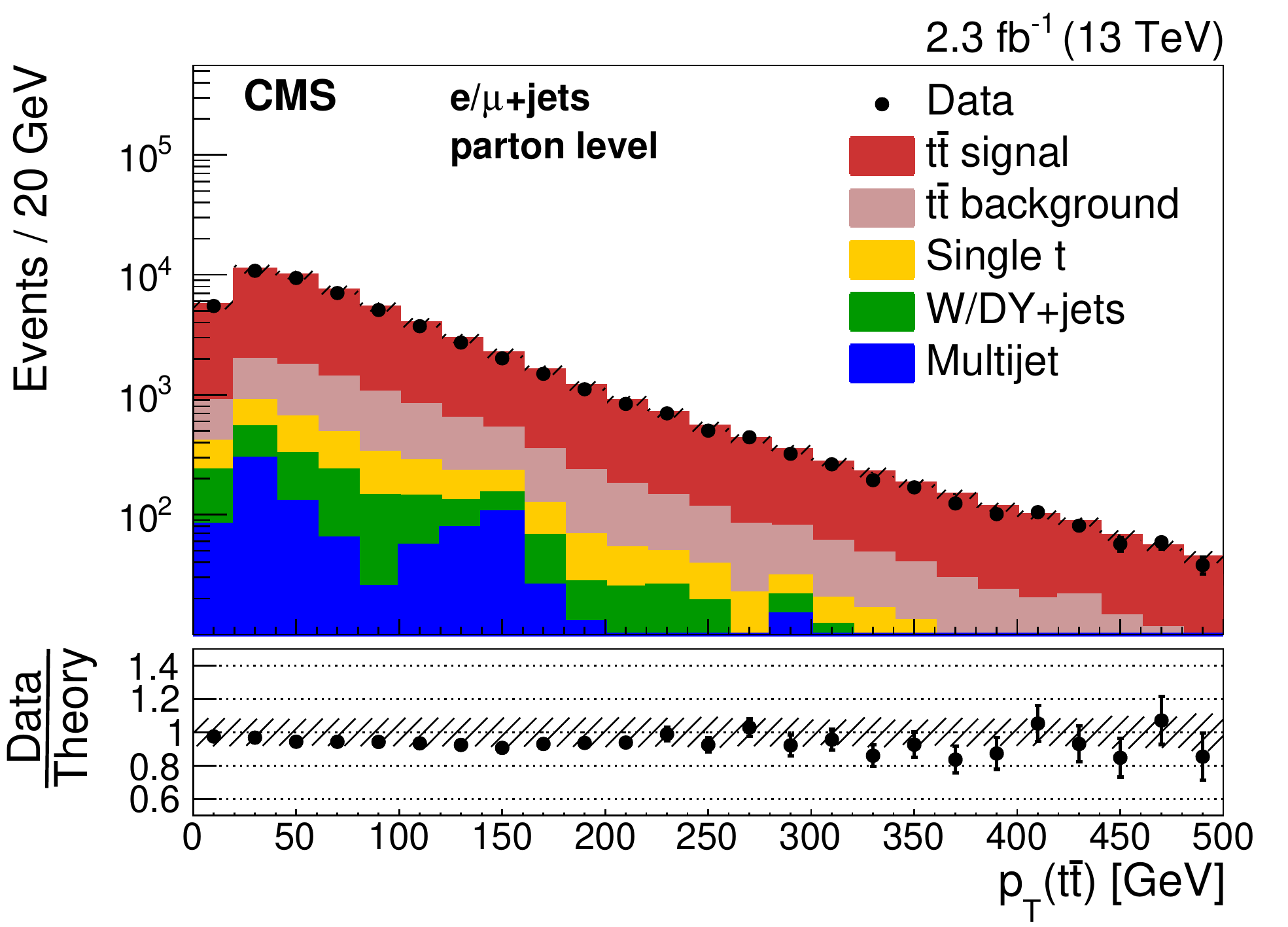}
\includegraphics[width=0.49\textwidth]{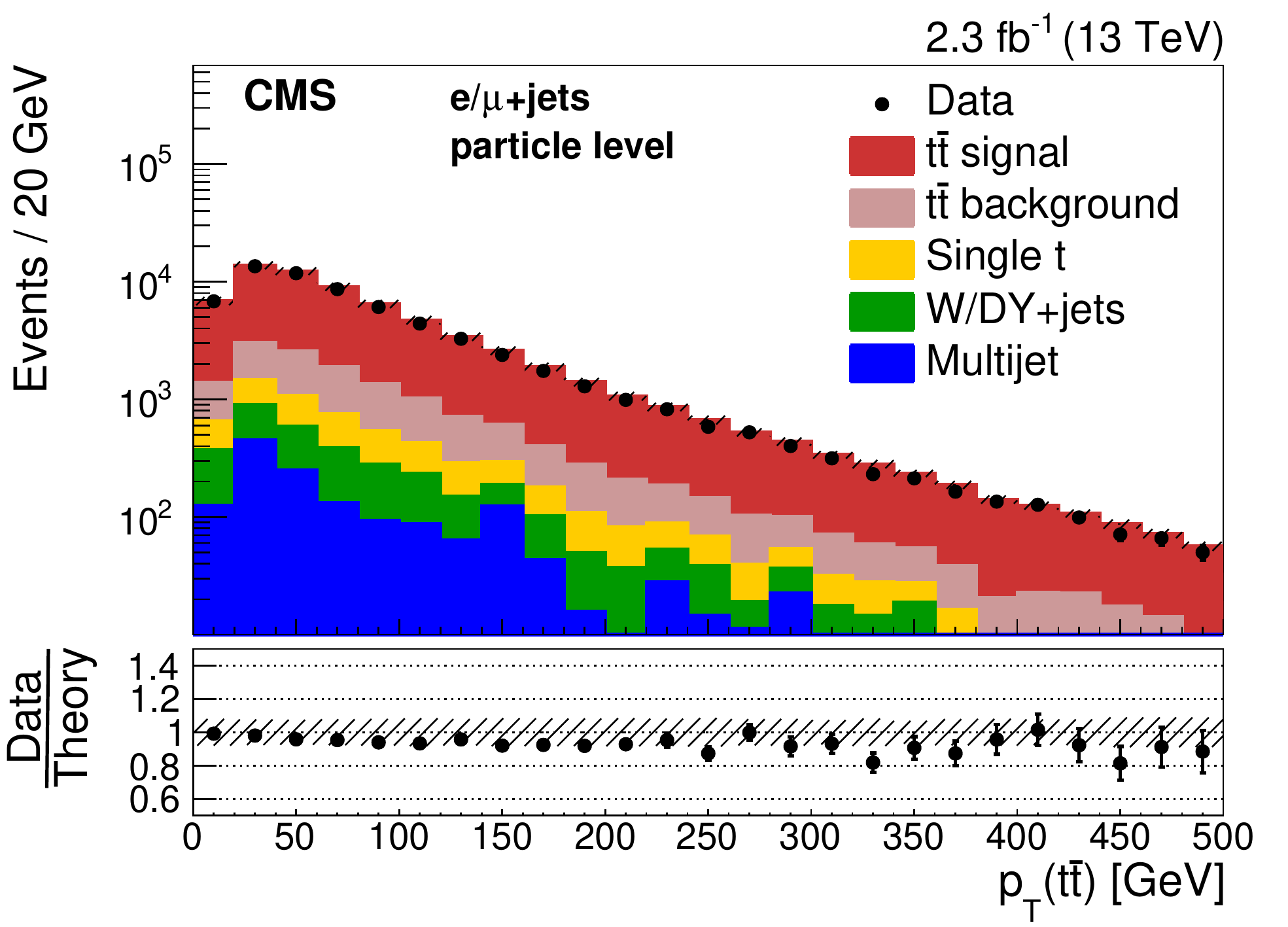}
\includegraphics[width=0.49\textwidth]{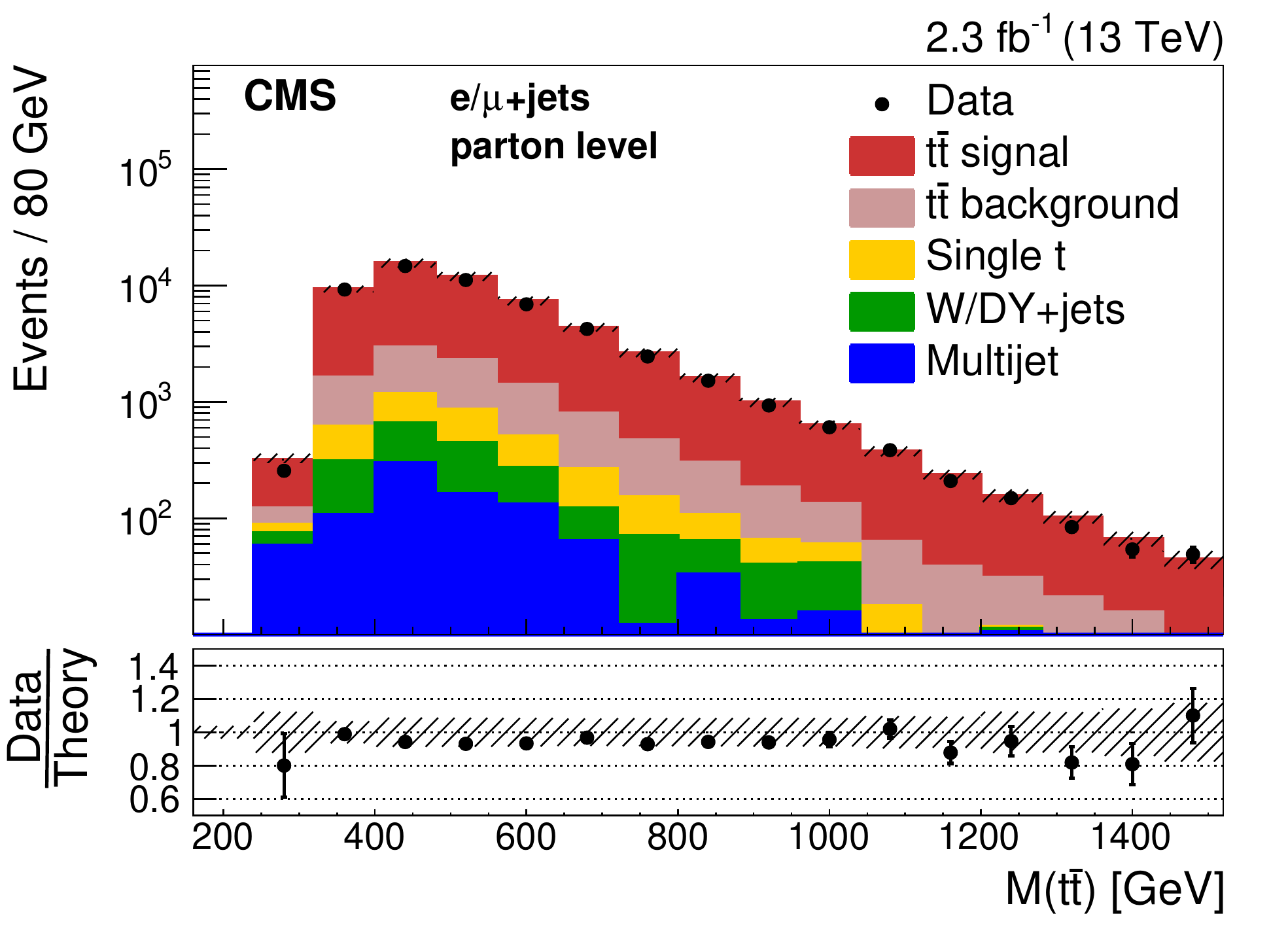}
\includegraphics[width=0.49\textwidth]{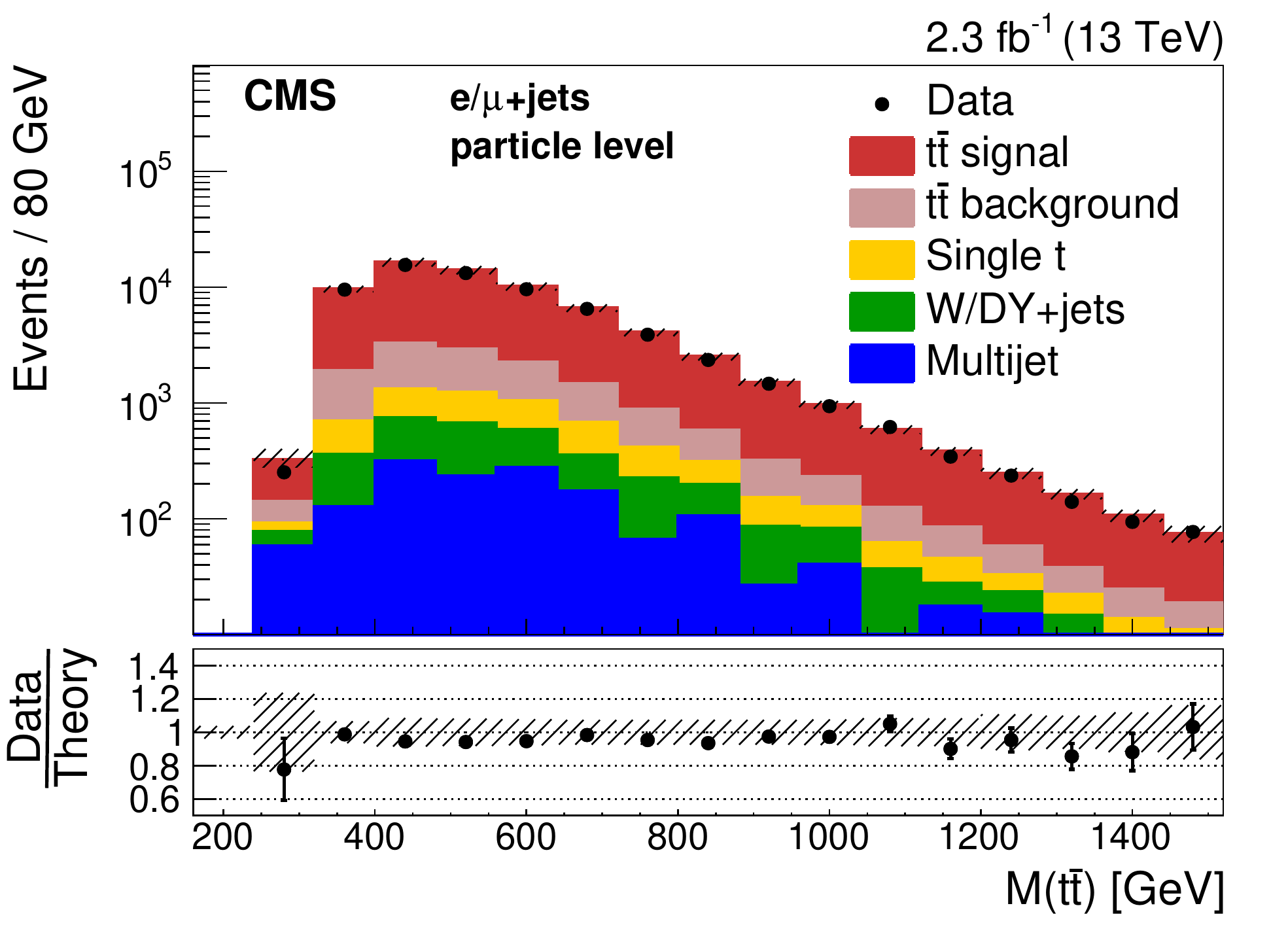}
\includegraphics[width=0.49\textwidth]{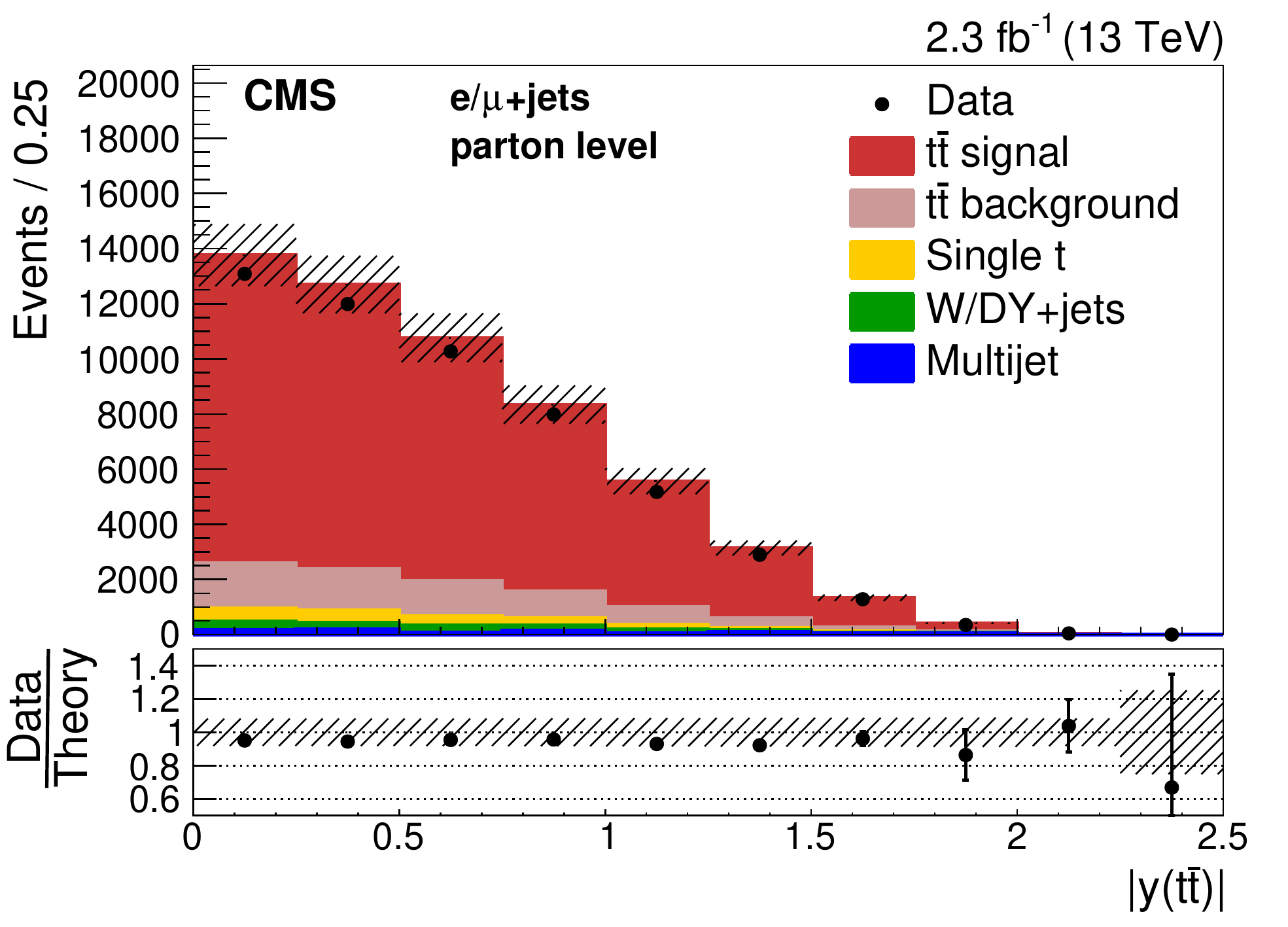}
\includegraphics[width=0.49\textwidth]{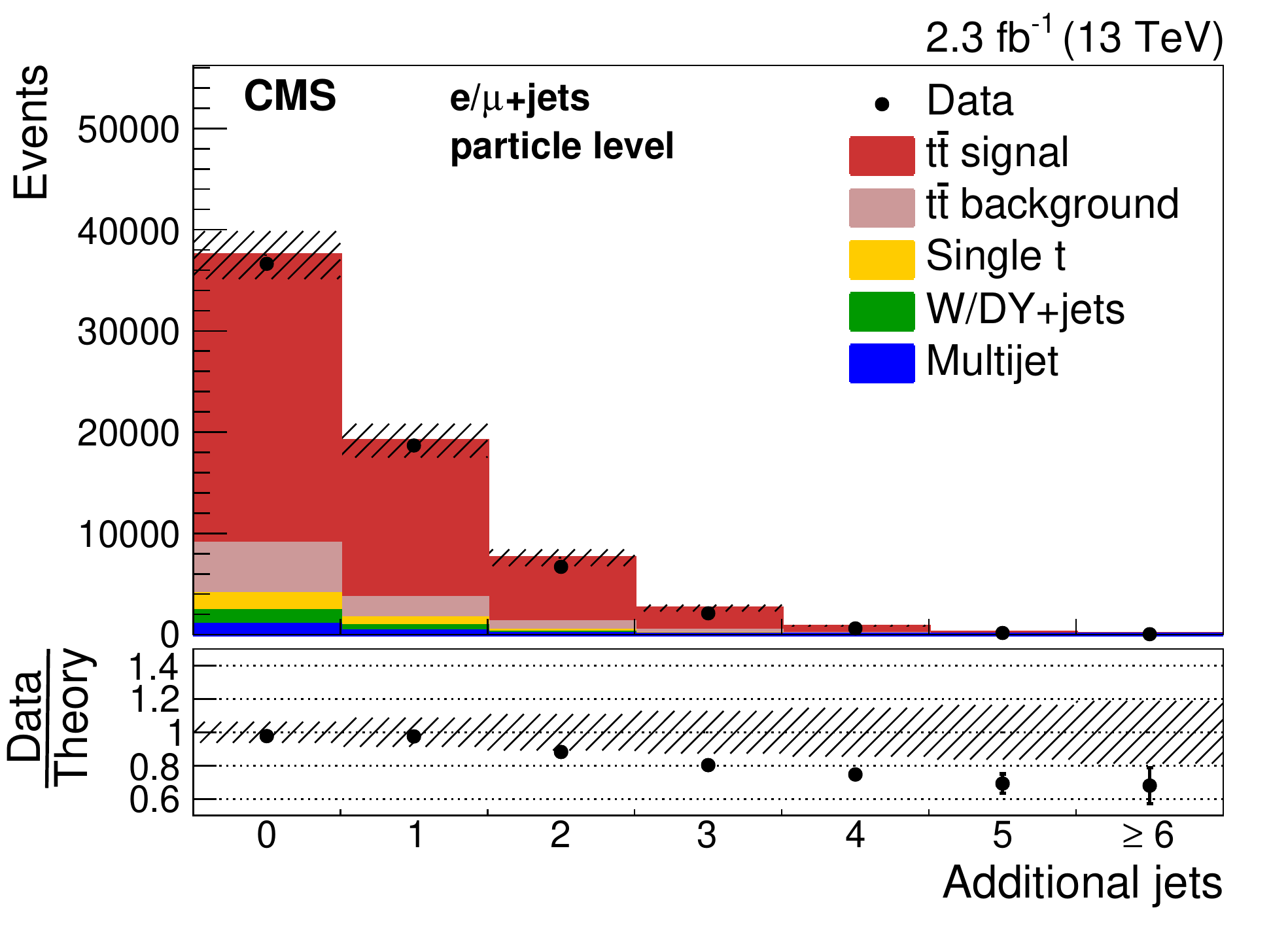}
\caption{Comparisons of the reconstructed distributions of $\pt(\ttbar)$ (top) and $M(\ttbar)$ (middle) for the parton- (left) and the particle- (right) level measurements in data and simulations. Bottom: distributions of $\abs{y(\ttbar)}$ (left) and the number of additional jets (right). The simulation of \POWHEG{}+\PYTHIAA is used to describe the \ttbar production. Experimental (cf. Section~\ref{UNC}) and statistical uncertainties (hatched area) are shown for the total simulated yield, which is normalized according to the measured integrated luminosity. The ratios of data to the expected yields are given at the bottom of each panel.}
\label{TTRECF4b}
\end{figure*}

In \FIG{TTRECF4a} the distributions of \pt and $\abs{y}$ of the reconstructed hadronically decaying top quarks for the parton- and particle-level measurements are compared to the simulation. In \FIG{TTRECF4b} the distributions of $\pt(\ttbar)$, $\abs{y(\ttbar)}$, $M(\ttbar)$, and the number of additional jets are shown. In general, good agreement is observed between the data and the simulation though the overall yield in the data is slightly lower, but within the experimental uncertainties. The observed jet multiplicities are lower than predicted.

\section{Background subtraction}
\label{BKG}
After the event selection and \ttbar reconstruction about 65\,000 (53\,000) events are observed in the particle- (parton-) level measurements. A small contribution of about 9\% of single top quark, DY, \W boson, and multijet events is expected. These have to be estimated and subtracted from the selected data.

The background from single top quark production is subtracted based on its simulation. Its overall contribution corresponds to about 4\% of the selected data. Single top quark production cross sections are calculated with precisions of a few percent~\cite{Kant:2014oha,Kidonakis:2012rm}. Since the calculations have a limited reliability after \ttbar selection we assume an overall uncertainty of 50\%. However, this conservative estimate has negligible impact on the final results and their accuracy.

The simulations of multijet, DY, and \W boson production contain limited numbers of events after the full selection. We extract the shapes of the distributions of these backgrounds from a control region in the data, similar to the signal region, but requiring no b-tagged jet in the event. In this selection the contribution of \ttbar events is estimated to be about 15\%. The remaining fraction consists of multijet, DY, and \W boson events. The reconstruction algorithm is exactly the same as for the signal selection. To estimate the shape dependency in the control region on the selection we vary the selection threshold of the $\PQb$ tagging discriminant. This changes the top quark contribution and the flavor composition, however, we find the observed shape variation to be negligible. For the background subtraction, the distributions extracted from the control region are normalized to the yield of multijet, DY, and \W boson events predicted by the simulation in the signal region. In the control region the expected and measured event yields agree within their statistical uncertainties. Taking into account the statistical uncertainty of the normalization factor and the shape differences between the signal and control regions in the simulation, we estimate an overall uncertainty of 20\% in this background estimation. The overall contribution to the selected data is about 5\%.

For the parton-level measurement, special care has to be taken with the contribution of nonsignal \ttbar events, i.e.,  dilepton, all-jets, and $\tau$+jets events. For the particle-level measurement care is needed with all \ttbar events for which no pair of particle-level top quarks exists. The behavior of this background depends on the \ttbar cross section and a subtraction according to the expected value can result in a bias of the measurement, especially if large differences between the simulation and the data are observed. However, the shapes of the distributions show an agreement within uncertainties between data and simulation and we subtract the predicted relative fractions from the remaining event yields.

\section{Unfolding}
\label{UNFO}

For the unfolding, the iterative D'Agostini method~\cite{D'Agostini:1994zf} is used. The migration matrices and the acceptances are needed as input. The migration matrix relates the quantities at particle (parton) level and at detector level. It accounts for the effects from the parton shower and hadronization as well as the detector response, where the former has a large impact on the parton-level measurement. For the central results the migration matrices and the acceptances are taken from the \POWHEG{}+\PYTHIAA simulation and other simulations are used to estimate the uncertainties. The binning in the unfolding is optimized based on the resolution in the simulation. We utilize for the minimal bin widths that, according to the resolution, at least 50\% of the events are reconstructed in the correct bin.

The iterative D'Agostini method takes the number of iterations as an input parameter to control the level of regularization. A small number of iterations corresponds to a large regularization, which may bias the unfolded results. The level of regularization and hence the bias decreases with the number of iterations -- but with the drawback of increasing variances in the unfolded spectra. To optimize the number of iterations, we chose the criterion that the compatibility between a model and the unfolded data at particle (parton) level is the same as the compatibility between the folded model and the data at detector level. The compatibilities are determined by $\chi^2$ tests at both levels based on all available simulations and several modified spectra obtained by reweighting the $\pt(\PQt)$, $\abs{y(\PQt)}$, or $\pt(\ttbar)$ distributions in the \POWHEG{}+\PYTHIAA simulation. The reweighted spectra are chosen in such a way that they cover the observed differences between the data and the unmodified simulation.

We find the above criterion fulfilled for the number of iterations such that a second $\chi^2$ test between the detector-level spectrum with its statistical uncertainty and the refolded spectrum exceeds a probability of 99.9\%. The refolded spectrum is obtained by inverting the unfolding step. This consists of a multiplication with the response matrix and does not need any regularization.

For the two-dimensional measurements with $n$ bins in one and $m$ bins in the other quantity the D'Agostini unfolding can be generalized using a vector of $n\cdot m$ entries of the form: ${b_{1,1},b_{2,1}\ldots b_{n,1},\ldots b_{1,m},b_{2,m}\ldots b_{n,m}}$ with a corresponding $(n\cdot m) \times (n\cdot m)$ migration matrix. The number of iterations is optimized in the same way.

\section{Systematic uncertainties}
\label{UNC}
We study several sources of experimental and theoretical uncertainty. Uncertainties in the jet and \ptvecmiss calibrations, in the pileup modeling, in the $\PQb$ tagging and lepton selection efficiencies, and in the integrated luminosity measurement fall into the first category.

Uncertainties in the jet energy calibration are estimated by shifting the energies of jets in the simulation up and down by their \pt- and $\eta$-dependent uncertainties of 3--7\%~\cite{JET}. At the same time \ptvecmiss is recalculated according to the rescaled jet energies. The recomputed backgrounds, response matrices, and acceptances are used to unfold the data. The observed differences between these and the original results are taken as an uncertainty in the unfolded event yields. The same technique is used to calculate the impact of the uncertainties in the jet energy resolution, the uncertainty in \ptvecmiss not related to the jet energy calibration, in the $\PQb$ tagging, and in the pileup modeling.

The $\PQb$ tagging efficiency in the simulation is corrected using scale factors determined from the data~\cite{BTV}. These have an uncertainty of about 2--5\% depending on the \pt of the $\PQb$ jet.

The effect on the measurement due to the uncertainty in the modeling of pileup in the simulation is estimated by varying the average number of pileup events per bunch crossing by 5\% and  reweighting the simulated events accordingly.

The trigger, reconstruction, and identification efficiencies of leptons are evaluated with tag-and-probe techniques using \Z boson dilepton decays~\cite{TNPREF}. The uncertainties in the scale factors, which are used to correct the simulation to match the data, take into account the different lepton selection efficiencies in events with high jet multiplicities. The overall uncertainty in the lepton reconstruction and selection efficiencies is 3\%.

The relative uncertainty in the integrated luminosity measurement is 2.3\%~\cite{LUMI}.

Uncertainties in the PDFs, the choice of factorization and renormalization scales, the modeling of the parton shower and hadronization, the effect of different NLO event generation methods, and the top quark mass fall into the second category of theoretical uncertainties.

The effects of these uncertainties are estimated either by using the various event weights introduced in Section~\ref{SIM}, e.g., in the case of PDFs, factorization scale, and renormalization scale, or by using a different \ttbar signal simulation. The \POWHEG simulation combined with \HERWIGpp is used to estimate the effect of different parton shower and hadronization models. In addition, \POWHEG{}+\PYTHIAA samples with a parton shower scale varied by a factor of two are used to study the parton shower modeling uncertainties. The result obtained with \AMCATNLO is used to estimate the effect of different NLO event generation methods. The effect due to uncertainties in the top quark mass is estimated using simulations with altered top quark masses. We quote as uncertainty the cross section differences observed for a top quark mass variation of 1\GeV around the central value of 172.5\GeV used in the central simulation.

The background predictions, response matrices, and acceptances obtained from these simulations are used to unfold the data. The observed deviations with respect to the original result are quoted as an uncertainty in the unfolded event yield.

For the PDF uncertainty only the variation in the acceptance is taken into account while variations due to migrations between bins are neglected. It is calculated according to the uncertainties in the NNPDF30\_nlo\_as\_0118~\cite{Ball:2014uwa} parametrization. In addition, the uncertainties obtained using the PDF sets derived with varied values of the strong coupling constant $\alpha_\mathrm{s} = 0.117$ and 0.119 are considered.

An overview of the uncertainties in the differential cross sections is provided in \TAB{UNCT1}, where the typical ranges of uncertainties in the bins are shown. In the double-differential measurements the jet energy scale uncertainty is about 15\% in bins of high jet multiplicities and the dominant uncertainties due to hadronization modeling and NLO calculation reach up to 30\% for the parton-level measurements.

\begin{table}[tbhp]
 \caption{Overview of the uncertainties in the differential cross section measurements at particle and at parton level. Typical ranges of uncertainties in the bins are shown.}
\centering\begin{scotch}{l|cc}
Source                   &  Particle& Parton\\
                         &  level\,[\%] & level\,[\%]\\\hline
Statistical uncertainty  &  1--5 & 1--5 \\\hline
Jet energy scale         &  5--8 & 6--8 \\
Jet energy resolution    &  $<$1 & $<$1 \\
\ptvecmiss (non jet)           &  $<$1 & $<$1 \\
b tagging                &  2--3 & 2--3 \\
Pileup                   &  $<$1 & $<$1 \\
Lepton selection         &  3 & 3 \\
Luminosity               &  2.3 & 2.3 \\
Background               &  1--3 & 1--3 \\
PDF                      &  $<$1 & $<$1 \\
Fact./ren. scale      &  $<$1 & $<$1 \\
Parton shower scale      &  2--5 & 2--9\\
\POWHEG{}+\PYTHIAA vs. \HERWIGpp  &  1--5 & 1--12\\
NLO event generation            &  1--5 & 1--10\\
\Mtop					&  1--2 & 1--3\\
\end{scotch}
\label{UNCT1}
\end{table}

\section{Cross section results}
\label{RES}
The cross section $\sigma$ in each bin is calculated as the ratio of the unfolded signal yield and the integrated luminosity. These are further divided by the bin width (the product of the two bin widths) to obtain the single- (double-) differential results.

\begin{figure*}[tbhp]
\centering
\includegraphics[width=0.49\textwidth]{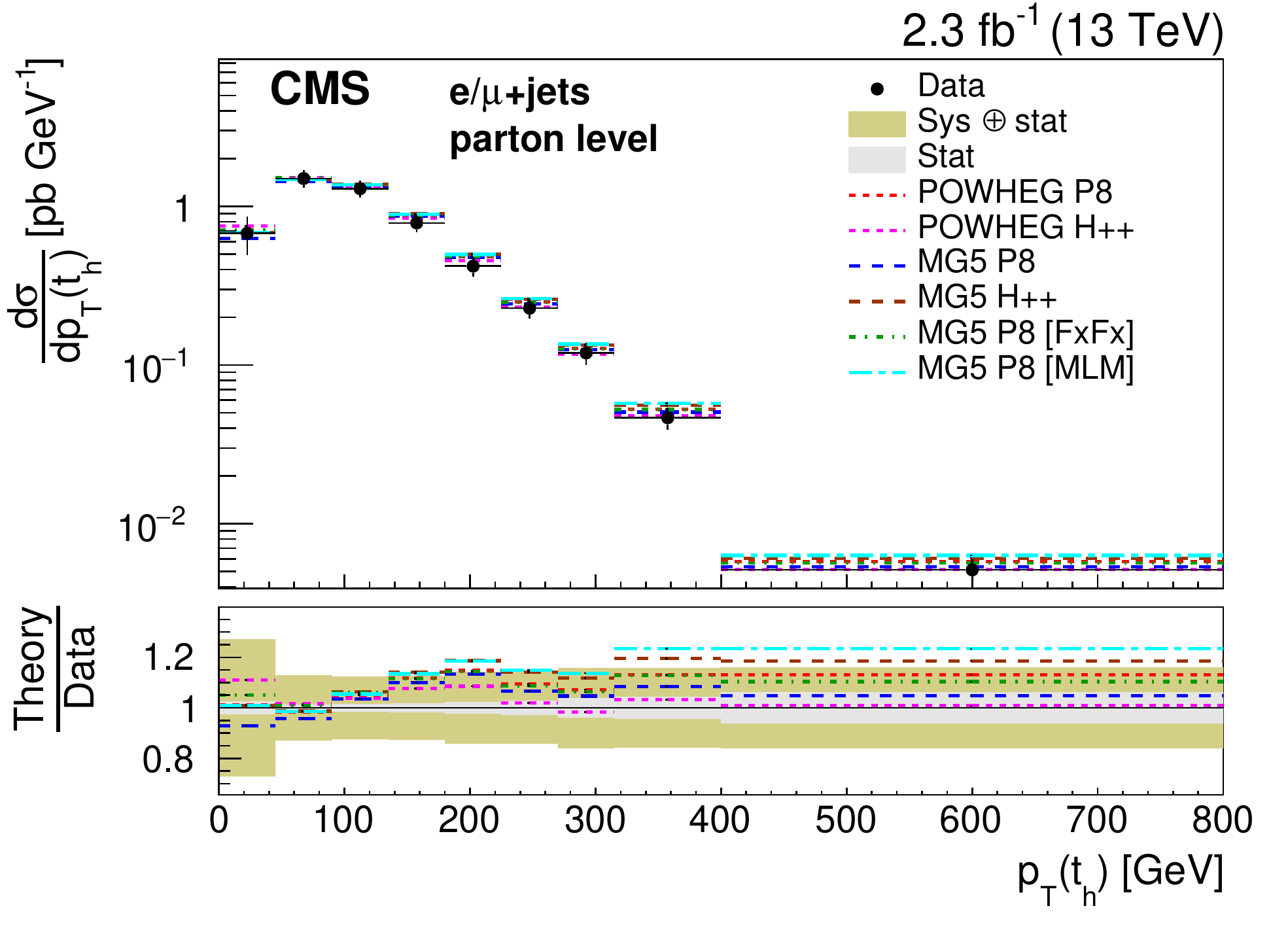}
\includegraphics[width=0.49\textwidth]{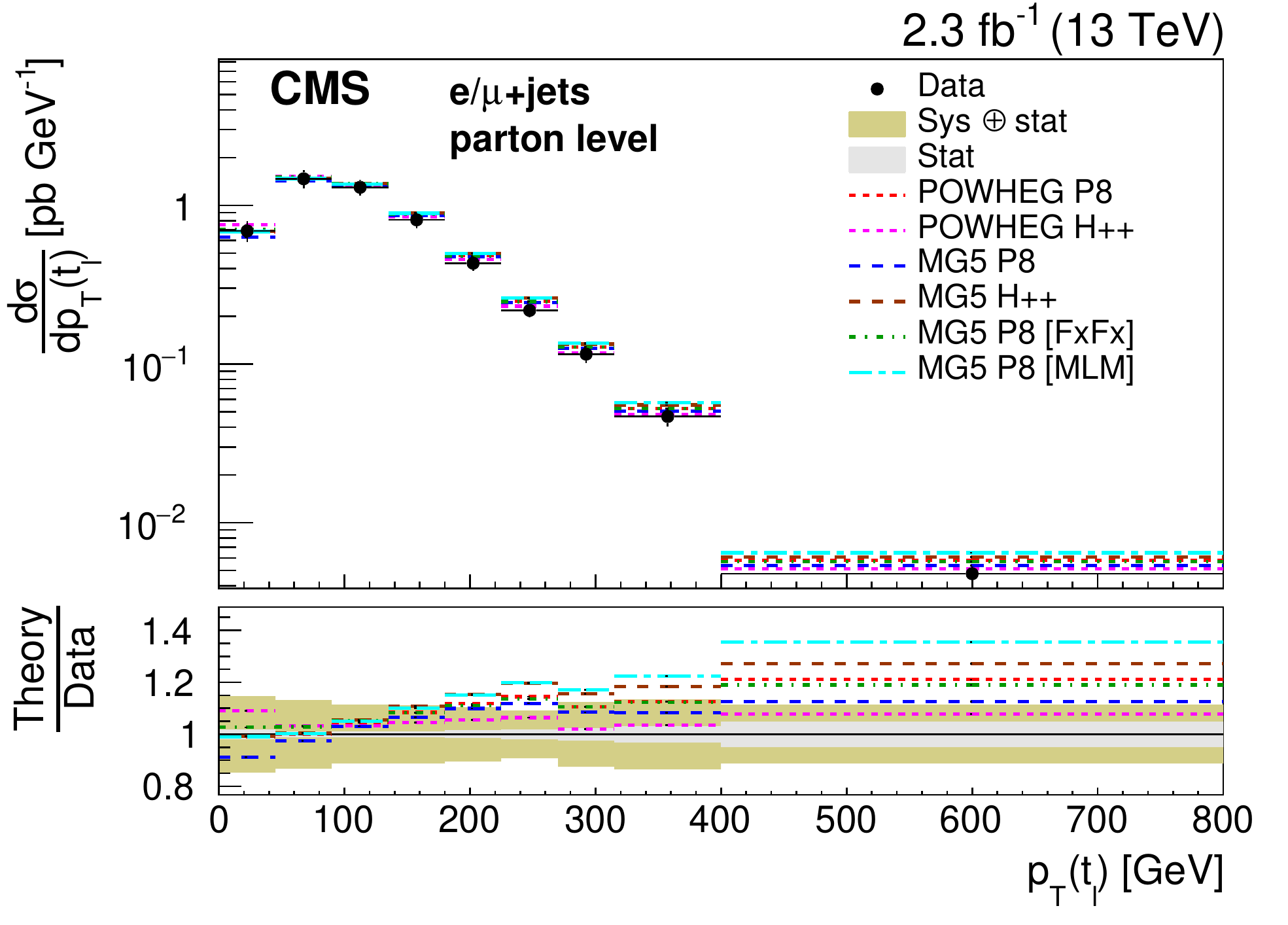}\\
\includegraphics[width=0.49\textwidth]{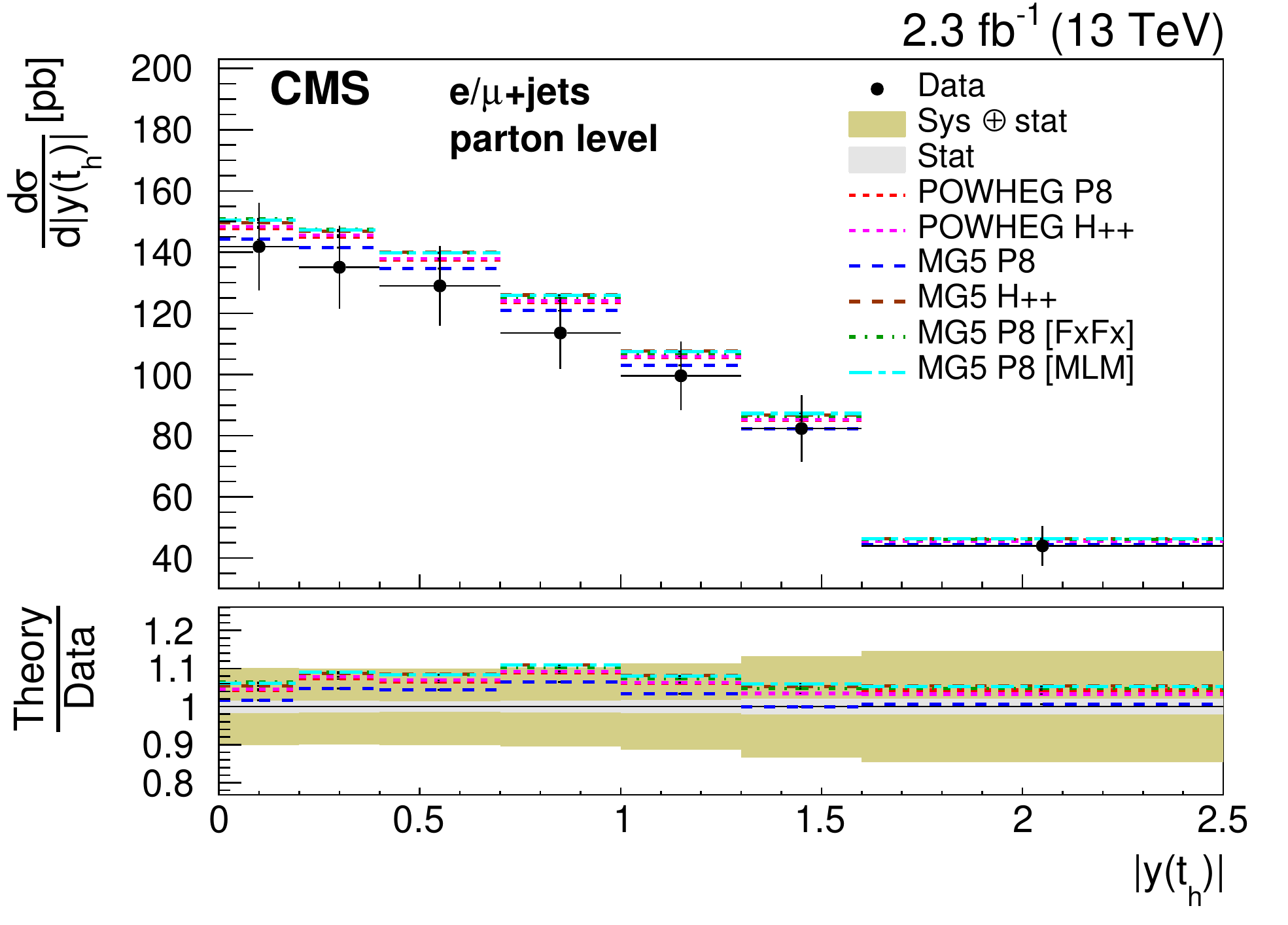}
\includegraphics[width=0.49\textwidth]{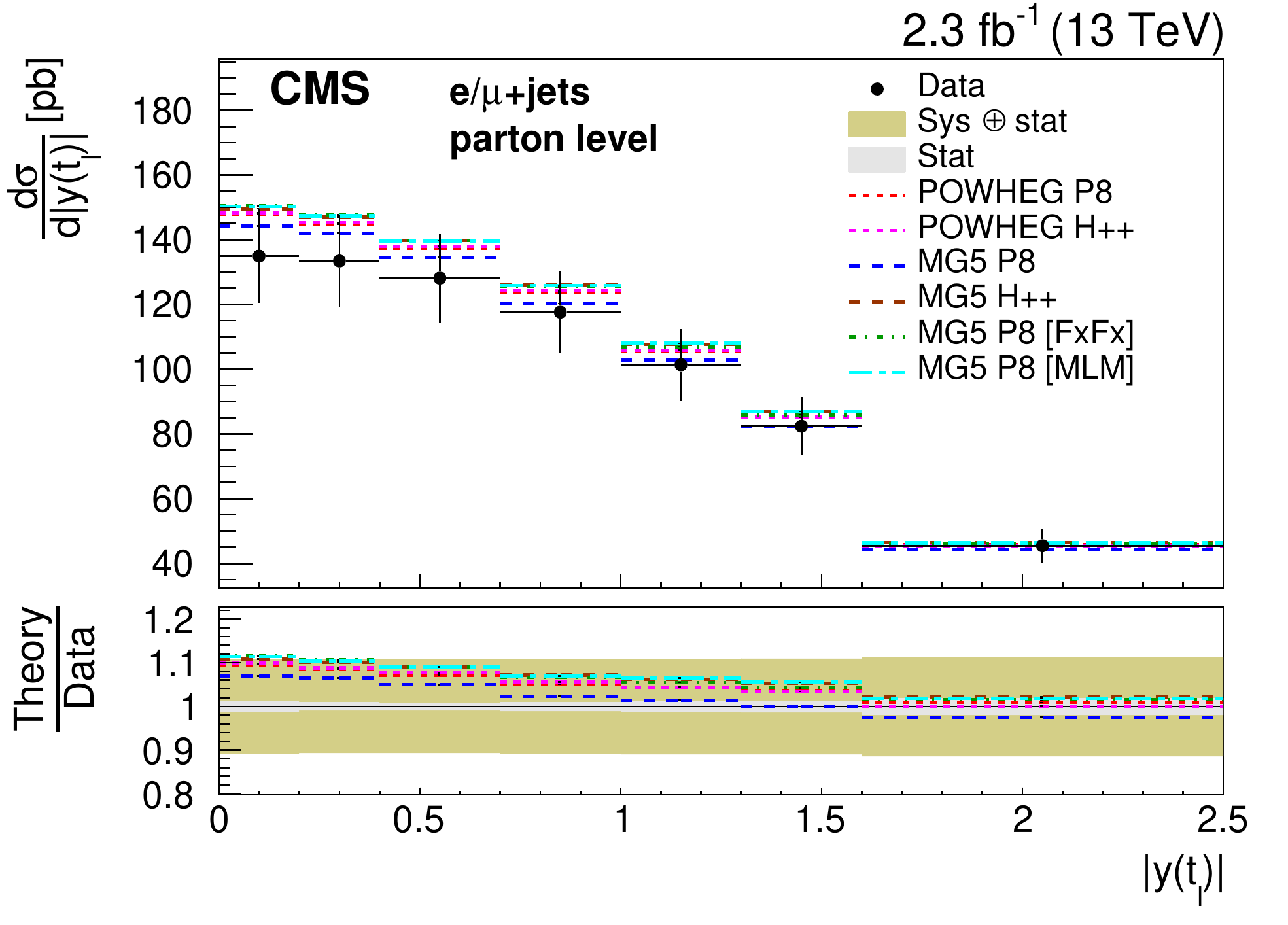}
\caption{Differential cross sections at parton level as a function of $\pt(\PQt)$ (top) and $\abs{y(\PQt)}$ (bottom) measured separately for the hadronically (left) and leptonically (right) decaying top quarks. The cross sections are compared to the predictions of \POWHEG and \AMCATNLO(MG5) combined with \PYTHIAA(P8) or \HERWIGpp(H++) and the multiparton simulations \AMCATNLO{}+\PYTHIAA MLM and \AMCATNLO{}+\PYTHIAA FxFx. The ratios of the various predictions to the measured cross sections are shown at the bottom of each panel together with the statistical and systematic uncertainties of the measurement.}
\label{XSECPA1}
\end{figure*}

\begin{figure*}[tbhp]
\centering
\includegraphics[width=0.49\textwidth]{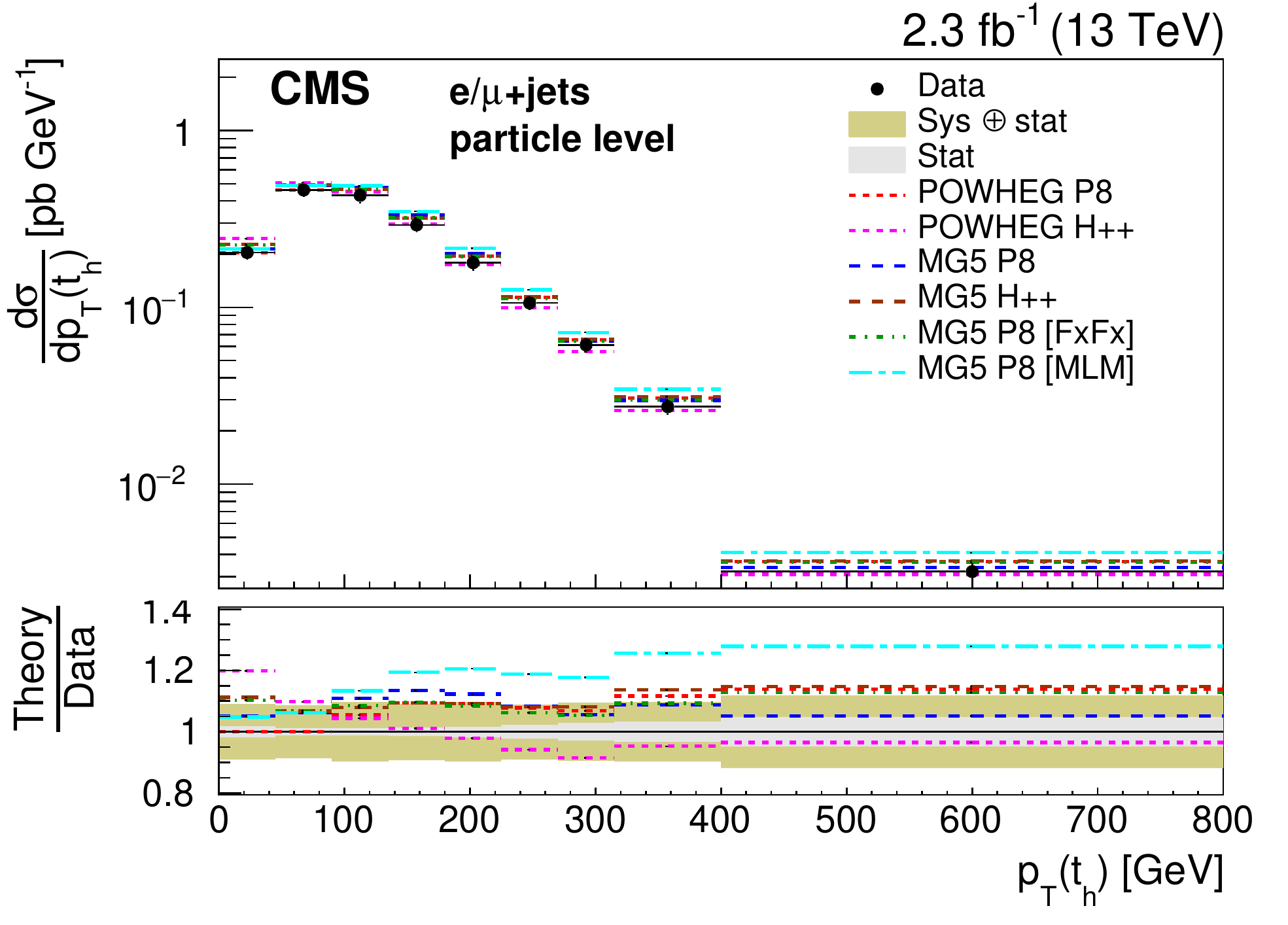}
\includegraphics[width=0.49\textwidth]{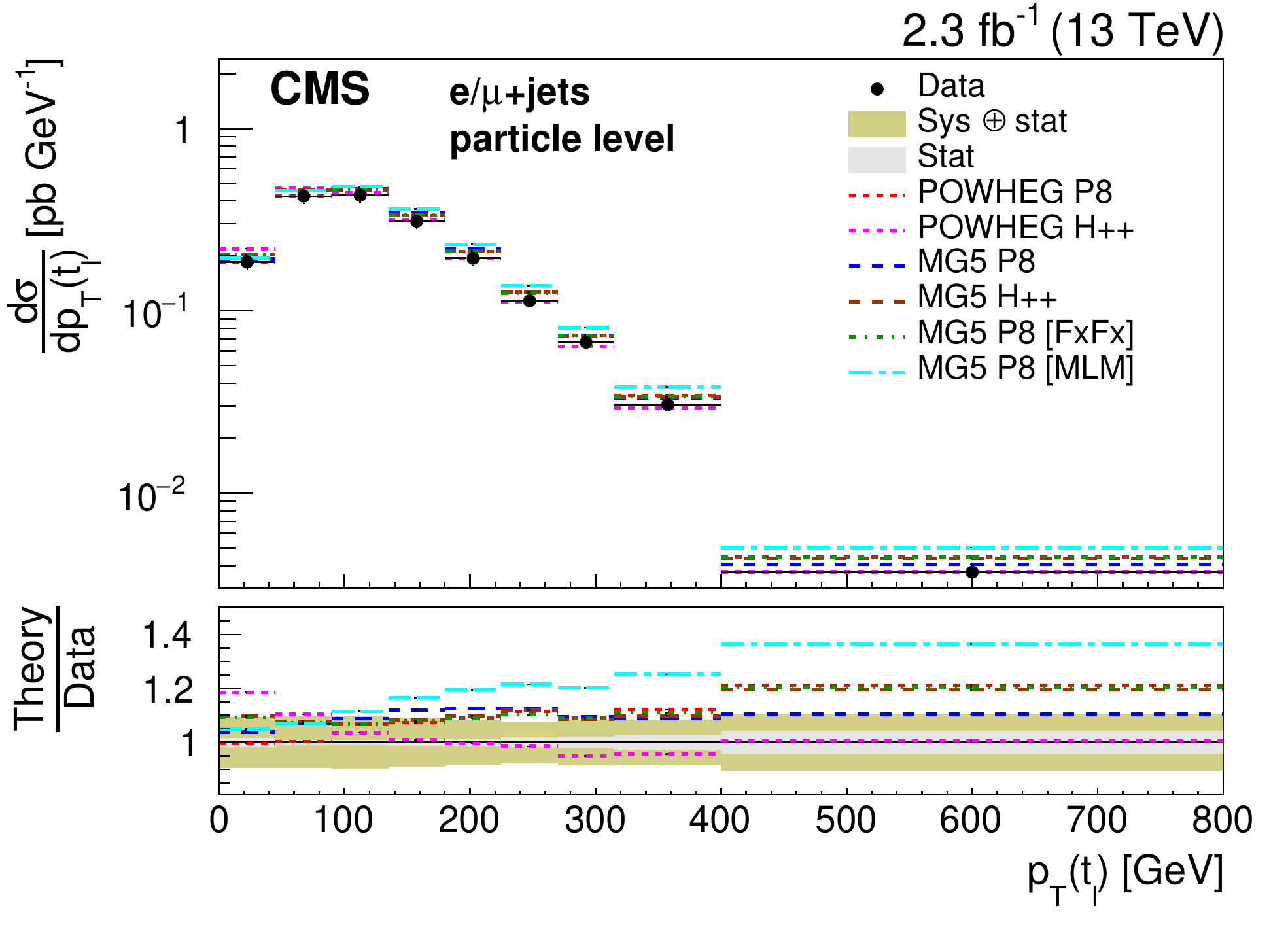}\\
\includegraphics[width=0.49\textwidth]{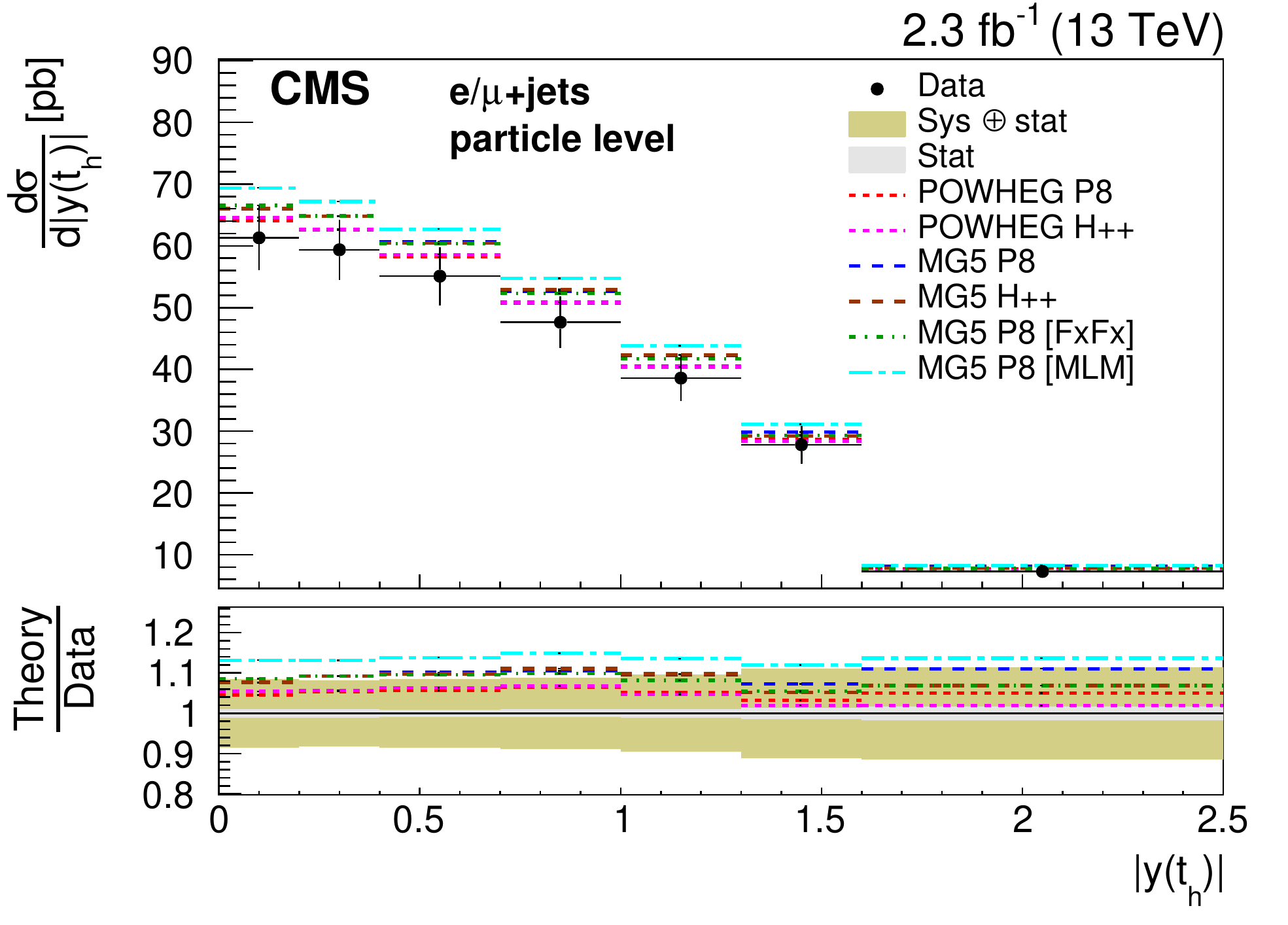}
\includegraphics[width=0.49\textwidth]{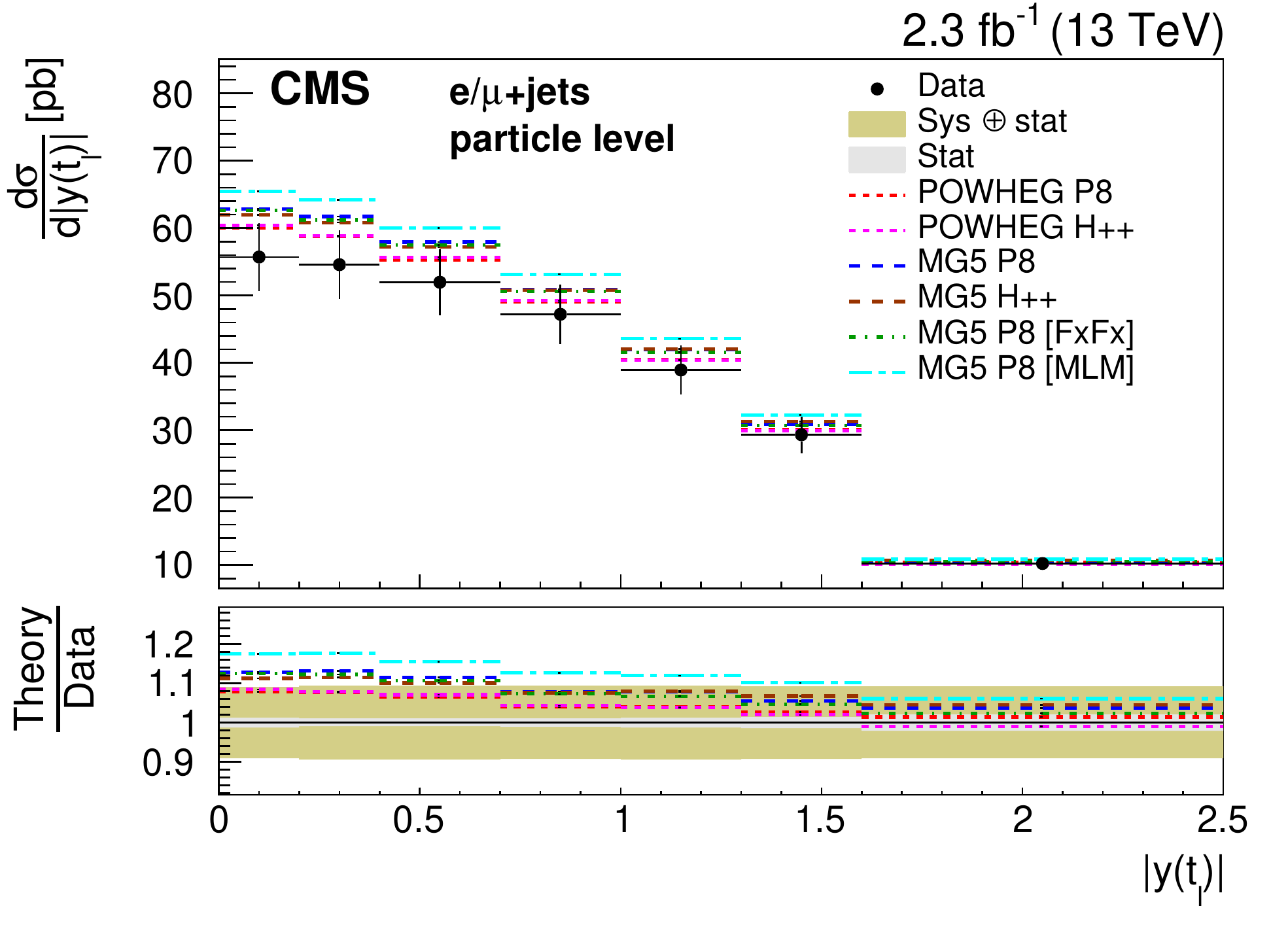}
\caption{Differential cross sections at particle level as a function of $\pt(\PQt)$ (top) and $\abs{y(\PQt)}$ (bottom) measured separately for the hadronically (left) and leptonically (right) decaying particle-level top quarks. The cross sections are compared to the predictions of \POWHEG and \AMCATNLO(MG5) combined with \PYTHIAA(P8) or \HERWIGpp(H++) and the multiparton simulations \AMCATNLO{}+\PYTHIAA MLM and \AMCATNLO{}+\PYTHIAA FxFx. The ratios of the various predictions to the measured cross sections are shown at the bottom of each panel together with the statistical and systematic uncertainties of the measurement.}
\label{XSECPS1}
\end{figure*}

\begin{figure*}[tbhp]
\centering
\includegraphics[width=0.49\textwidth]{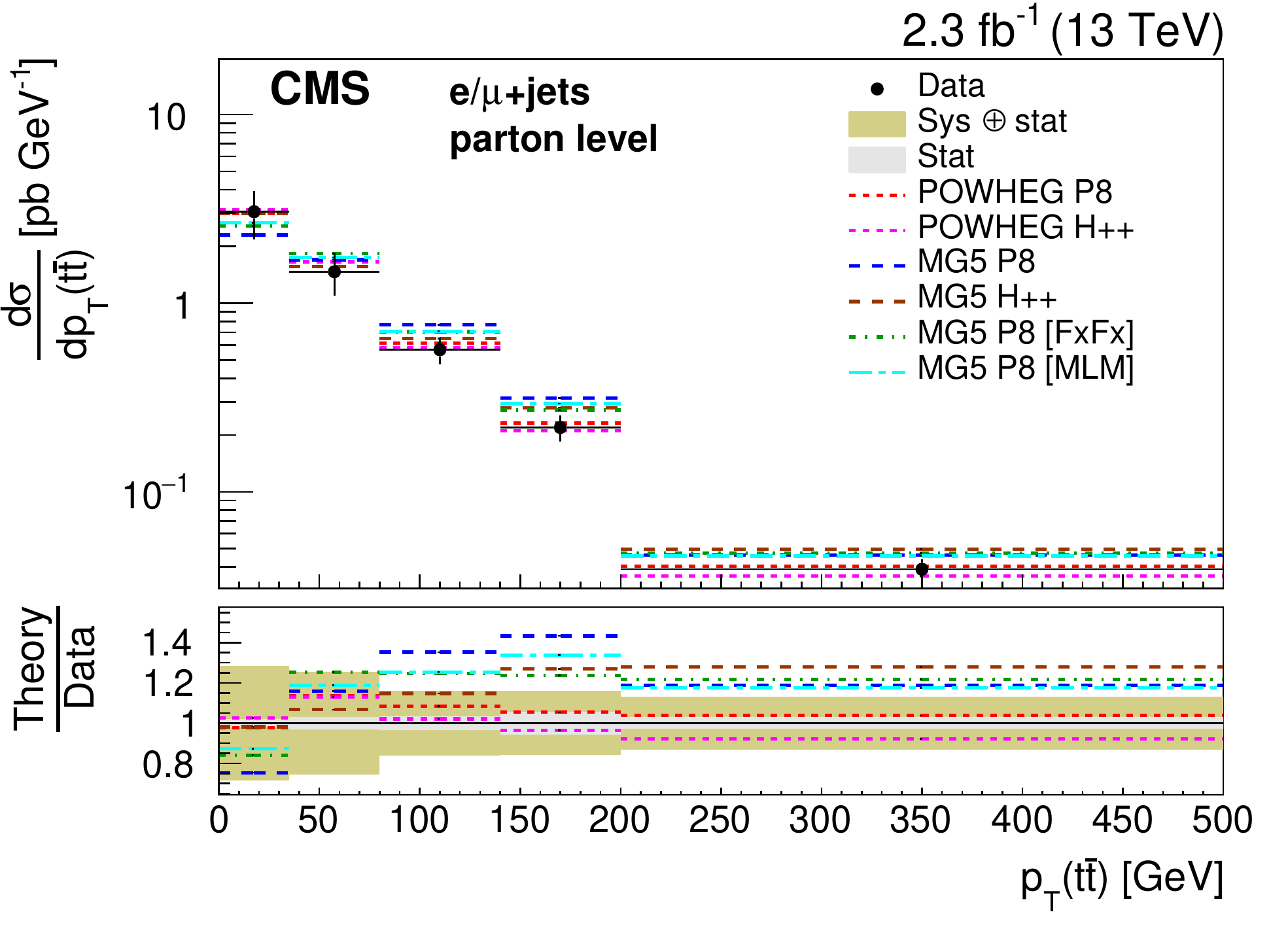}
\includegraphics[width=0.49\textwidth]{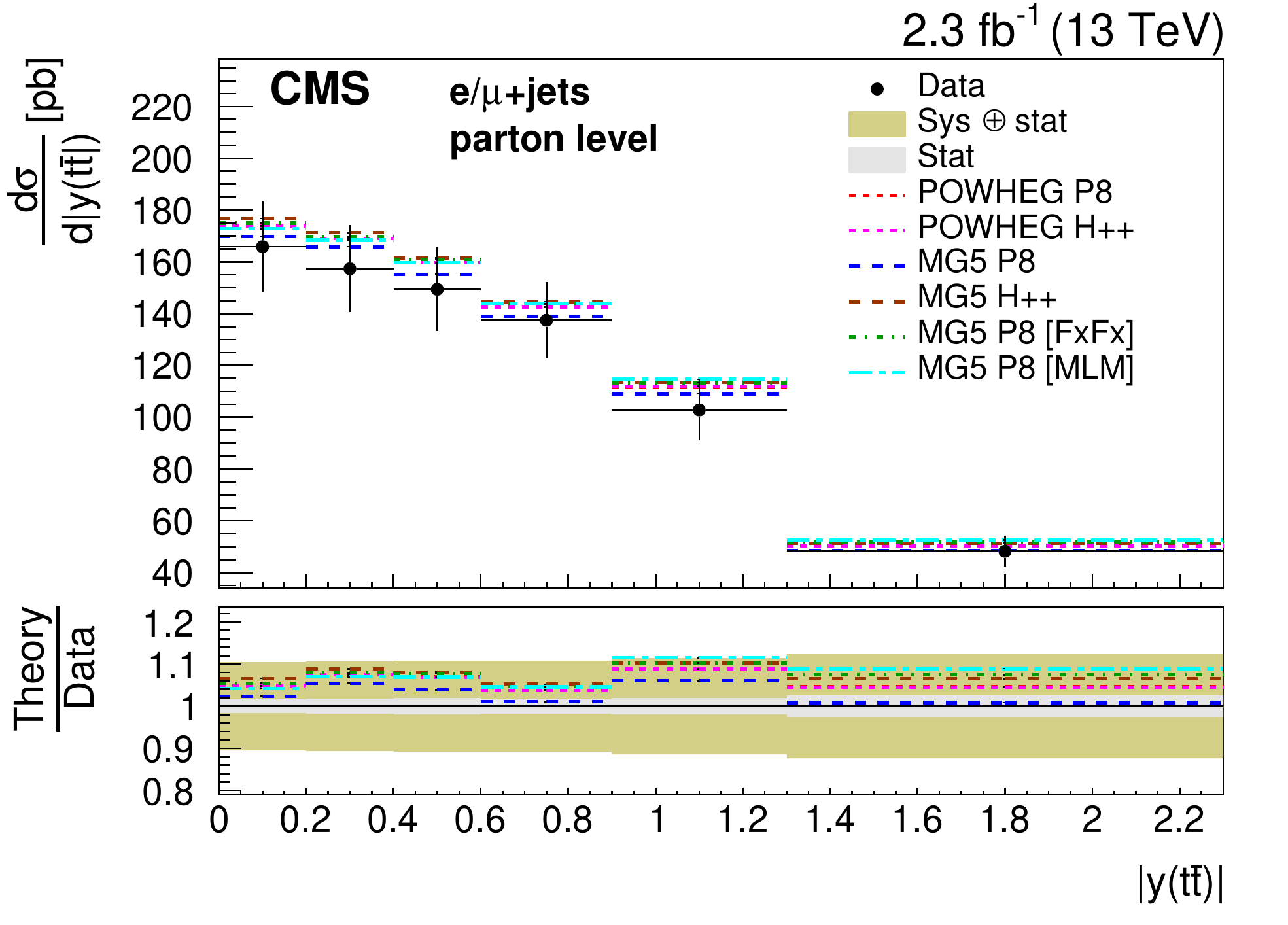}\\
\includegraphics[width=0.49\textwidth]{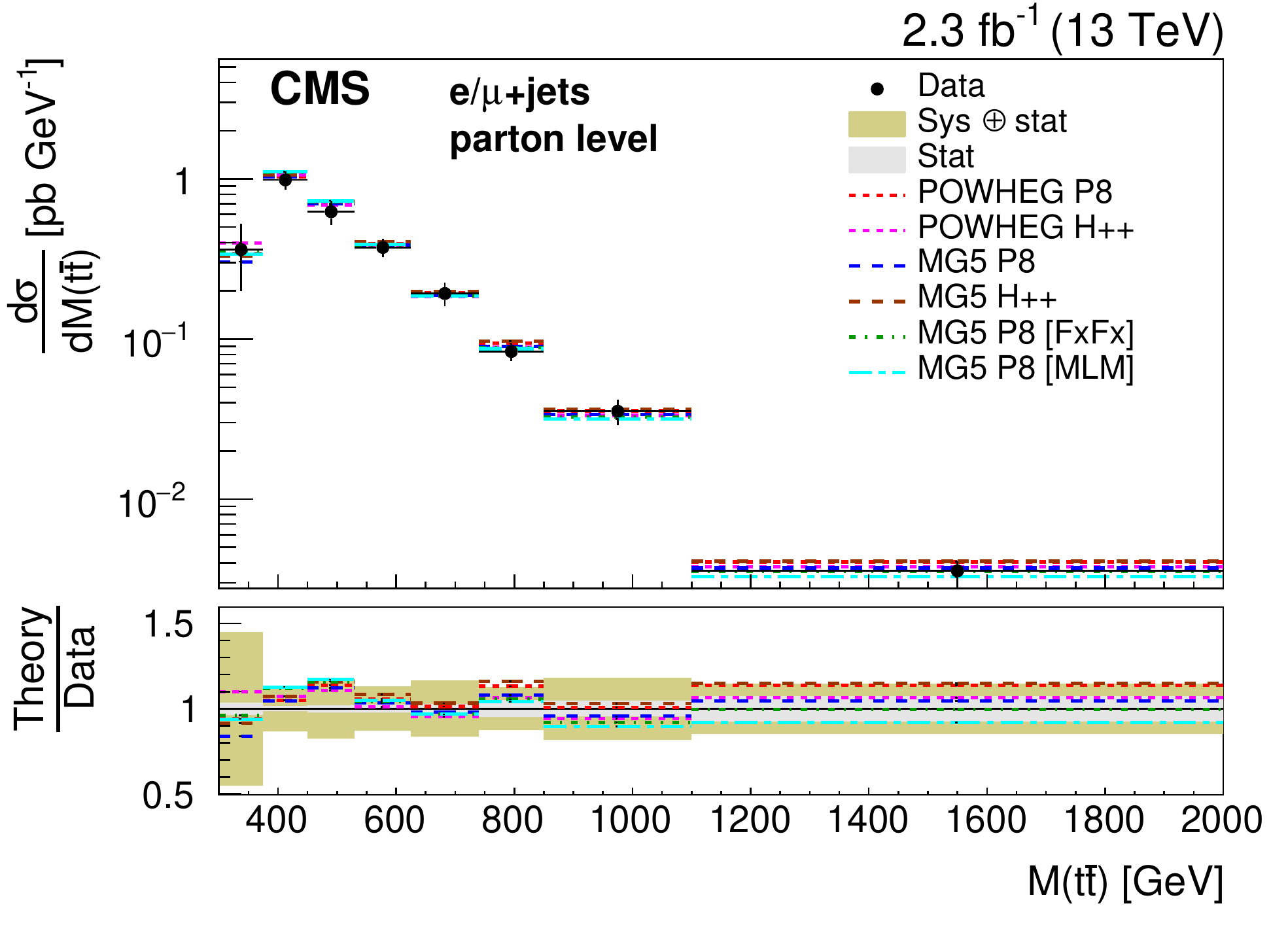}
\includegraphics[width=0.49\textwidth]{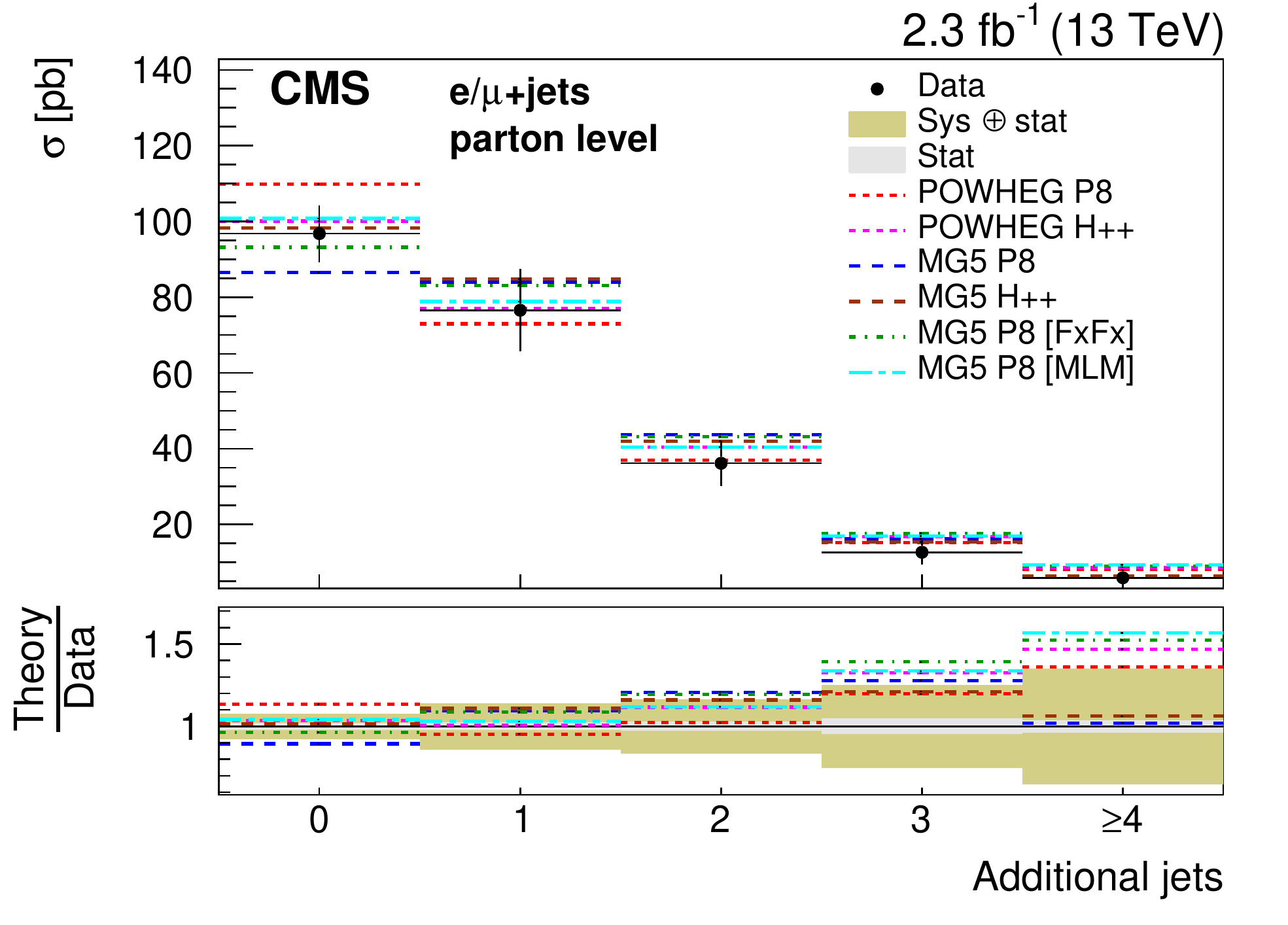}
\caption{Differential cross sections at parton level as a function of $\pt(\ttbar)$, $\abs{y(\ttbar)}$, $M(\ttbar)$, and cross sections as a function of the number of additional jets compared to the predictions of \POWHEG and \AMCATNLO(MG5) combined with \PYTHIAA(P8) or \HERWIGpp(H++) and the multiparton simulations \AMCATNLO{}+\PYTHIAA MLM  and \AMCATNLO{}+\PYTHIAA FxFx. The ratios of the various predictions to the measured cross sections are shown at the bottom of each panel together with the statistical and systematic uncertainties of the measurement.}
\label{XSECPA2}
\end{figure*}

\begin{figure*}[tbhp]
\centering
\includegraphics[width=0.49\textwidth]{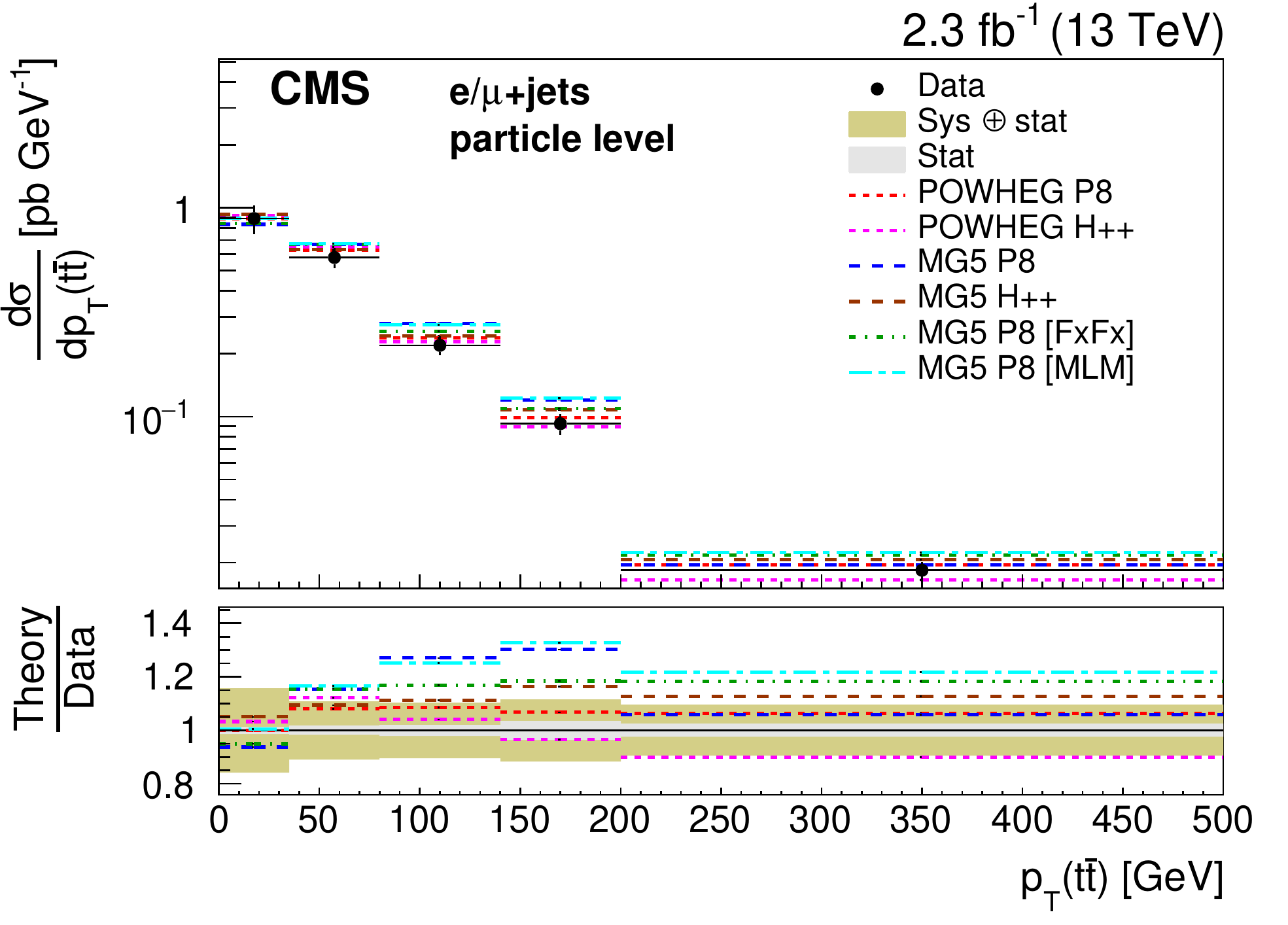}
\includegraphics[width=0.49\textwidth]{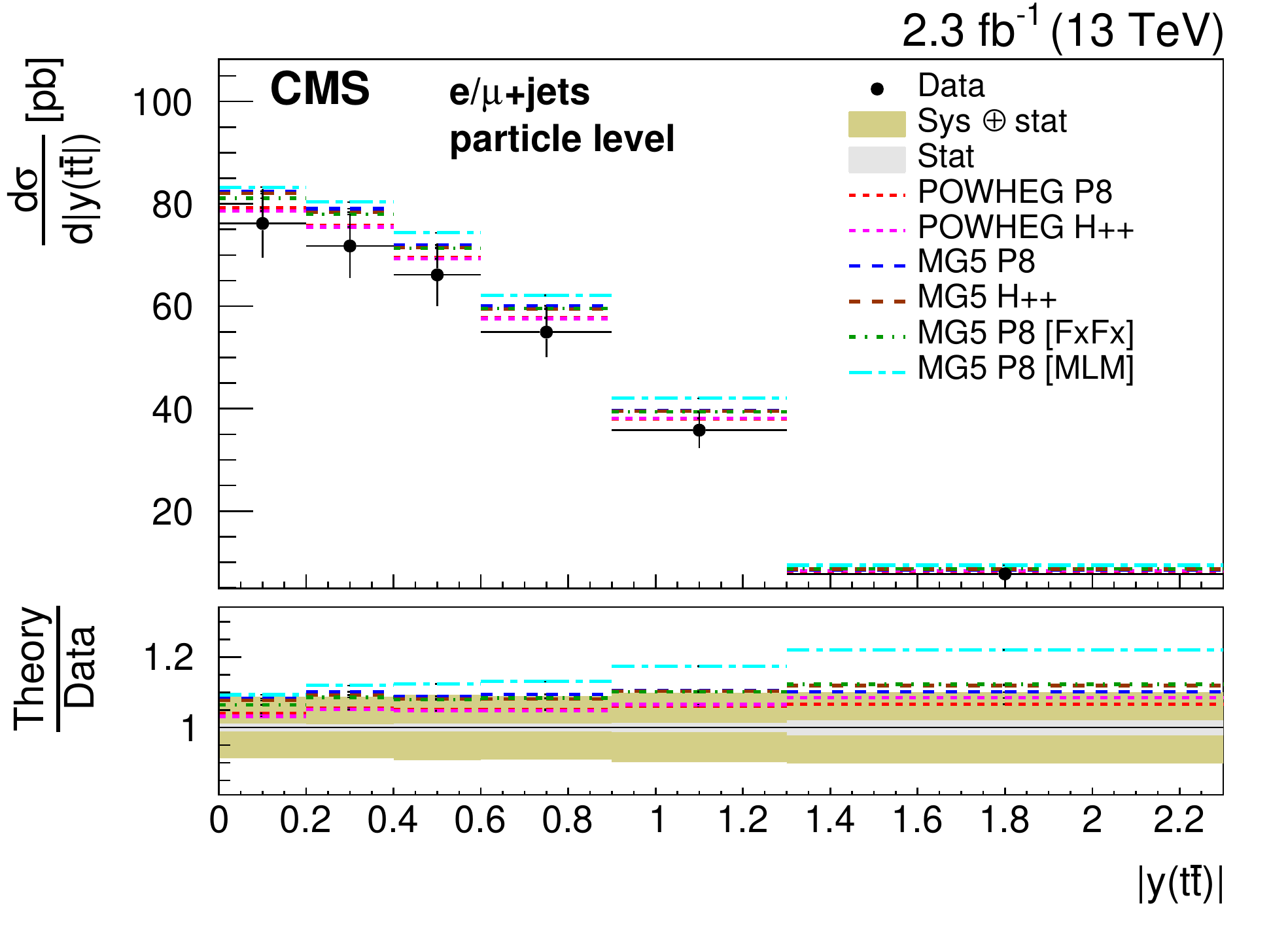}\\
\includegraphics[width=0.49\textwidth]{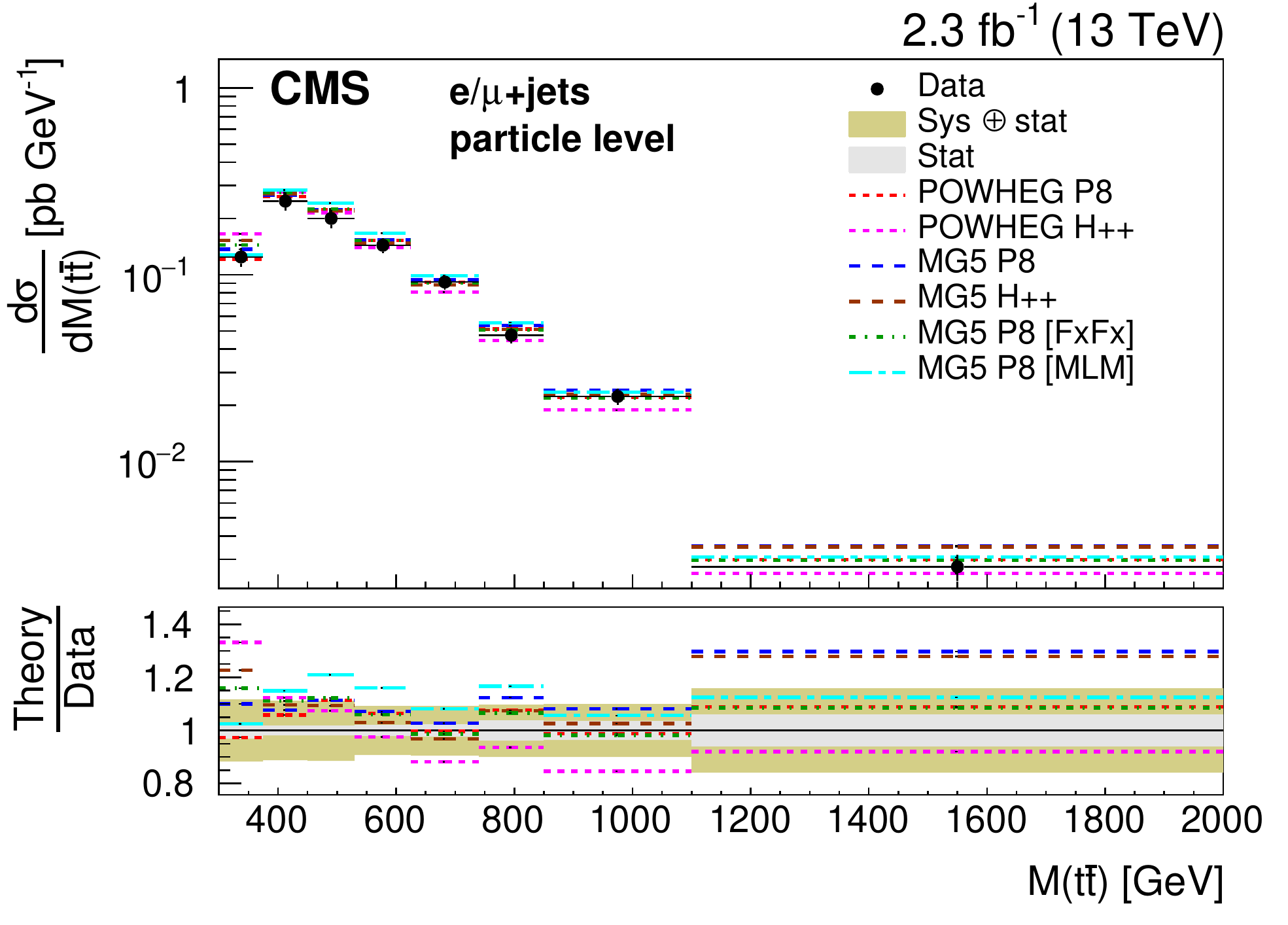}
\includegraphics[width=0.49\textwidth]{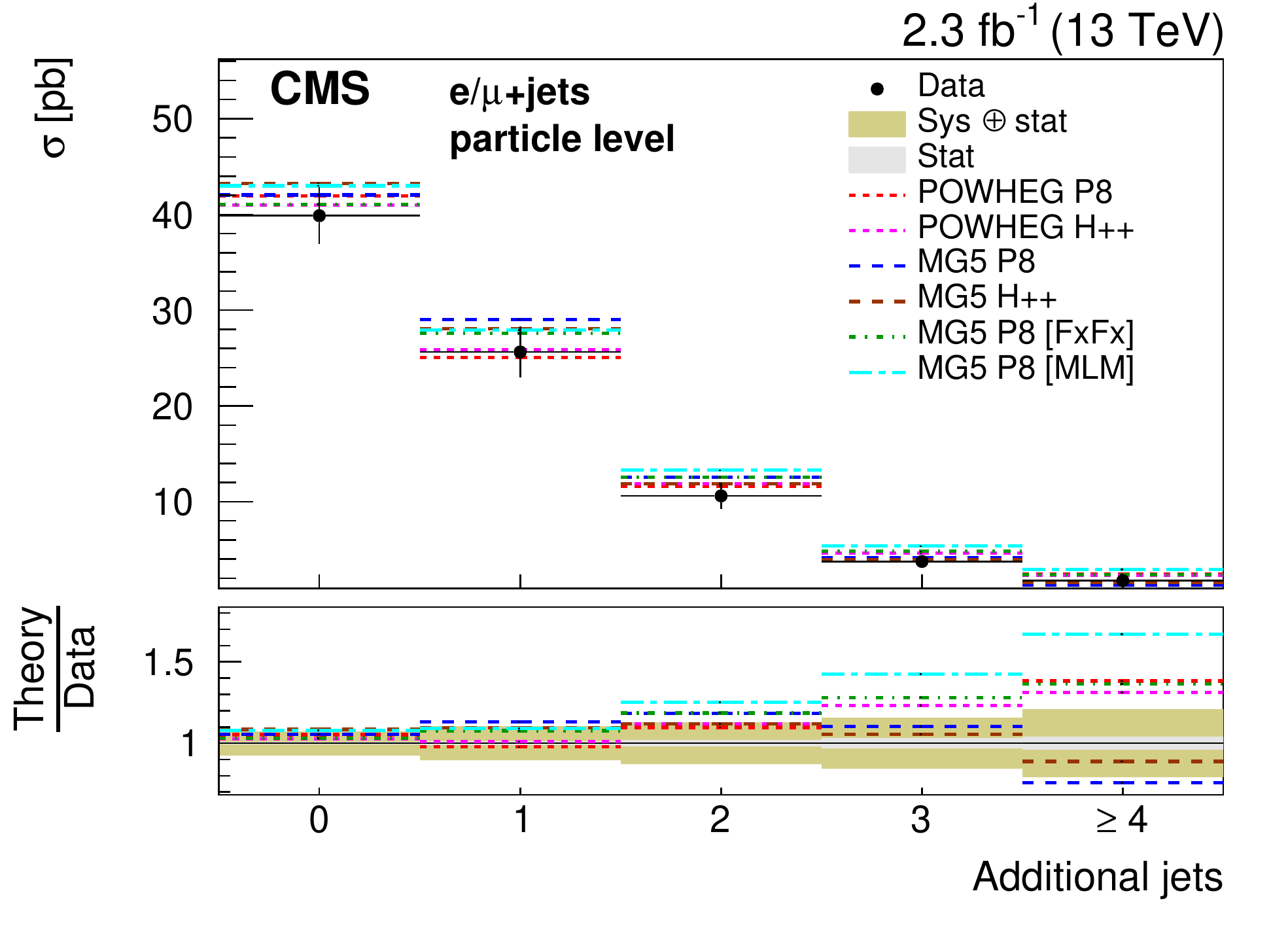}
\caption{Differential cross sections at particle level as a function of $\pt(\ttbar)$, $\abs{y(\ttbar)}$, $M(\ttbar)$, and cross sections as a function of the number of additional jets compared to the predictions of \POWHEG and \AMCATNLO(MG5) combined with \PYTHIAA(P8) or \HERWIGpp(H++) and the multiparton simulations \AMCATNLO{}+\PYTHIAA MLM  and \AMCATNLO{}+\PYTHIAA FxFx. The ratios of the various predictions to the measured cross sections are shown at the bottom of each panel together with the statistical and systematic uncertainties of the measurement.}
\label{XSECPS2}
\end{figure*}

{\tolerance=1200
The measured differential cross sections are compared to the predictions of \POWHEG and \AMCATNLO, each combined with the parton shower simulations of \PYTHIAA and \HERWIGpp. In addition, the \ttbar multiparton simulations of \AMCATNLO at LO and NLO with a \PYTHIAA parton shower are shown in Fig.~\ref{XSECPA1} (\ref{XSECPS1}) as a function of the top quark \pt and $\abs{y}$ at parton (particle) level. In Figs.~\ref{XSECPA2} and \ref{XSECPS2} the cross sections as a function of kinematic variables of the \ttbar system and the number of additional jets are compared to the same theoretical predictions.
\par}

\begin{figure*}[tbhp]
\centering
\includegraphics[width=0.49\textwidth]{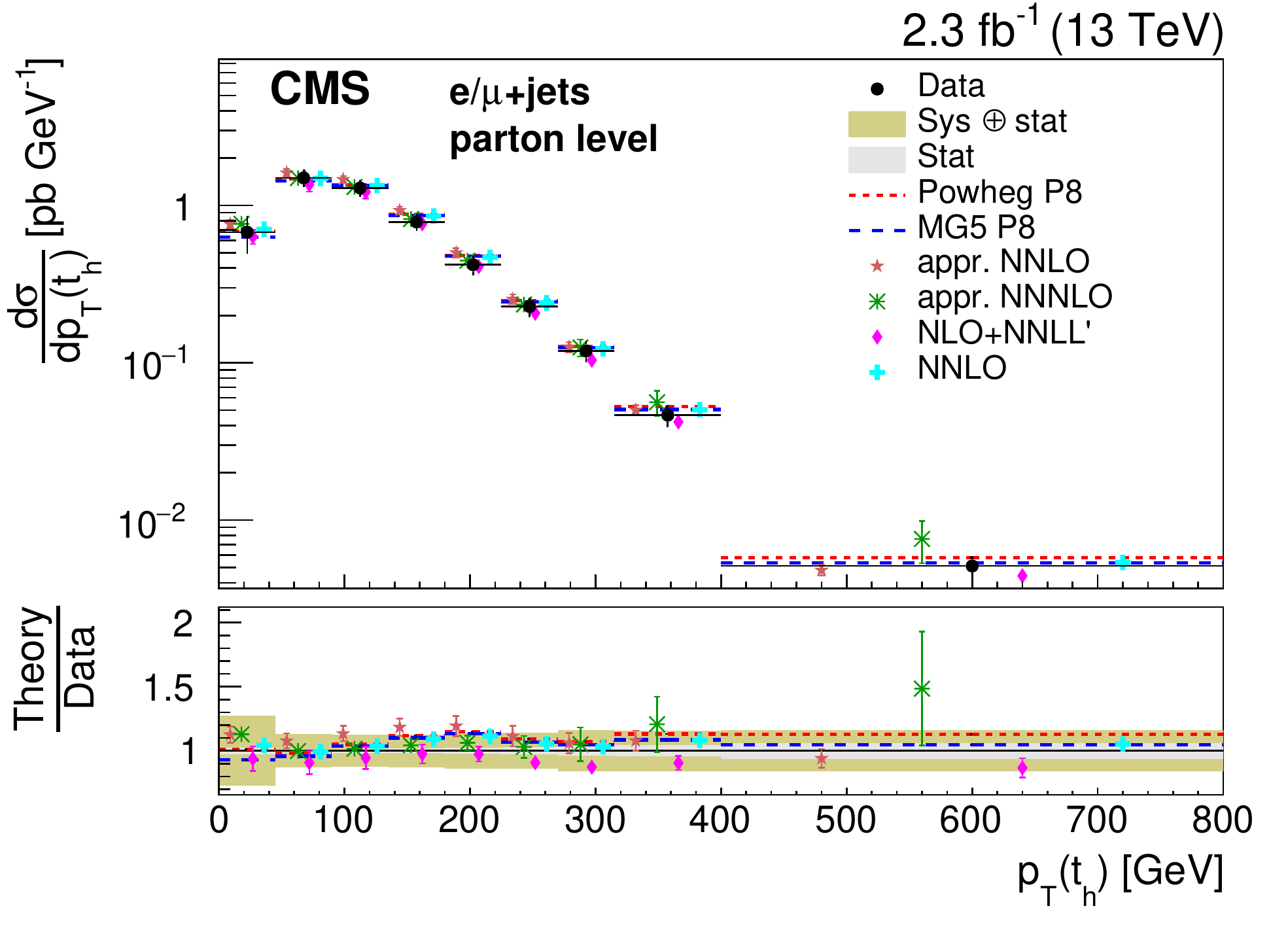}
\includegraphics[width=0.49\textwidth]{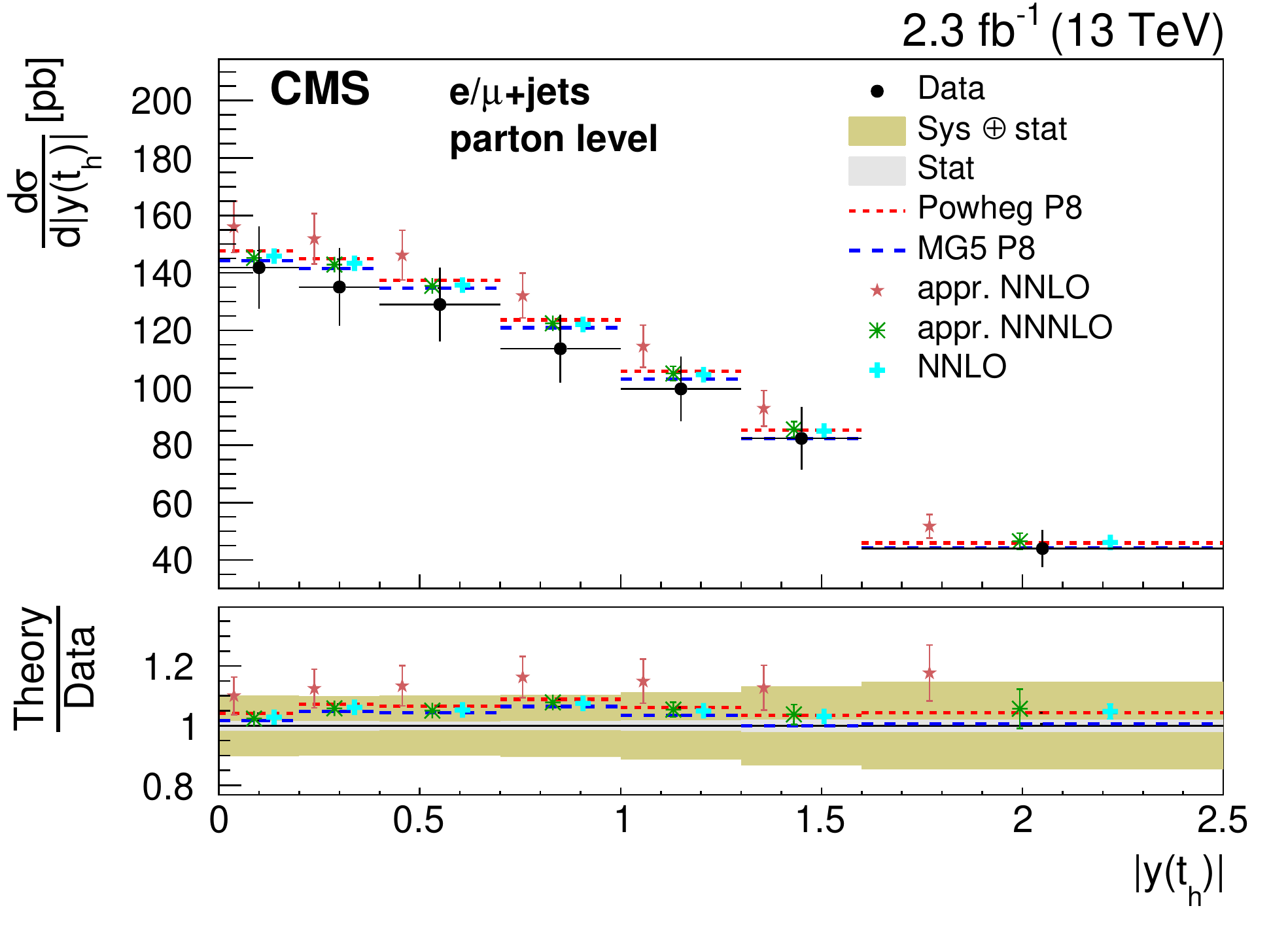}\\
\includegraphics[width=0.49\textwidth]{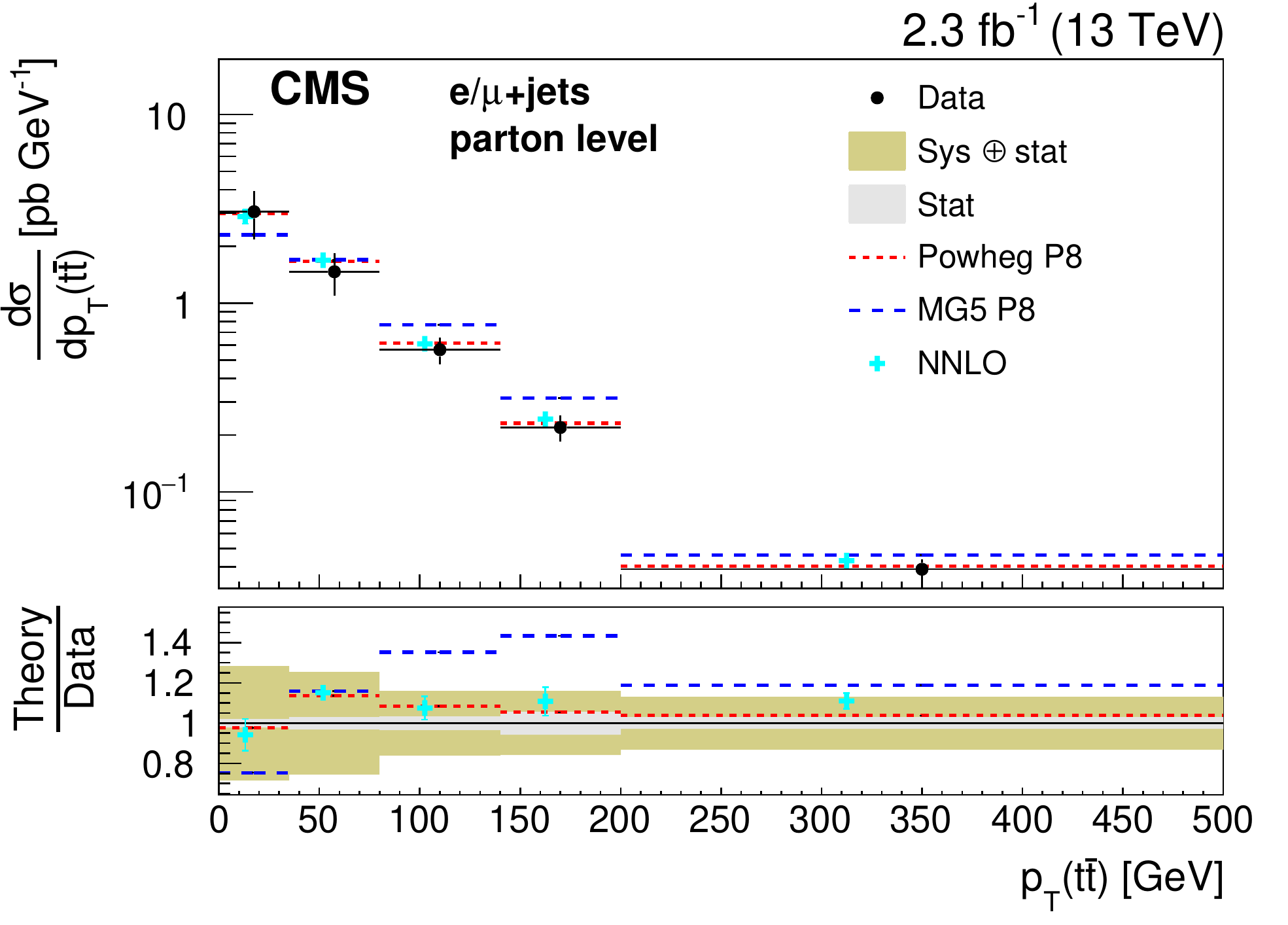}
\includegraphics[width=0.49\textwidth]{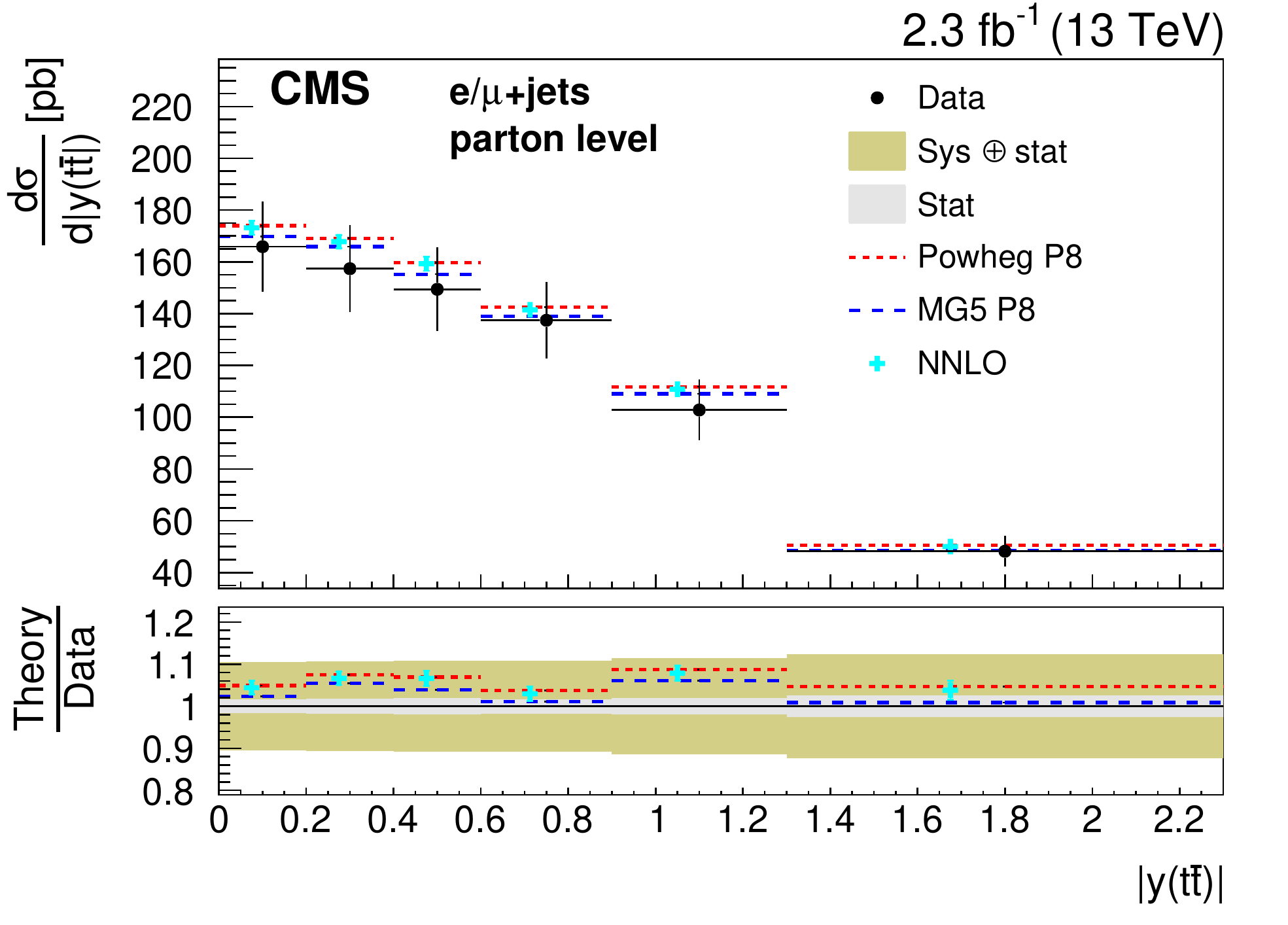}\\
\includegraphics[width=0.49\textwidth]{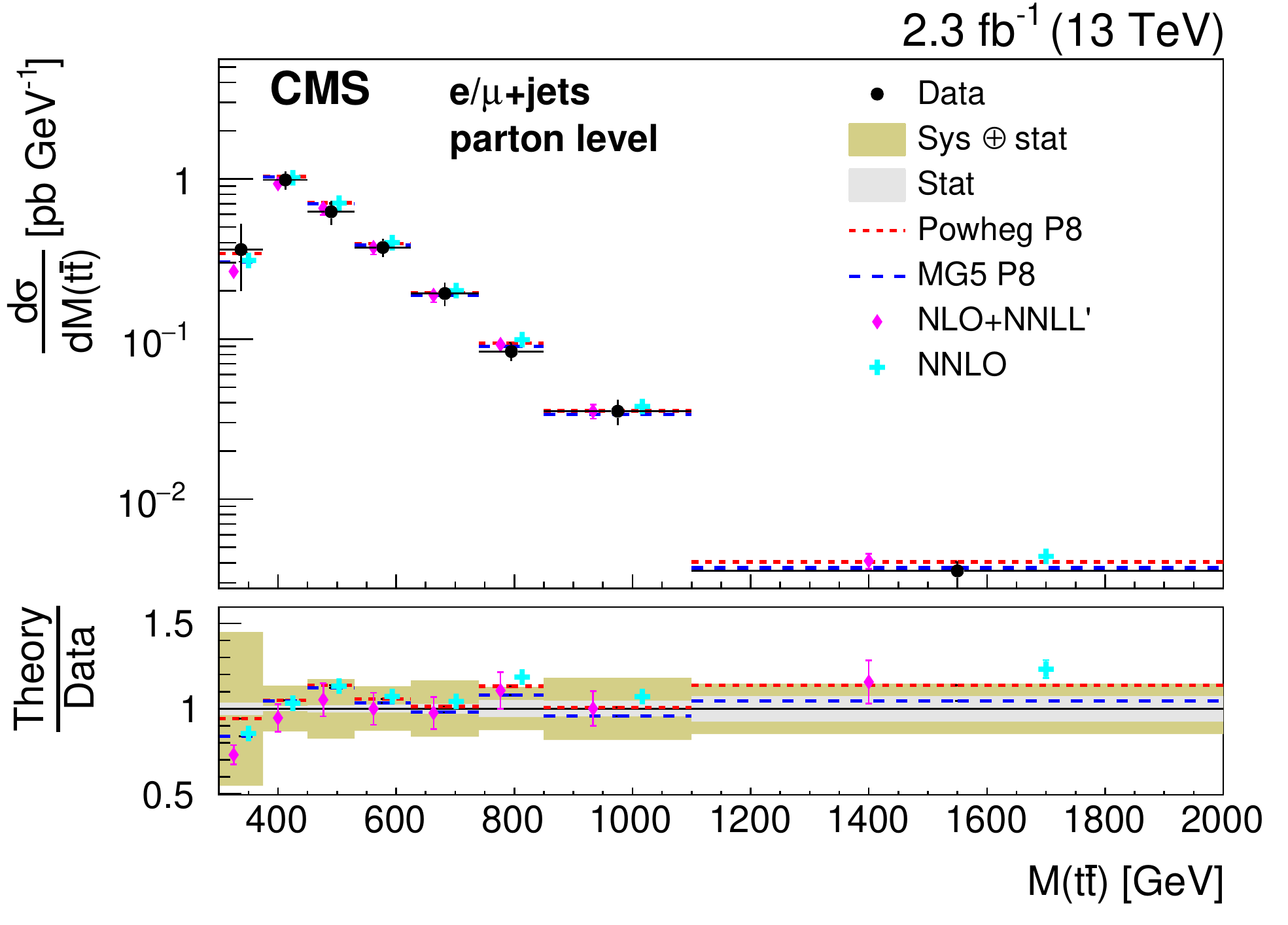}
\caption{Differential cross sections at parton level as a function of $\pt(\PQt)$, $\abs{y(\PQt)}$, $\pt(\ttbar)$, $\abs{y(\ttbar)}$, and $M(\ttbar)$ compared to the available predictions of an approximate NNLO calculation~\cite{Guzzi:2014wia}, an approximate NNNLO calculation~\cite{ANNNLO, ANNNLOdiff}, a NLO+NNLL' calculation~\cite{NLONNLL}, and a full NNLO calculation~\cite{NNLO}. For these models uncertainties due to the choices of scales are shown. To improve the visibility the theoretical predictions are horizontally shifted. The ratios of the various predictions to the measured cross sections are shown at the bottom of each panel together with the statistical and systematic uncertainties of the measurement.}
\label{XSECPA1t}
\end{figure*}

In \FIG{XSECPA1t} the parton-level results are compared to theoretical predictions of various accuracies. The first is an approximate NNLO~\cite{Guzzi:2014wia} QCD calculation using the CT14\,NNLO~\cite{Dulat:2015mca} PDF and $\Mtop = 172.5$\GeV. The factorization and renormalization scales are fixed at \Mtop. The second is an approximate next-to-NNLO (NNNLO)~\cite{ANNNLO, ANNNLOdiff} QCD calculation using the MSTW2008nnlo~\cite{Martin:2009iq} PDF, $\Mtop = 172.5$\GeV and factorization and renormalization scales fixed at \Mtop. The third combines the NLO QCD calculation with an improved NNLL QCD calculation (NLO+NNLL')~\cite{NLONNLL} using the MSTW2008nnlo PDF, $\Mtop=173.2$\GeV, and the renormalization and factorization scales of $M_\mathrm{T} = \sqrt{\Mtop^2 + \pt^2(\PQt)}$ for the $\pt(\PQt)$ calculation and $M(\ttbar)/2$ for the $M(\ttbar)$ calculation. The fourth is a full NNLO~\cite{NNLO} QCD calculation using the NNPDF3.0 PDF, $\Mtop = 173.3$\GeV, and the renormalization and factorization scales of $M_\mathrm{T}/2$ for the $\pt(\PQt)$ calculation and one-fourth of the sum of the \pt of all partons for the other distributions.The displayed uncertainties come from varying the scales up and down by a factor of two. Only the uncertainties in the approximate NNLO calculation include PDF uncertainties and a \Mtop variation of 1\GeV.

\begin{figure*}[tbhp]
\centering
\includegraphics[width=0.85\textwidth]{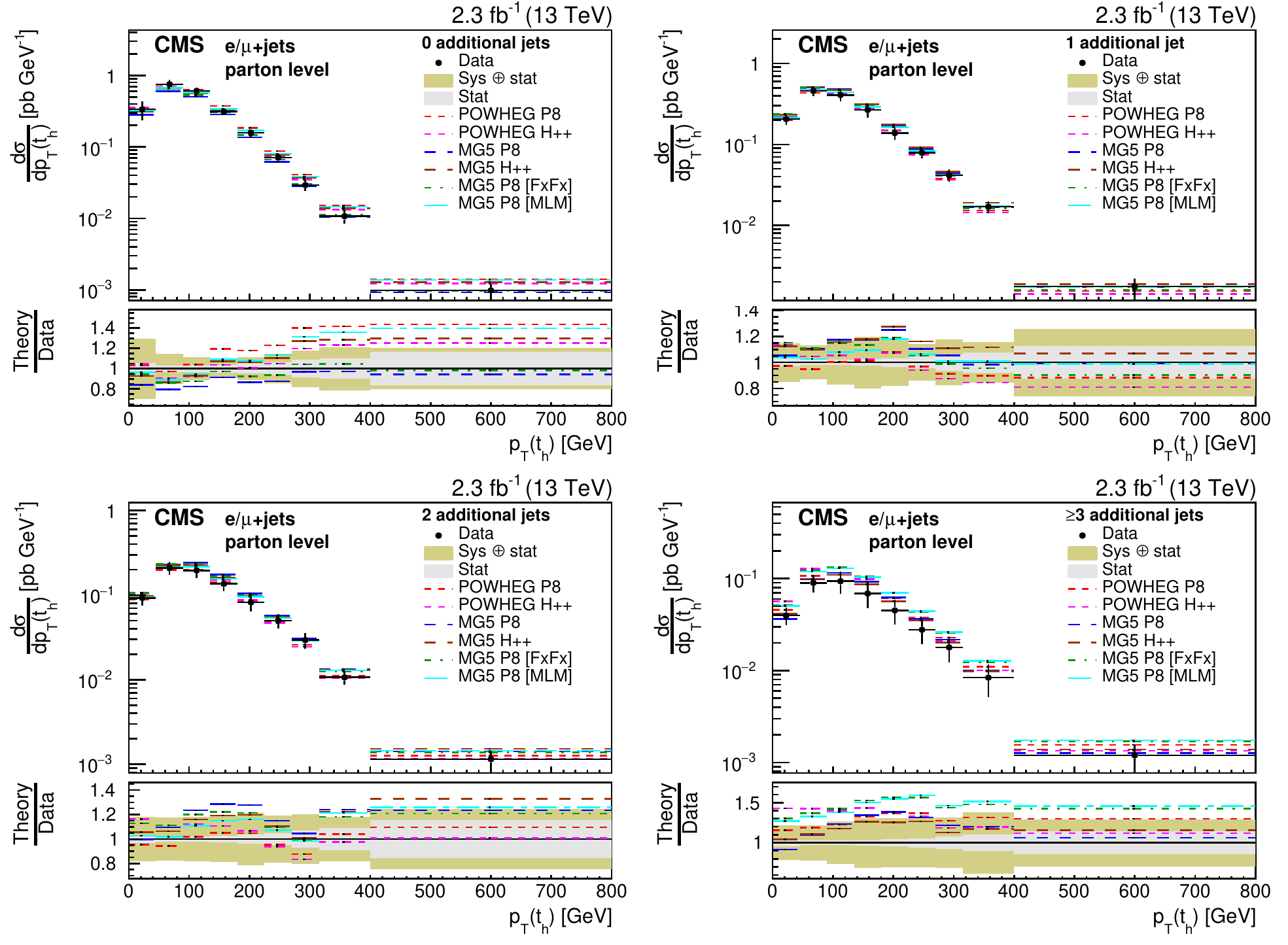}\\
\includegraphics[width=0.85\textwidth]{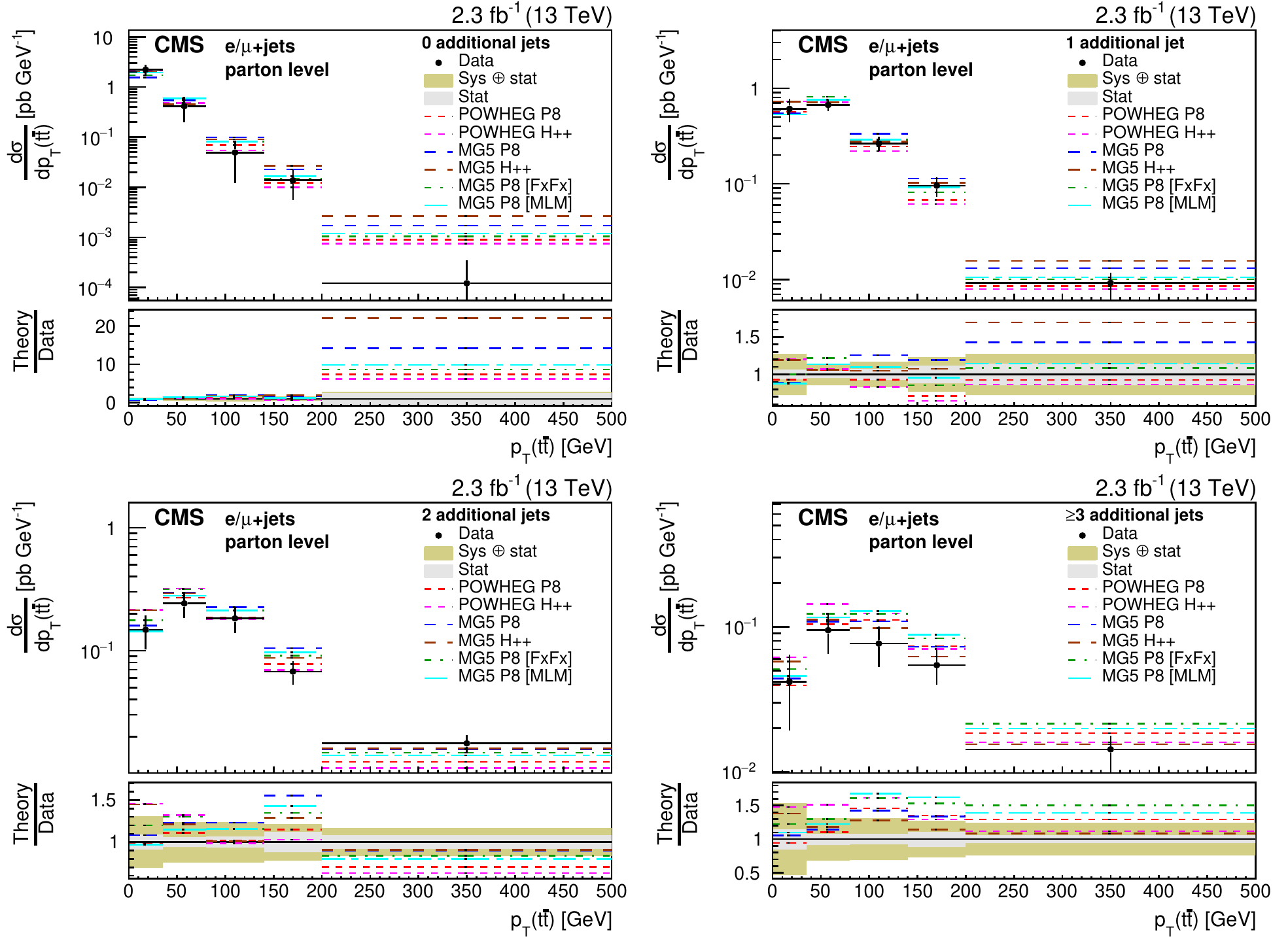}
\caption{Differential cross sections at parton level as a function of $\pt(\tqh)$ (upper two rows) and $\pt(\ttbar)$ (lower two rows) in bins of the number of additional jets. The measurements are compared to the predictions of \POWHEG and \AMCATNLO(MG5) combined with \PYTHIAA(P8) or \HERWIGpp(H++) and the multiparton simulations \AMCATNLO{}+\PYTHIAA MLM  and \AMCATNLO{}+\PYTHIAA FxFx. The ratios of the predictions to the measured cross sections are shown at the bottom of each panel together with the statistical and systematic uncertainties of the measurement.}
\label{XSECPA2D1}
\end{figure*}

\begin{figure*}[tbhp]
\centering
\includegraphics[width=0.85\textwidth]{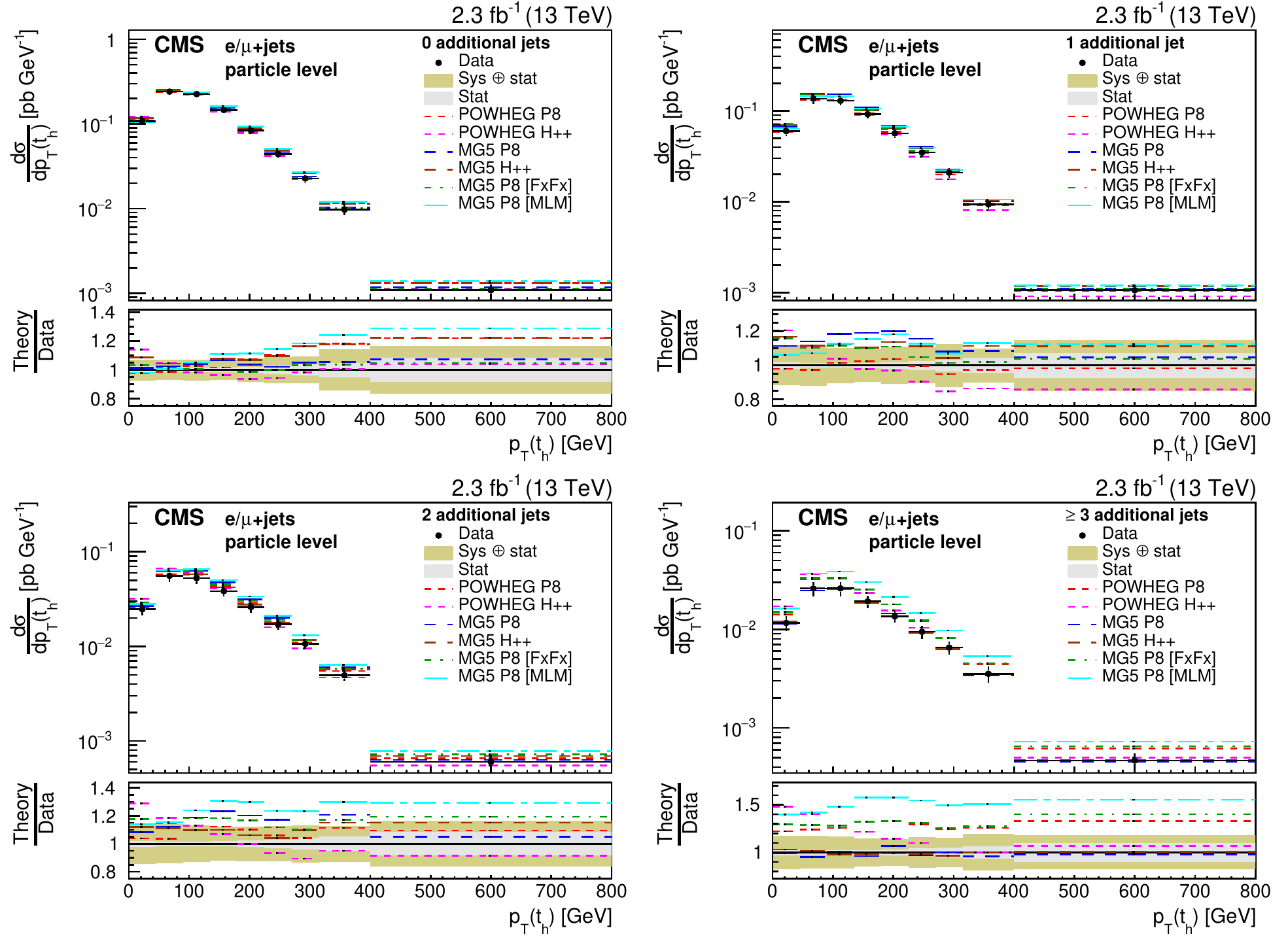}\\
\includegraphics[width=0.85\textwidth]{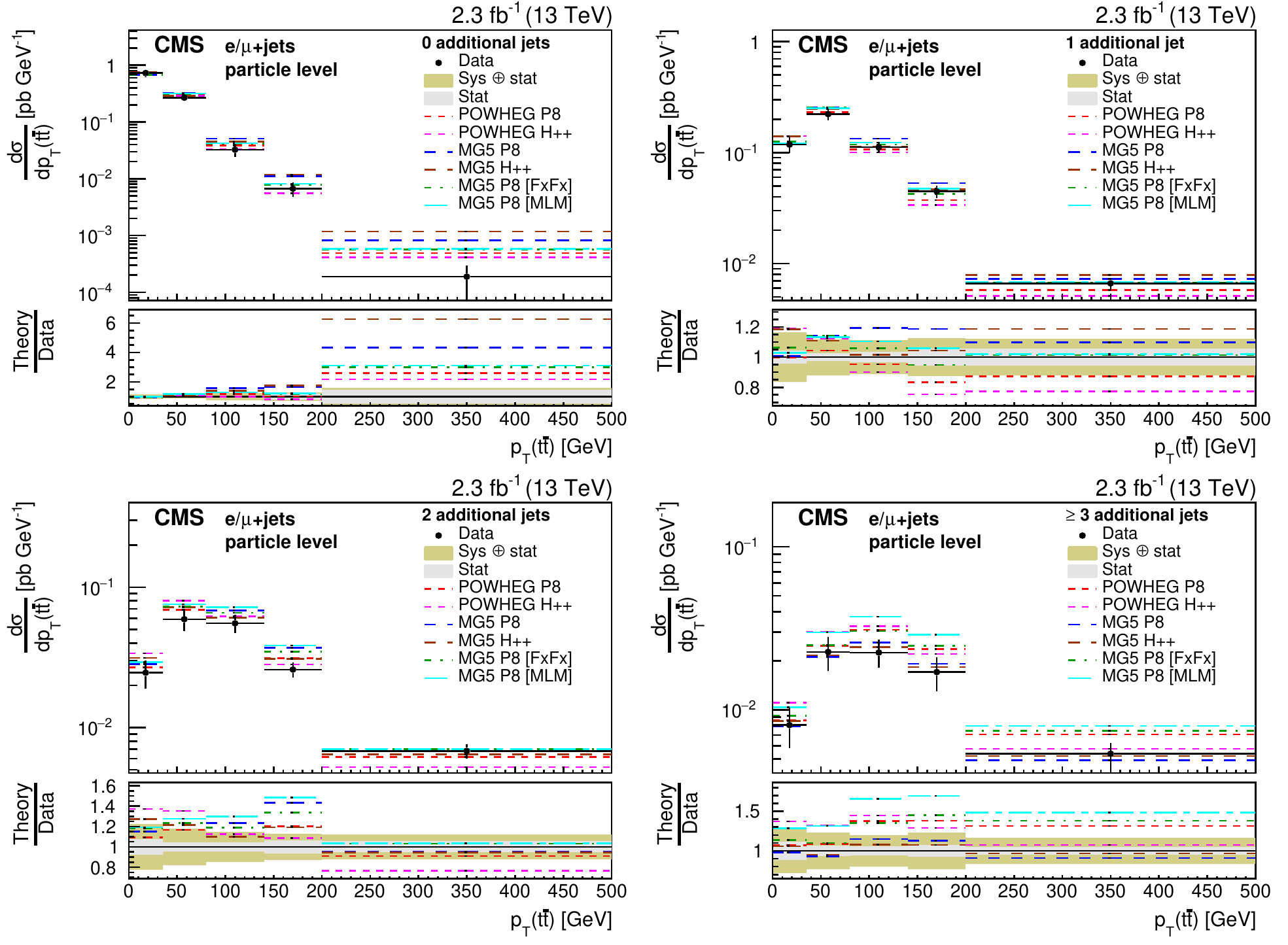}
\caption{Differential cross sections at particle level as a function of $\pt(\tqh)$ (upper two rows) and $\pt(\ttbar)$ (lower two rows) in bins of the number of additional jets. The measurements are compared to the predictions of \POWHEG and \AMCATNLO(MG5) combined with \PYTHIAA(P8) or \HERWIGpp(H++) and the multiparton simulations \AMCATNLO{}+\PYTHIAA MLM  and \AMCATNLO{}+\PYTHIAA FxFx. The ratios of the predictions to the measured cross sections are shown at the bottom of each panel together with the statistical and systematic uncertainties of the measurement.}
\label{XSECPS2D1}
\end{figure*}

\begin{figure*}[tbhp]
\centering
\includegraphics[width=0.85\textwidth]{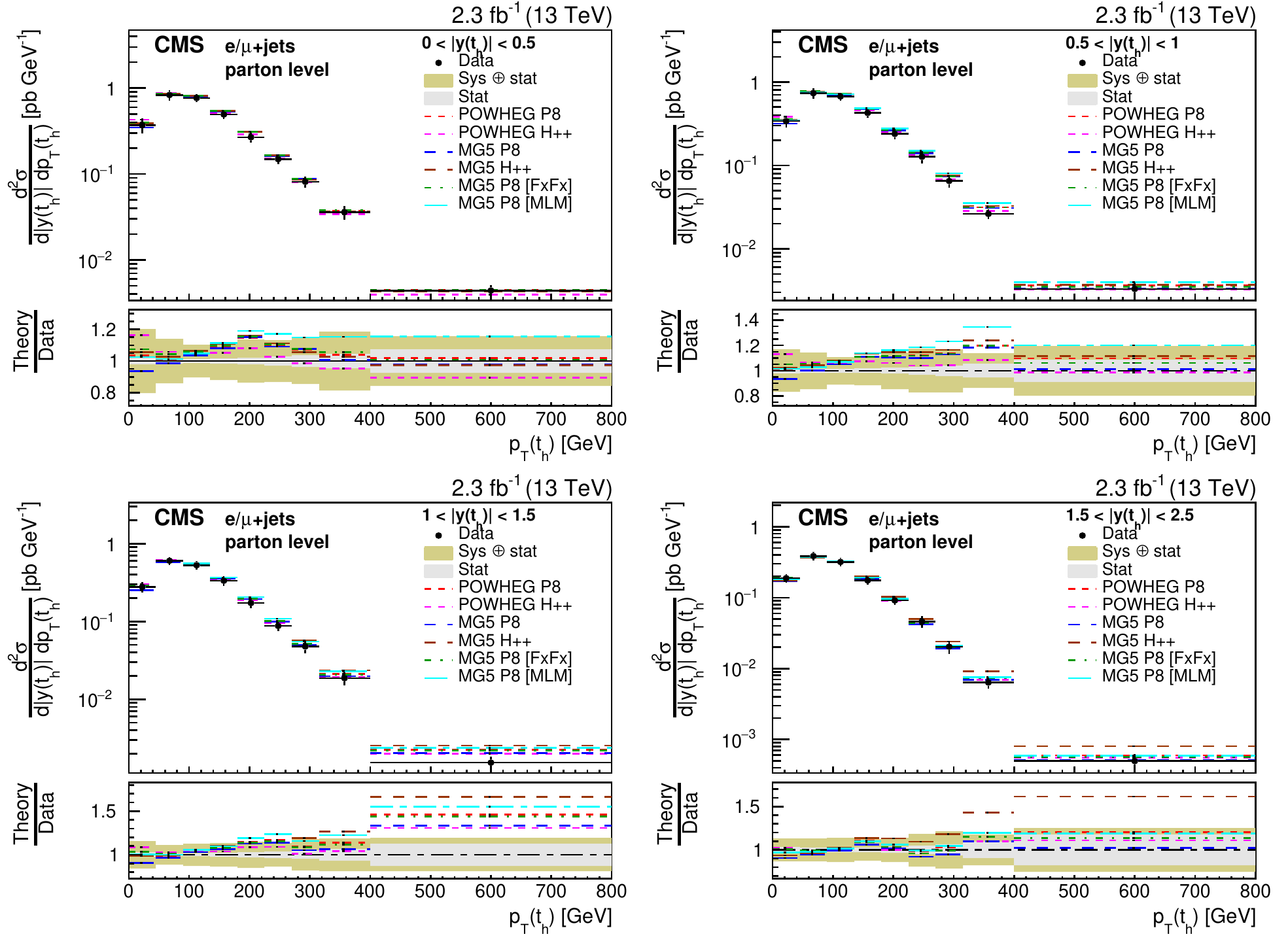}
\includegraphics[width=0.85\textwidth]{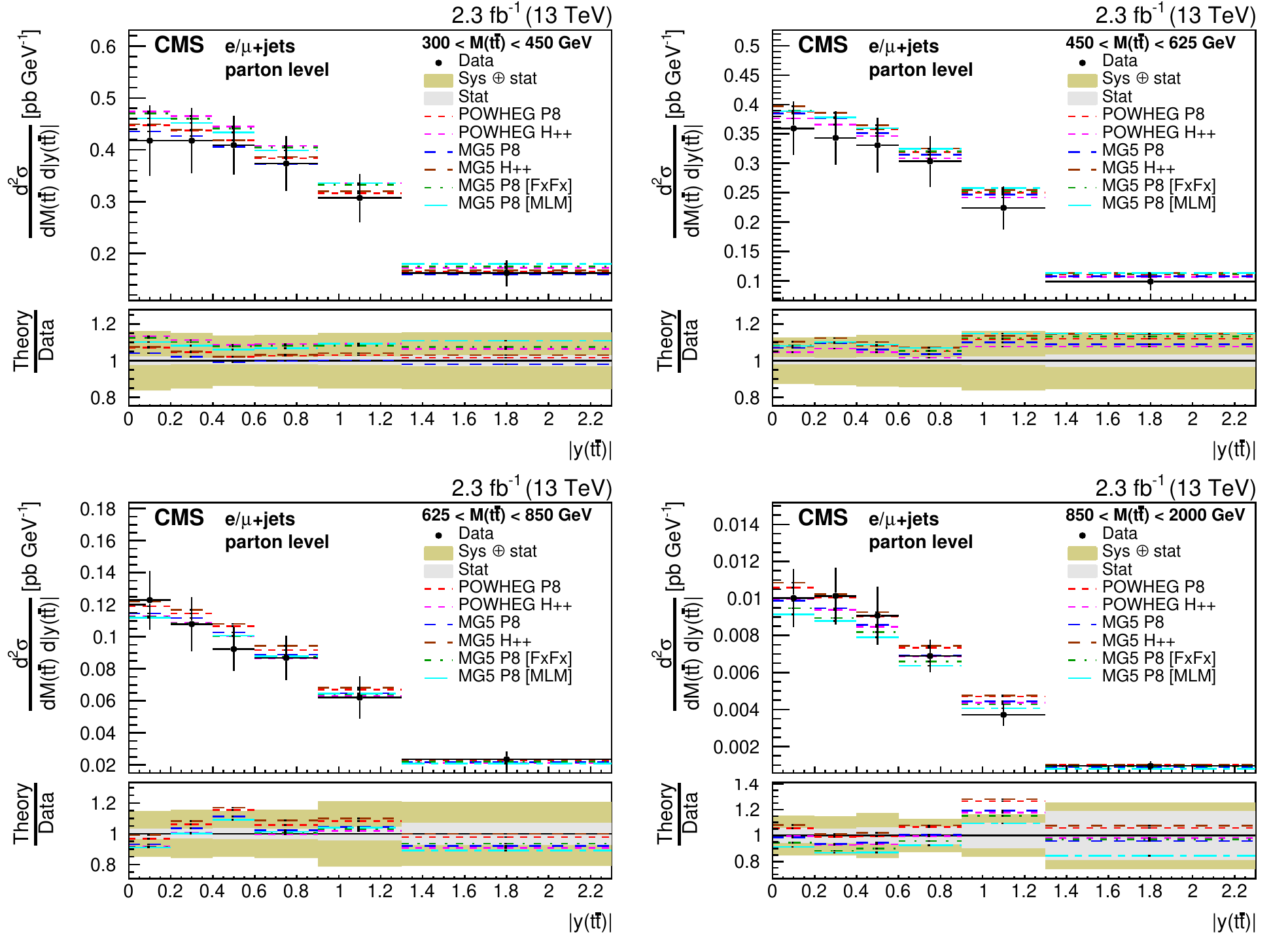}
\caption{Double-differential cross sections at parton level as a function of $\abs{y(\tqh)}$ \vs $\pt(\tqh)$ (upper two rows) and $M(\ttbar)$ \vs $\abs{y(\ttbar)}$ (lower two rows). The measurements are compared to the predictions of \POWHEG and \AMCATNLO(MG5) combined with \PYTHIAA(P8) or \HERWIGpp(H++) and the multiparton simulations \AMCATNLO{}+\PYTHIAA MLM  and \AMCATNLO{}+\PYTHIAA FxFx. The ratios of the predictions to the measured cross sections are shown at the bottom of each panel together with the statistical and systematic uncertainties of the measurement.}
\label{XSECPA2D3}
\end{figure*}

\begin{figure*}[tbhp]
\centering
\includegraphics[width=0.85\textwidth]{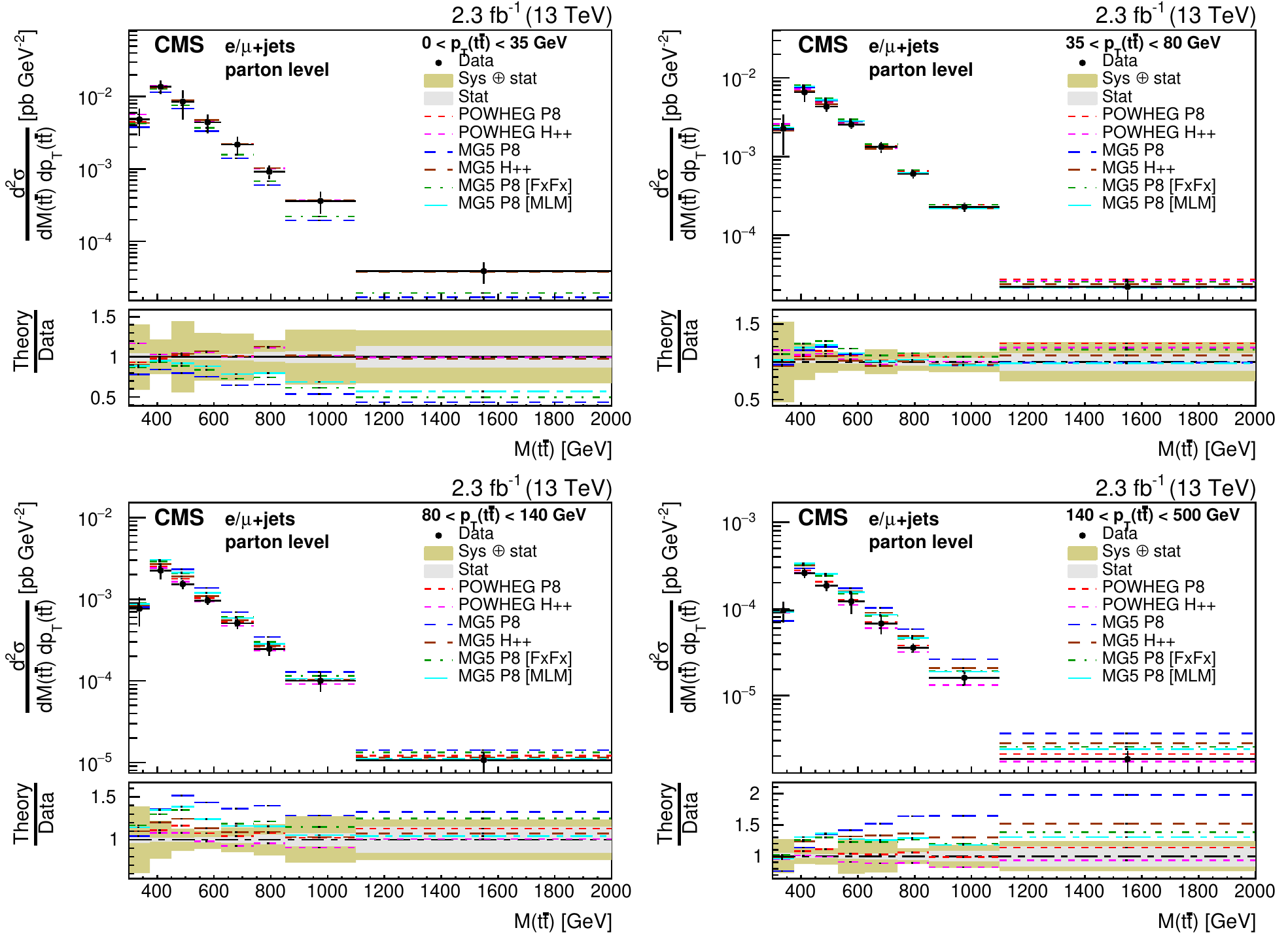}
\caption{Double-differential cross section at parton level as a function of $\pt(\ttbar)$ \vs $M(\ttbar)$. The measurements are compared to the predictions of \POWHEG and \AMCATNLO(MG5) combined with \PYTHIAA(P8) or \HERWIGpp(H++) and the multiparton simulations \AMCATNLO{}+\PYTHIAA MLM  and \AMCATNLO{}+\PYTHIAA FxFx. The ratios of the predictions to the measured cross sections are shown at the bottom of each panel together with the statistical and systematic uncertainties of the measurement.}
\label{XSECPA2D5}
\end{figure*}

\begin{figure*}[tbhp]
\centering
\includegraphics[width=0.85\textwidth]{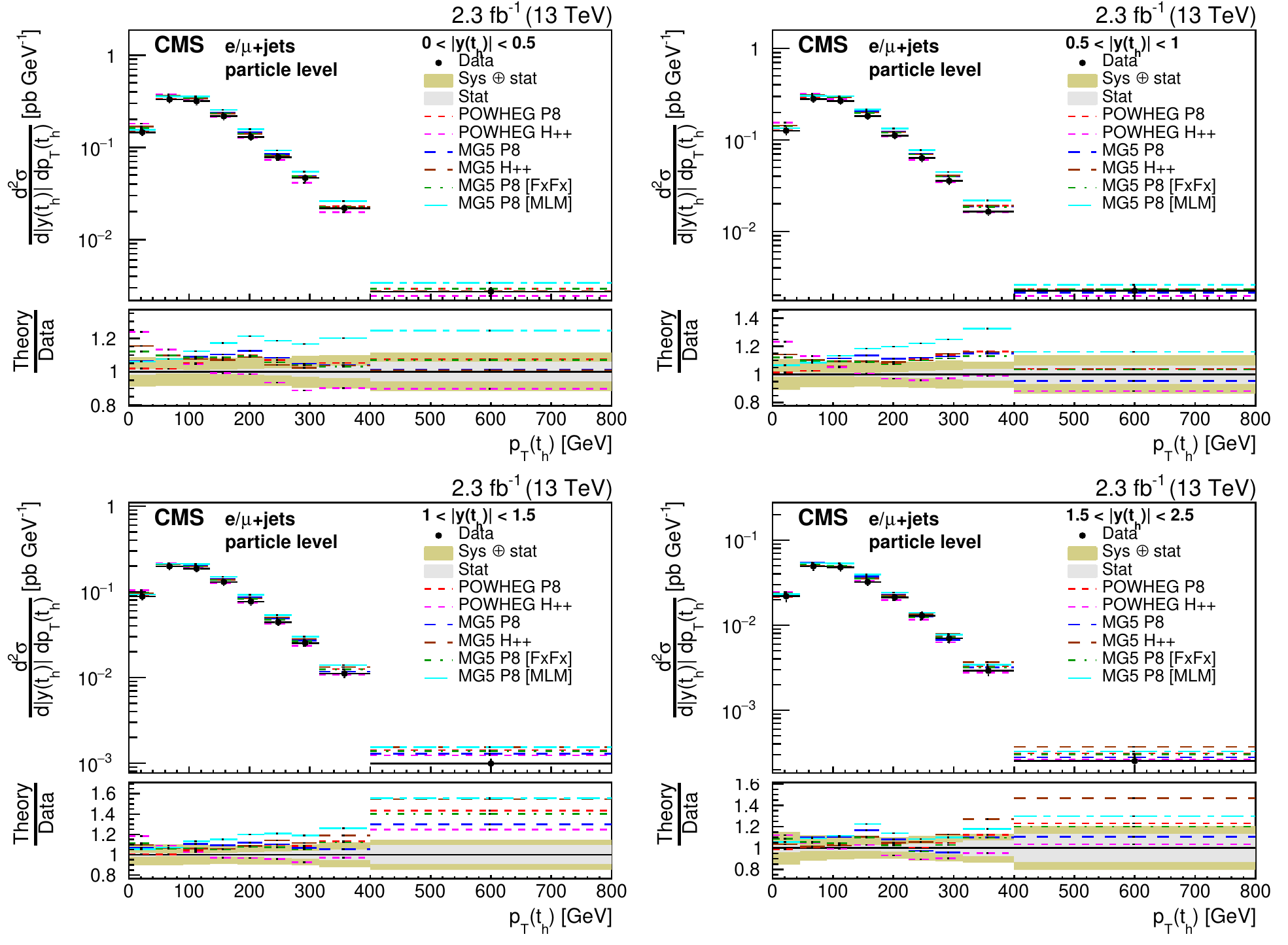}
\includegraphics[width=0.85\textwidth]{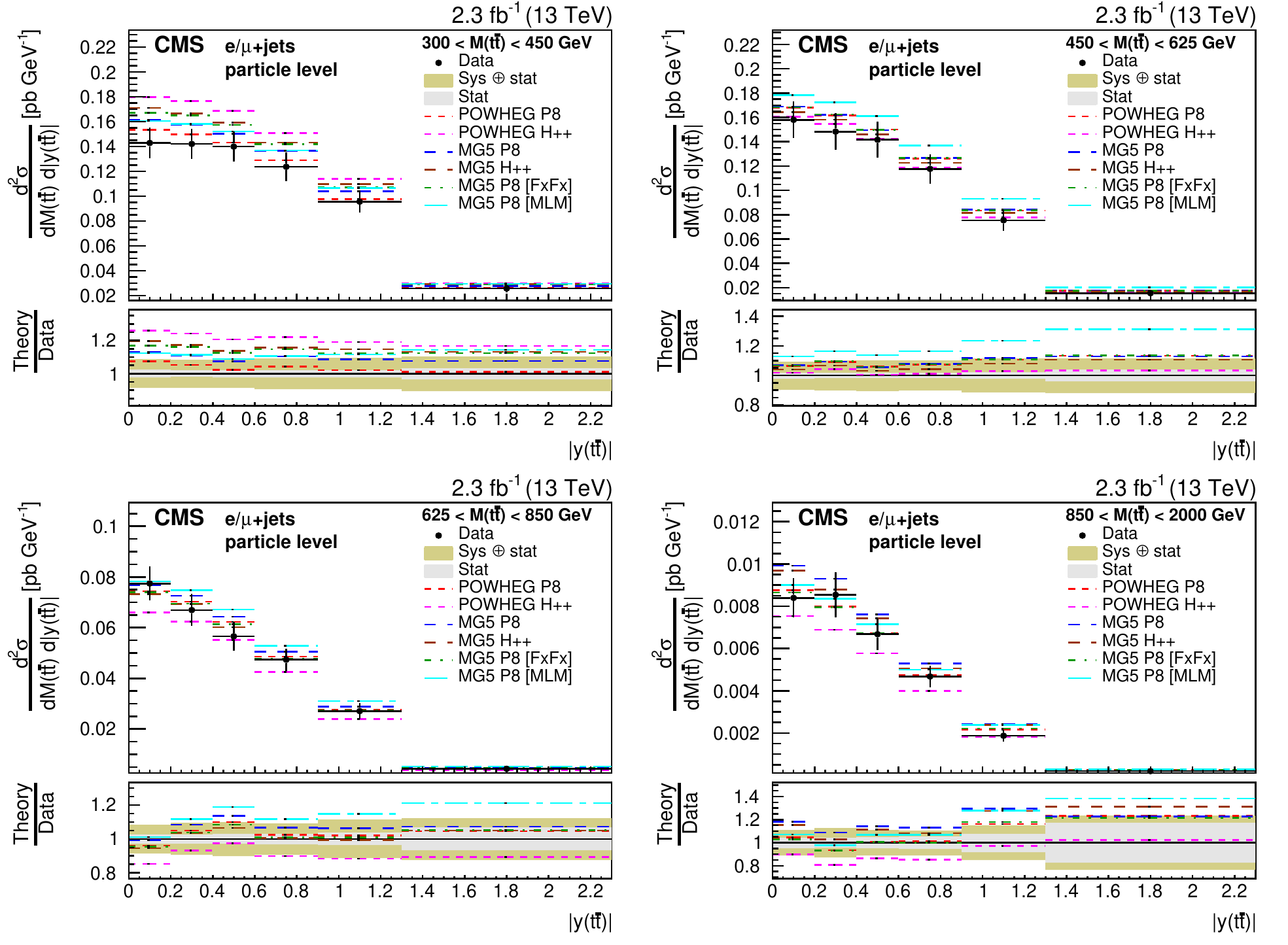}
\caption{Double-differential cross sections at particle level as a function of $\abs{y(\tqh)}$ \vs $\pt(\tqh)$ (upper two rows) and $M(\ttbar)$ \vs $\abs{y(\ttbar)}$ (lower two rows). The measurements are compared to the predictions of \POWHEG and \AMCATNLO(MG5) combined with \PYTHIAA(P8) or \HERWIGpp(H++) and the multiparton simulations \AMCATNLO{}+\PYTHIAA MLM  and \AMCATNLO{}+\PYTHIAA FxFx. The ratios of the predictions to the measured cross sections are shown at the bottom of each panel together with the statistical and systematic uncertainties of the measurement.}
\label{XSECPS2D3}
\end{figure*}

\begin{figure*}[tbhp]
\centering
\includegraphics[width=0.85\textwidth]{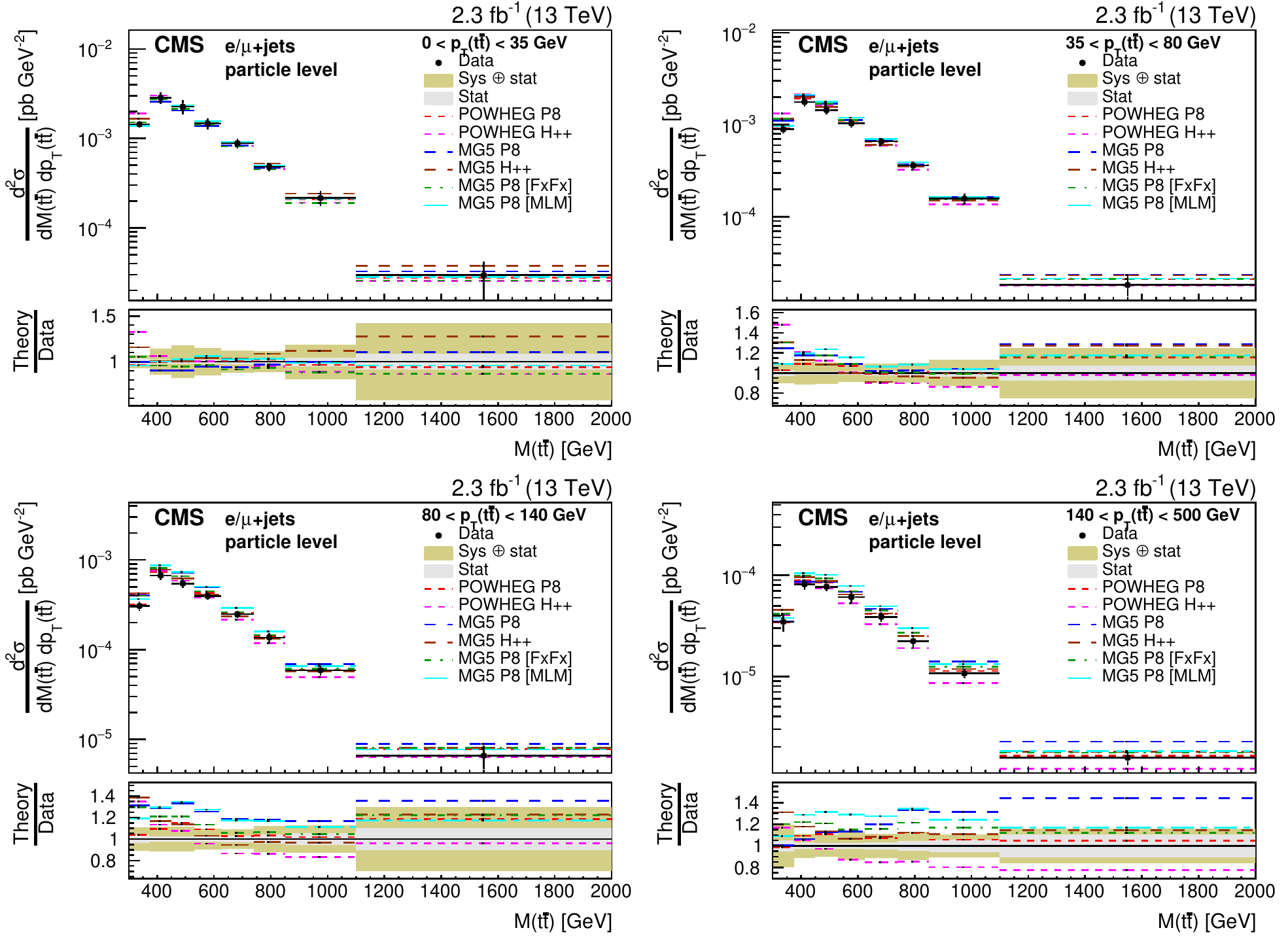}
\caption{Double-differential cross section at particle level as a function of $\pt(\ttbar)$ \vs $M(\ttbar)$. The measurements are compared to the predictions of \POWHEG and \AMCATNLO(MG5) combined with \PYTHIAA(P8) or \HERWIGpp(H++) and the multiparton simulations \AMCATNLO{}+\PYTHIAA MLM  and \AMCATNLO{}+\PYTHIAA FxFx. The ratios of the predictions to the measured cross sections are shown at the bottom of each panel together with the statistical and systematic uncertainties of the measurement.}
\label{XSECPS2D5}
\end{figure*}

The differential cross sections as a function of $\pt(\tqh)$ and $\pt(\ttbar)$ in bins of the number of additional jets are shown in \FIG{XSECPA2D1} (\ref{XSECPS2D1}) at parton (particle) level. The double-differential cross sections as a function of $\abs{y(\tqh)}$ \vs $\pt(\tqh)$, $M(\ttbar)$ \vs $\abs{y(\ttbar)}$, and $\pt(\ttbar)$ \vs $M(\ttbar)$ are shown at parton level in Figs.~\ref{XSECPA2D3}--\ref{XSECPA2D5} and at particle level in Figs.~\ref{XSECPS2D3}--\ref{XSECPS2D5}. The results are compared to the predictions of the event generators. All cross section values together with their statistical and systematic uncertainties are listed in Appendices~\ref{APP1} and \ref{APP2} for the parton- and particle-level measurements, respectively.

The precision of the measurement is limited by systematic uncertainties, dominated by jet energy scale uncertainties on the experimental side and parton shower and hadronization modeling uncertainties on the theoretical side. As expected, the theoretical uncertainties are reduced in the particle-level measurements since these are less dependent on theory-based extrapolations.

We evaluate the level of agreement between the measured differential cross sections and the various theoretical predictions using $\chi^2$ tests. In these tests we take into account the full covariance matrix obtained from the unfolding procedure for the statistical uncertainty. For each of the studied systematic uncertainties we assume a full correlation among all bins. No uncertainties in the theoretical predictions are considered for this comparison. However, these uncertainties are known to be large. Typically, differences between the various models are used to assess their uncertainties. From the $\chi^2$ values and the numbers of degrees of freedom, which corresponds to the number of bins in the distributions, the p-values are calculated. The results are shown in \TAB{REST1} for the parton-level and in \TAB{REST2} for the particle-level measurements.

\begin{table*}[tbhp]
\caption{Comparison between the measured distributions at parton level and the predictions of \POWHEG and \AMCATNLO combined with \PYTHIAA(P8) or \HERWIGpp(H++) and the multiparton simulations \AMCATNLO MLM  and \AMCATNLO FxFx, as well as the predictions of an approximate NNNLO calculation~\cite{ANNNLO, ANNNLOdiff}, a NLO+NNLL' calculation~\cite{NLONNLL}, and a full NNLO calculation~\cite{NNLO}. We list the results of the $\chi^2$ tests together with the numbers of degrees of freedom (dof) and the corresponding p-values. For the comparison no uncertainties in the theoretical predictions are taken into account.}
\centering
\cmsTable{
\renewcommand{\arraystretch}{1.2}
\begin{scotch}{l|r@{\hspace{4mm}}lr@{\hspace{4mm}}lr@{\hspace{4mm}}l}
Distribution & $\chi^2/\mathrm{dof}$ & p-value & $\chi^2/\mathrm{dof}$ & p-value & $\chi^2/\mathrm{dof}$ & p-value\\\hline
 & \multicolumn{2}{c}{\POWHEG{}+P8} & \multicolumn{2}{c}{\POWHEG{}+H++} & \multicolumn{2}{c}{\AMCATNLO{}+P8 MLM}\\
 & \multicolumn{2}{c}{Order: NLO} & \multicolumn{2}{c}{Order: NLO} & \multicolumn{2}{c}{Order: LO, up to 3 add. partons}\\
$\pt(\tqh)$ & 10.7/9&0.295 & 8.01/9&0.533 & 19.0/9&0.025\\
$\abs{y(\tqh)}$ & 3.91/7&0.790 & 4.33/7&0.741 & 4.49/7&0.721\\
$\pt(\tql)$ & 14.9/9&0.093 & 9.03/9&0.435 & 41.8/9&$<0.01$\\
$\abs{y(\tql)}$ & 11.4/7&0.121 & 13.1/7&0.070 & 12.0/7&0.100\\
$M(\ttbar)$ & 5.61/8&0.691 & 10.9/8&0.206 & 45.0/8&$<0.01$\\
$\pt(\ttbar)$ & 0.941/5&0.967 & 4.34/5&0.501 & 16.8/5&$<0.01$\\
$\abs{y(\ttbar)}$ & 1.95/6&0.924 & 2.04/6&0.916 & 5.55/6&0.476\\
Additional jets & 8.22/5&0.145 & 6.88/5&0.230 & 5.82/5&0.324\\
Additional jets \vs $\pt(\ttbar)$ & 85.3/20&$<0.01$ & 132/20&$<0.01$ & 135/20&$<0.01$\\
Additional jets \vs $\pt(\tqh)$ & 89.0/36&$<0.01$ & 43.1/36&0.193 & 71.7/36&$<0.01$\\
$\abs{y(\tqh)}$ \vs $\pt(\tqh)$ & 55.3/36&0.021 & 52.4/36&0.038 & 60.7/36&$<0.01$\\
$M(\ttbar)$ \vs $\abs{y(\ttbar)}$ & 19.3/24&0.734 & 18.3/24&0.788 & 49.4/24&$<0.01$\\
$\pt(\ttbar)$ \vs $M(\ttbar)$ & 14.5/32&0.997 & 26.2/32&0.755 & 100/32&$<0.01$\\\hline
 & \multicolumn{2}{c}{\AMCATNLO{}+P8} & \multicolumn{2}{c}{\AMCATNLO{}+H++} & \multicolumn{2}{c}{\AMCATNLO{}+P8 FxFx}\\
 & \multicolumn{2}{c}{Order: NLO} & \multicolumn{2}{c}{Order: NLO} & \multicolumn{2}{c}{Order: NLO, up to 2 add. partons}\\
$\pt(\tqh)$ & 8.68/9&0.467 & 15.3/9&0.084 & 9.35/9&0.406\\
$\abs{y(\tqh)}$ & 4.11/7&0.767 & 5.42/7&0.608 & 3.91/7&0.790\\
$\pt(\tql)$ & 13.0/9&0.162 & 26.8/9&$<0.01$ & 11.7/9&0.228\\
$\abs{y(\tql)}$ & 14.3/7&0.046 & 10.7/7&0.151 & 16.4/7&0.022\\
$M(\ttbar)$ & 9.91/8&0.271 & 5.93/8&0.655 & 28.0/8&$<0.01$\\
$\pt(\ttbar)$ & 31.1/5&$<0.01$ & 24.6/5&$<0.01$ & 18.4/5&$<0.01$\\
$\abs{y(\ttbar)}$ & 1.97/6&0.923 & 2.04/6&0.916 & 2.49/6&0.870\\
Additional jets & 21.5/5&$<0.01$ & 4.21/5&0.520 & 7.98/5&0.158\\
Additional jets \vs $\pt(\ttbar)$ & 319/20&$<0.01$ & 259/20&$<0.01$ & 121/20&$<0.01$\\
Additional jets \vs $\pt(\tqh)$ & 90.9/36&$<0.01$ & 45.0/36&0.145 & 52.5/36&0.037\\
$\abs{y(\tqh)}$ \vs $\pt(\tqh)$ & 73.1/36&$<0.01$ & 111/36&$<0.01$ & 48.1/36&0.086\\
$M(\ttbar)$ \vs $\abs{y(\ttbar)}$ & 26.1/24&0.347 & 17.8/24&0.811 & 36.7/24&0.047\\
$\pt(\ttbar)$ \vs $M(\ttbar)$ & 229/32&$<0.01$ & 71.5/32&$<0.01$ & 97.6/32&$<0.01$\\\hline
 & \multicolumn{2}{c}{appr. NNLO} & \multicolumn{2}{c}{appr. NNNLO} & \multicolumn{2}{c}{NLO+NNLL'}\\
$\pt(\tqh)$ & 14.3/9&0.111 & 36.7/9&$<0.01$ & 6.29/9&0.710\\
$\abs{y(\tqh)}$ & 5.30/7&0.623 & 2.59/7&0.920 & \NA & \NA\\
$\pt(\tql)$ & 12.1/9&0.209 & 92.1/9&$<0.01$ & 3.06/9&0.962\\
$\abs{y(\tql)}$ & 3.77/7&0.805 & 4.34/7&0.739 & \NA & \NA\\
$M(\ttbar)$ & \NA & \NA & \NA & \NA & 6.70/8&0.569\\\hline
 & \multicolumn{2}{c}{NNLO} & \multicolumn{4}{c}{}\\
$\pt(\tqh)$ & 5.78/9&0.762&\multicolumn{4}{c}{}\\
$\abs{y(\tqh)}$ & 2.20/7&0.948&\multicolumn{4}{c}{}\\
$\pt(\tql)$ & 5.54/9&0.785&\multicolumn{4}{c}{}\\
$\abs{y(\tql)}$ & 6.48/7&0.485&\multicolumn{4}{c}{}\\
$M(\ttbar)$ & 5.88/8&0.660&\multicolumn{4}{c}{}\\
$\pt(\ttbar)$ & 3.50/5&0.623&\multicolumn{4}{c}{}\\
$\abs{y(\ttbar)}$ & 1.42/6&0.965&\multicolumn{4}{c}{}\\
\end{scotch}
}
\label{REST1}
\end{table*}

\begin{table*}[tbhp]
\caption{Comparison between the measured distributions at particle level and the predictions of \POWHEG and \AMCATNLO combined with \PYTHIAA(P8) or \HERWIGpp(H++) and the multiparton simulations \AMCATNLO MLM  and \AMCATNLO FxFx. We list the results of the $\chi^2$ tests together with the numbers of degrees of freedom (dof) and the corresponding p-values. For the comparison no uncertainties in the theoretical predictions are taken into account.}
\centering
\cmsTable{
\renewcommand{\arraystretch}{1.2}
\begin{scotch}{l|r@{\hspace{4mm}}lr@{\hspace{4mm}}lr@{\hspace{4mm}}l}
Distribution & $\chi^2/\mathrm{dof}$ & p-value & $\chi^2/\mathrm{dof}$ & p-value & $\chi^2/\mathrm{dof}$ & p-value\\\hline
 & \multicolumn{2}{c}{\POWHEG{}+P8} & \multicolumn{2}{c}{\POWHEG{}+H++} & \multicolumn{2}{c}{\AMCATNLO{}+P8 MLM}\\
 & \multicolumn{2}{c}{Order: NLO} & \multicolumn{2}{c}{Order: NLO} & \multicolumn{2}{c}{Order: LO, up to 3 add. partons}\\
$\pt(\tqh)$ & 14.2/9&0.115 & 24.0/9&$<0.01$ & 32.8/9&$<0.01$\\
$\abs{y(\tqh)}$ & 3.47/7&0.838 & 5.66/7&0.579 & 6.64/7&0.468\\
$\pt(\tql)$ & 20.8/9&0.013 & 38.2/9&$<0.01$ & 49.7/9&$<0.01$\\
$\abs{y(\tql)}$ & 6.37/7&0.497 & 9.69/7&0.207 & 16.1/7&0.025\\
$M(\ttbar)$ & 9.03/8&0.340 & 148/8&$<0.01$ & 12.0/8&0.151\\
$\pt(\ttbar)$ & 2.15/5&0.829 & 29.4/5&$<0.01$ & 49.2/5&$<0.01$\\
$\abs{y(\ttbar)}$ & 0.869/6&0.990 & 2.06/6&0.914 & 13.2/6&0.040\\
Additional jets & 28.2/5&$<0.01$ & 17.2/5&$<0.01$ & 36.8/5&$<0.01$\\
Additional jets \vs $\pt(\ttbar)$ & 70.7/20&$<0.01$ & 86.1/20&$<0.01$ & 161/20&$<0.01$\\
Additional jets \vs $\pt(\tqh)$ & 91.6/36&$<0.01$ & 200/36&$<0.01$ & 162/36&$<0.01$\\
$\abs{y(\tqh)}$ \vs $\pt(\tqh)$ & 56.2/36&0.017 & 197/36&$<0.01$ & 114/36&$<0.01$\\
$M(\ttbar)$ \vs $\abs{y(\ttbar)}$ & 26.6/24&0.324 & 263/24&$<0.01$ & 38.1/24&0.034\\
$\pt(\ttbar)$ \vs $M(\ttbar)$ & 13.4/32&0.998 & 459/32&$<0.01$ & 89.0/32&$<0.01$\\\hline
 & \multicolumn{2}{c}{\AMCATNLO{}+P8} & \multicolumn{2}{c}{\AMCATNLO{}+H++} & \multicolumn{2}{c}{\AMCATNLO{}+P8 FxFx}\\
 & \multicolumn{2}{c}{Order: NLO} & \multicolumn{2}{c}{Order: NLO} & \multicolumn{2}{c}{Order: NLO, up to 2 add. partons}\\
$\pt(\tqh)$ & 11.9/9&0.221 & 5.51/9&0.788 & 4.17/9&0.900\\
$\abs{y(\tqh)}$ & 7.34/7&0.394 & 10.6/7&0.156 & 5.93/7&0.547\\
$\pt(\tql)$ & 11.0/9&0.274 & 6.37/9&0.702 & 6.51/9&0.688\\
$\abs{y(\tql)}$ & 12.3/7&0.092 & 6.04/7&0.535 & 14.3/7&0.047\\
$M(\ttbar)$ & 9.57/8&0.296 & 28.7/8&$<0.01$ & 28.5/8&$<0.01$\\
$\pt(\ttbar)$ & 37.1/5&$<0.01$ & 7.92/5&0.161 & 29.6/5&$<0.01$\\
$\abs{y(\ttbar)}$ & 1.75/6&0.942 & 1.98/6&0.922 & 2.87/6&0.825\\
Additional jets & 29.6/5&$<0.01$ & 12.2/5&0.032 & 11.6/5&0.041\\
Additional jets \vs $\pt(\ttbar)$ & 197/20&$<0.01$ & 163/20&$<0.01$ & 85.3/20&$<0.01$\\
Additional jets \vs $\pt(\tqh)$ & 151/36&$<0.01$ & 57.7/36&0.012 & 40.4/36&0.282\\
$\abs{y(\tqh)}$ \vs $\pt(\tqh)$ & 36.6/36&0.441 & 82.5/36&$<0.01$ & 42.2/36&0.222\\
$M(\ttbar)$ \vs $\abs{y(\ttbar)}$ & 21.4/24&0.612 & 47.9/24&$<0.01$ & 52.3/24&$<0.01$\\
$\pt(\ttbar)$ \vs $M(\ttbar)$ & 119/32&$<0.01$ & 164/32&$<0.01$ & 107/32&$<0.01$\\
\end{scotch}
}
 \label{REST2}
\end{table*}

The observed cross sections are slightly lower than expected. However, taking into account the systematic uncertainties, that are highly correlated among the bins, there is no significant deviation. In general, the measured distributions are in agreement with the predictions of the event generators with some exceptions in the $\pt(\ttbar)$ and $M(\ttbar)$ distributions. The jet multiplicities are lower than predicted by almost all simulations. The measured \pt of the top quarks is slightly softer than predicted. Such an effect has already been observed in previous measurements~\cite{Aad:2015eia,Chatrchyan:2012saa,Aad:2015mbv,Khachatryan:2015oqa}. However, the comparison between the \HERWIGpp and \PYTHIAA simulations together with the same matrix-element calculations show the large impact of the parton shower and hadronization modeling. The parton-level results are well described by the matrix-element calculations. Especially, the soft \pt of the top quarks is predicted by the NNLO and NLO+NNLL' QCD calculation.

\section{Summary}
Measurements of the differential and double-differential cross sections for \ttbar production in proton-proton collisions at 13\TeV have been presented. The data correspond to an integrated luminosity of 2.3\,\fbinv recorded by the CMS experiment. The \ttbar production cross section is measured in the lepton+jets channel as a function of transverse momentum \pt and rapidity $\abs{y}$ of the top quarks; \pt, $\abs{y}$, and invariant mass of the \ttbar system; and the number of additional jets. The measurement at parton level is dominated by the uncertainties in the parton shower and hadronization modeling. The dependence on these theoretical models is reduced for the particle-level measurement, for which the experimental uncertainties of jet energy calibration and $\PQb$ tagging efficiency are dominant.

The results are compared to several standard model predictions that use different methods and approximations for their calculations. In general, the measured cross sections are slightly lower than predicted, but within the uncertainty compatible with the expectation. The measured distributions are in agreement with the predictions of the event generators with some exceptions in the $\pt(\ttbar)$ and $M(\ttbar)$ distributions. The number of additional jets is lower and the measured \pt of the top quarks is slightly softer than predicted by most of the event generators. A softer \pt of the top quarks has already been observed in previous measurements and is predicted by the NNLO and the NLO+NNLL' QCD calculation.

\begin{acknowledgments}
\hyphenation{Bundes-ministerium Forschungs-gemeinschaft Forschungs-zentren Rachada-pisek} We congratulate our colleagues in the CERN accelerator departments for the excellent performance of the LHC and thank the technical and administrative staffs at CERN and at other CMS institutes for their contributions to the success of the CMS effort. In addition, we gratefully acknowledge the computing centers and personnel of the Worldwide LHC Computing Grid for delivering so effectively the computing infrastructure essential to our analyses. Finally, we acknowledge the enduring support for the construction and operation of the LHC and the CMS detector provided by the following funding agencies: the Austrian Federal Ministry of Science, Research and Economy and the Austrian Science Fund; the Belgian Fonds de la Recherche Scientifique, and Fonds voor Wetenschappelijk Onderzoek; the Brazilian Funding Agencies (CNPq, CAPES, FAPERJ, and FAPESP); the Bulgarian Ministry of Education and Science; CERN; the Chinese Academy of Sciences, Ministry of Science and Technology, and National Natural Science Foundation of China; the Colombian Funding Agency (COLCIENCIAS); the Croatian Ministry of Science, Education and Sport, and the Croatian Science Foundation; the Research Promotion Foundation, Cyprus; the Secretariat for Higher Education, Science, Technology and Innovation, Ecuador; the Ministry of Education and Research, Estonian Research Council via IUT23-4 and IUT23-6 and European Regional Development Fund, Estonia; the Academy of Finland, Finnish Ministry of Education and Culture, and Helsinki Institute of Physics; the Institut National de Physique Nucl\'eaire et de Physique des Particules~/~CNRS, and Commissariat \`a l'\'Energie Atomique et aux \'Energies Alternatives~/~CEA, France; the Bundesministerium f\"ur Bildung und Forschung, Deutsche Forschungsgemeinschaft, and Helmholtz-Gemeinschaft Deutscher Forschungszentren, Germany; the General Secretariat for Research and Technology, Greece; the National Scientific Research Foundation, and National Innovation Office, Hungary; the Department of Atomic Energy and the Department of Science and Technology, India; the Institute for Studies in Theoretical Physics and Mathematics, Iran; the Science Foundation, Ireland; the Istituto Nazionale di Fisica Nucleare, Italy; the Ministry of Science, ICT and Future Planning, and National Research Foundation (NRF), Republic of Korea; the Lithuanian Academy of Sciences; the Ministry of Education, and University of Malaya (Malaysia); the Mexican Funding Agencies (BUAP, CINVESTAV, CONACYT, LNS, SEP, and UASLP-FAI); the Ministry of Business, Innovation and Employment, New Zealand; the Pakistan Atomic Energy Commission; the Ministry of Science and Higher Education and the National Science Centre, Poland; the Funda\c{c}\~ao para a Ci\^encia e a Tecnologia, Portugal; JINR, Dubna; the Ministry of Education and Science of the Russian Federation, the Federal Agency of Atomic Energy of the Russian Federation, Russian Academy of Sciences, and the Russian Foundation for Basic Research; the Ministry of Education, Science and Technological Development of Serbia; the Secretar\'{\i}a de Estado de Investigaci\'on, Desarrollo e Innovaci\'on and Programa Consolider-Ingenio 2010, Spain; the Swiss Funding Agencies (ETH Board, ETH Zurich, PSI, SNF, UniZH, Canton Zurich, and SER); the Ministry of Science and Technology, Taipei; the Thailand Center of Excellence in Physics, the Institute for the Promotion of Teaching Science and Technology of Thailand, Special Task Force for Activating Research and the National Science and Technology Development Agency of Thailand; the Scientific and Technical Research Council of Turkey, and Turkish Atomic Energy Authority; the National Academy of Sciences of Ukraine, and State Fund for Fundamental Researches, Ukraine; the Science and Technology Facilities Council, UK; the US Department of Energy, and the US National Science Foundation.

Individuals have received support from the Marie-Curie program and the European Research Council and EPLANET (European Union); the Leventis Foundation; the A. P. Sloan Foundation; the Alexander von Humboldt Foundation; the Belgian Federal Science Policy Office; the Fonds pour la Formation \`a la Recherche dans l'Industrie et dans l'Agriculture (FRIA-Belgium); the Agentschap voor Innovatie door Wetenschap en Technologie (IWT-Belgium); the Ministry of Education, Youth and Sports (MEYS) of the Czech Republic; the Council of Science and Industrial Research, India; the HOMING PLUS program of the Foundation for Polish Science, cofinanced from European Union, Regional Development Fund, the Mobility Plus program of the Ministry of Science and Higher Education, the National Science Center (Poland), contracts Harmonia 2014/14/M/ST2/00428, Opus 2013/11/B/ST2/04202, 2014/13/B/ST2/02543 and 2014/15/B/ST2/03998, Sonata-bis 2012/07/E/ST2/01406; the Thalis and Aristeia programs cofinanced by EU-ESF and the Greek NSRF; the National Priorities Research Program by Qatar National Research Fund; the Programa Clar\'in-COFUND del Principado de Asturias; the Rachadapisek Sompot Fund for Postdoctoral Fellowship, Chulalongkorn University and the Chulalongkorn Academic into Its 2nd Century Project Advancement Project (Thailand); and the Welch Foundation, contract C-1845.
\end{acknowledgments}
\clearpage
\bibliography{auto_generated}

\providecommand{\href}[2]{#2}\begingroup\raggedright\begin{thebibliography}{10}%
\makeatletter
\providecommand{\hrefCMSnoop }[0]{\@secondoftwo}%
\makeatother
\providecommand{\doi}{\texttt{doi:}\begingroup \urlstyle{tt}\Url}

\bibitem{LUMI}
\href {https://cds.cern.ch/record/2138682}{{CMS Collaboration}, ``CMS
  Luminosity Measurement for the 2015 Data Taking Period'',} CMS Physics
  Analysis Summary CMS-PAS-LUM-15-001, 2017.

\bibitem{Chatrchyan:2012saa}
\hrefCMSnoop {}{{CMS} Collaboration, ``{Measurement of differential top-quark
  pair production cross sections in pp colisions at $\sqrt{s}=7$\,TeV}'',}
  \textit{ Eur. Phys. J. C} \textbf{ 73} (2013) 2339,
  \href{http://dx.doi.org/10.1140/epjc/s10052-013-2339-4}{\doi{10.1140/epjc/s10052-013-2339-4}},
\href{http://www.arXiv.org/abs/1211.2220}{\texttt{arXiv:1211.2220}}.

\bibitem{Aad:2015eia}
\hrefCMSnoop {}{{ATLAS} Collaboration, ``{Differential top-antitop
  cross-section measurements as a function of observables constructed from
  final-state particles using pp collisions at $\sqrt{s}=7$\,TeV in the ATLAS
  detector}'',} \textit{ JHEP} \textbf{ 06} (2015) 100,
  \href{http://dx.doi.org/10.1007/JHEP06(2015)100}{\doi{10.1007/JHEP06(2015)100}},
\href{http://www.arXiv.org/abs/1502.05923}{\texttt{arXiv:1502.05923}}.

\bibitem{Khachatryan:2015oqa}
\hrefCMSnoop {}{{CMS} Collaboration, ``{Measurement of the differential cross
  section for top quark pair production in pp collisions at $\sqrt{s} =
  8\,\text {TeV} $}'',} \textit{ Eur. Phys. J. C} \textbf{ 75} (2015) 542,
  \href{http://dx.doi.org/10.1140/epjc/s10052-015-3709-x}{\doi{10.1140/epjc/s10052-015-3709-x}},
\href{http://www.arXiv.org/abs/1505.04480}{\texttt{arXiv:1505.04480}}.

\bibitem{Aad:2015mbv}
\hrefCMSnoop {}{{ATLAS} Collaboration, ``{Measurements of top-quark pair
  differential cross-sections in the lepton+jets channel in pp collisions at
  $\sqrt{s}=8$\,TeV using the ATLAS detector}'',} \textit{ Eur. Phys. J. C}
  \textbf{ 76} (2016) 538,
  \href{http://dx.doi.org/10.1140/epjc/s10052-016-4366-4}{\doi{10.1140/epjc/s10052-016-4366-4}},
\href{http://www.arXiv.org/abs/1511.04716}{\texttt{arXiv:1511.04716}}.

\bibitem{Aad:2015hna}
\hrefCMSnoop {}{{ATLAS} Collaboration, ``{Measurement of the differential
  cross-section of highly boosted top quarks as a function of their transverse
  momentum in $\sqrt{s}$ = 8\,TeV proton-proton collisions using the ATLAS
  detector}'',} \textit{ Phys. Rev. D} \textbf{ 93} (2016) 032009,
  \href{http://dx.doi.org/10.1103/PhysRevD.93.032009}{\doi{10.1103/PhysRevD.93.032009}},
\href{http://www.arXiv.org/abs/1510.03818}{\texttt{arXiv:1510.03818}}.

\bibitem{Khachatryan:2015fwh}
\hrefCMSnoop {}{{CMS} Collaboration, ``{Measurement of the $\ttbar$ production
  cross section in the all-jets final state in pp collisions at
  $\sqrt{s}=8\,\text{TeV}$}'',} \textit{ Eur. Phys. J. C} \textbf{ 76} (2016)
  128,
  \href{http://dx.doi.org/10.1140/epjc/s10052-016-3956-5}{\doi{10.1140/epjc/s10052-016-3956-5}},
\href{http://www.arXiv.org/abs/1509.06076}{\texttt{arXiv:1509.06076}}.

\bibitem{Khachatryan:2149620}
\hrefCMSnoop {}{{CMS Collaboration}, ``{Measurement of the integrated and
  differential $\mathrm{t \bar{t}}$ production cross sections for
  high-$p_{\mathrm{T}}$ top quarks in pp collisions at $ \sqrt{s} = $ 8
  TeV}'',} \textit{ Phys. Rev. D} \textbf{ 94} (2016) 072002,
  \href{http://dx.doi.org/10.1103/PhysRevD.94.072002}{\doi{10.1103/PhysRevD.94.072002}},
  \href{http://www.arXiv.org/abs/1605.00116}{\texttt{arXiv:1605.00116}}.

\bibitem{Nason:2004rx}
\hrefCMSnoop {}{P.~Nason, ``{A new method for combining NLO QCD with shower
  Monte Carlo algorithms}'',} \textit{ JHEP} \textbf{ 11} (2004) 040,
  \href{http://dx.doi.org/10.1088/1126-6708/2004/11/040}{\doi{10.1088/1126-6708/2004/11/040}},
\href{http://www.arXiv.org/abs/hep-ph/0409146}{\texttt{arXiv:hep-ph/0409146}}.

\bibitem{Frixione:2007vw}
\hrefCMSnoop {}{S.~Frixione, P.~Nason, and C.~Oleari, ``{Matching NLO QCD
  computations with Parton Shower simulations: the POWHEG method}'',} \textit{
  JHEP} \textbf{ 11} (2007) 070,
  \href{http://dx.doi.org/10.1088/1126-6708/2007/11/070}{\doi{10.1088/1126-6708/2007/11/070}},
\href{http://www.arXiv.org/abs/0709.2092}{\texttt{arXiv:0709.2092}}.

\bibitem{Alioli:2010xd}
\hrefCMSnoop {}{S.~Alioli, P.~Nason, C.~Oleari, and E.~Re, ``{A general
  framework for implementing NLO calculations in shower Monte Carlo programs:
  the POWHEG BOX}'',} \textit{ JHEP} \textbf{ 06} (2010) 043,
  \href{http://dx.doi.org/10.1007/JHEP06(2010)043}{\doi{10.1007/JHEP06(2010)043}},
\href{http://www.arXiv.org/abs/1002.2581}{\texttt{arXiv:1002.2581}}.

\bibitem{Campbell:2014kua}
\hrefCMSnoop {}{J.~M. Campbell, R.~K. Ellis, P.~Nason, and E.~Re, ``{Top-pair
  production and decay at NLO matched with parton showers}'',} \textit{ JHEP}
  \textbf{ 04} (2015) 114,
  \href{http://dx.doi.org/10.1007/JHEP04(2015)114}{\doi{10.1007/JHEP04(2015)114}},
\href{http://www.arXiv.org/abs/1412.1828}{\texttt{arXiv:1412.1828}}.

\bibitem{Alwall:2014hca}
J.~Alwall\hrefCMSnoop {}{ {et~al.}, ``{The automated computation of tree-level
  and next-to-leading order differential cross sections, and their matching to
  parton shower simulations}'',} \textit{ JHEP} \textbf{ 07} (2014) 079,
  \href{http://dx.doi.org/10.1007/JHEP07(2014)079}{\doi{10.1007/JHEP07(2014)079}},
\href{http://www.arXiv.org/abs/1405.0301}{\texttt{arXiv:1405.0301}}.

\bibitem{Sjostrand:2006za}
\hrefCMSnoop {}{T.~Sj{\"o}strand, S.~Mrenna, and P.~Skands, ``{PYTHIA} 6.4
  physics and manual'',} \textit{ JHEP} \textbf{ 05} (2006) 026,
  \href{http://dx.doi.org/10.1088/1126-6708/2006/05/026}{\doi{10.1088/1126-6708/2006/05/026}},
\href{http://www.arXiv.org/abs/hep-ph/0603175}{\texttt{arXiv:hep-ph/0603175}}.

\bibitem{Sjostrand:2007gs}
\hrefCMSnoop {}{T.~Sj{\"o}strand, S.~Mrenna, and P.~Z. Skands, ``A brief
  introduction to {PYTHIA 8.1}'',} \textit{ Comput. Phys. Commun.} \textbf{
  178} (2008) 852,
  \href{http://dx.doi.org/10.1016/j.cpc.2008.01.036}{\doi{10.1016/j.cpc.2008.01.036}},
\href{http://www.arXiv.org/abs/0710.3820}{\texttt{arXiv:0710.3820}}.

\bibitem{Skands:2014pea}
\hrefCMSnoop {}{P.~Skands, S.~Carrazza, and J.~Rojo, ``{Tuning PYTHIA 8.1: the
  Monash 2013 Tune}'',} \textit{ Eur. Phys. J. C} \textbf{ 74} (2014) 3024,
  \href{http://dx.doi.org/10.1140/epjc/s10052-014-3024-y}{\doi{10.1140/epjc/s10052-014-3024-y}},
\href{http://www.arXiv.org/abs/1404.5630}{\texttt{arXiv:1404.5630}}.

\bibitem{MLM}
J.~Alwall\hrefCMSnoop {}{ {et~al.}, ``Comparative study of various algorithms
  for the merging of parton showers and matrix elements in hadronic
  collisions'',} \textit{ Eur. Phys. J. C} \textbf{ 53} (2008) 473,
  \href{http://dx.doi.org/10.1140/epjc/s10052-007-0490-5}{\doi{10.1140/epjc/s10052-007-0490-5}},
  \href{http://www.arXiv.org/abs/0706.2569}{\texttt{arXiv:0706.2569}}.

\bibitem{Frederix:2012ps}
\hrefCMSnoop {}{R.~Frederix and S.~Frixione, ``{Merging meets matching in
  MC@NLO}'',} \textit{ JHEP} \textbf{ 12} (2012) 061,
  \href{http://dx.doi.org/10.1007/JHEP12(2012)061}{\doi{10.1007/JHEP12(2012)061}},
\href{http://www.arXiv.org/abs/1209.6215}{\texttt{arXiv:1209.6215}}.

\bibitem{Ball:2014uwa}
\hrefCMSnoop {}{{NNPDF} Collaboration, ``{Parton distributions for the LHC Run
  II}'',} \textit{ JHEP} \textbf{ 04} (2015) 040,
  \href{http://dx.doi.org/10.1007/JHEP04(2015)040}{\doi{10.1007/JHEP04(2015)040}},
\href{http://www.arXiv.org/abs/1410.8849}{\texttt{arXiv:1410.8849}}.

\bibitem{Czakon:2011xx}
\hrefCMSnoop {}{M.~Czakon and A.~Mitov, ``{Top++: A Program for the Calculation
  of the Top-Pair Cross-Section at Hadron Colliders}'',} \textit{ Comput. Phys.
  Commun.} \textbf{ 185} (2014) 2930,
  \href{http://dx.doi.org/10.1016/j.cpc.2014.06.021}{\doi{10.1016/j.cpc.2014.06.021}},
\href{http://www.arXiv.org/abs/1112.5675}{\texttt{arXiv:1112.5675}}.

\bibitem{Bahr:2008pv}
M.~B{\"a}hr\hrefCMSnoop {}{ {et~al.}, ``Herwig++ physics and manual'',}
  \textit{ Eur. Phys. J. C} \textbf{ 58} (2008) 639,
  \href{http://dx.doi.org/10.1140/epjc/s10052-008-0798-9}{\doi{10.1140/epjc/s10052-008-0798-9}},
\href{http://www.arXiv.org/abs/0803.0883}{\texttt{arXiv:0803.0883}}.

\bibitem{Seymour:2013qka}
\hrefCMSnoop {}{M.~H. Seymour and A.~Siodmok, ``{Constraining MPI models using
  $\sigma_{eff}$ and recent Tevatron and LHC Underlying Event data}'',}
  \textit{ JHEP} \textbf{ 10} (2013) 113,
  \href{http://dx.doi.org/10.1007/JHEP10(2013)113}{\doi{10.1007/JHEP10(2013)113}},
\href{http://www.arXiv.org/abs/1307.5015}{\texttt{arXiv:1307.5015}}.

\bibitem{Li:2012wna}
\hrefCMSnoop {}{Y.~Li and F.~Petriello, ``{Combining QCD and electroweak
  corrections to dilepton production in FEWZ}'',} \textit{ Phys. Rev. D}
  \textbf{ 86} (2012) 094034,
  \href{http://dx.doi.org/10.1103/PhysRevD.86.094034}{\doi{10.1103/PhysRevD.86.094034}},
\href{http://www.arXiv.org/abs/1208.5967}{\texttt{arXiv:1208.5967}}.

\bibitem{Kant:2014oha}
P.~Kant\hrefCMSnoop {}{ {et~al.}, ``{HatHor for single top-quark production:
  Updated predictions and uncertainty estimates for single top-quark production
  in hadronic collisions}'',} \textit{ Comput. Phys. Commun.} \textbf{ 191}
  (2015) 74,
  \href{http://dx.doi.org/10.1016/j.cpc.2015.02.001}{\doi{10.1016/j.cpc.2015.02.001}},
\href{http://www.arXiv.org/abs/1406.4403}{\texttt{arXiv:1406.4403}}.

\bibitem{Kidonakis:2012rm}
\hrefCMSnoop {}{N.~Kidonakis, ``{NNLL threshold resummation for top-pair and
  single-top production}'',} \textit{ Phys. Part. Nucl.} \textbf{ 45} (2014)
  714,
  \href{http://dx.doi.org/10.1134/S1063779614040091}{\doi{10.1134/S1063779614040091}},
\href{http://www.arXiv.org/abs/1210.7813}{\texttt{arXiv:1210.7813}}.

\bibitem{Allison:2006ve}
\hrefCMSnoop {}{J.~Allison {et~al.}, ``{Geant4 developments and
  applications}'',} \textit{ IEEE Trans. Nucl. Sci.} \textbf{ 53} (2006) 270,
\href{http://dx.doi.org/10.1109/TNS.2006.869826}{\doi{10.1109/TNS.2006.869826}}.

\bibitem{Cacciari:2008gp}
\hrefCMSnoop {}{M.~Cacciari, G.~P. Salam, and G.~Soyez, ``The
  anti-$k_\mathrm{t}$ jet clustering algorithm'',} \textit{ JHEP} \textbf{ 04}
  (2008) 063,
  \href{http://dx.doi.org/10.1088/1126-6708/2008/04/063}{\doi{10.1088/1126-6708/2008/04/063}},
  \href{http://www.arXiv.org/abs/0802.1189}{\texttt{arXiv:0802.1189}}.

\bibitem{Cacciari:2011ma}
\hrefCMSnoop {}{M.~Cacciari, G.~P. Salam, and G.~Soyez, ``{FastJet user
  manual}'',} \textit{ Eur. Phys. J. C} \textbf{ 72} (2012) 1896,
  \href{http://dx.doi.org/10.1140/epjc/s10052-012-1896-2}{\doi{10.1140/epjc/s10052-012-1896-2}},
\href{http://www.arXiv.org/abs/1111.6097}{\texttt{arXiv:1111.6097}}.

\bibitem{Chatrchyan:2008zzk}
\hrefCMSnoop {}{{CMS} Collaboration, ``The {CMS} experiment at the {CERN}
  {LHC}'',} \textit{ JINST} \textbf{ 3} (2008) S08004,
\href{http://dx.doi.org/10.1088/1748-0221/3/08/S08004}{\doi{10.1088/1748-0221/3/08/S08004}}.

\bibitem{CMS-PAS-PFT-09-001}
\href {http://cdsweb.cern.ch/record/1194487}{{CMS} Collaboration,
  ``{Particle-flow event reconstruction in {CMS} and performance for jets,
  taus, and {\MET}}'',} CMS Physics Analysis Summary CMS-PAS-PFT-09-001, 2009.

\bibitem{CMS-PAS-PFT-10-001}
\href {http://cdsweb.cern.ch/record/1247373}{{CMS} Collaboration,
  ``{Commissioning of the particle-flow event reconstruction with the first
  {LHC} collisions recorded in the {CMS} detector}'',} CMS Physics Analysis
  Summary CMS-PAS-PFT-10-001, 2010.

\bibitem{Chatrchyan:2012xi}
\hrefCMSnoop {}{{CMS} Collaboration, ``{Performance of CMS muon reconstruction
  in pp collision events at $\sqrt{s} = 7$\TeV}'',} \textit{ JINST} \textbf{ 7}
  (2012) P10002,
  \href{http://dx.doi.org/10.1088/1748-0221/7/10/P10002}{\doi{10.1088/1748-0221/7/10/P10002}},
\href{http://www.arXiv.org/abs/1206.4071}{\texttt{arXiv:1206.4071}}.

\bibitem{TNPREF}
\hrefCMSnoop {}{{CMS} Collaboration, ``{Measurement of inclusive W and Z boson
  production cross sections in $pp$ collisions at $\sqrt{s} = 8\TeV$}'',}
  \textit{ Phys. Rev. Lett.} \textbf{ 112} (2014) 191802,
  \href{http://dx.doi.org/10.1103/PhysRevLett.112.191802}{\doi{10.1103/PhysRevLett.112.191802}},
\href{http://www.arXiv.org/abs/1402.0923}{\texttt{arXiv:1402.0923}}.

\bibitem{Cacciari:2007}
\hrefCMSnoop {}{M.~Cacciari and G.~P. Salam, ``Pileup subtraction using jet
  areas'',} \textit{ Phys. Lett. B} \textbf{ 659} (2008) 119,
  \href{http://dx.doi.org/10.1016/j.physletb.2007.09.077}{\doi{10.1016/j.physletb.2007.09.077}},
  \href{http://www.arXiv.org/abs/0707.1378}{\texttt{arXiv:0707.1378}}.

\bibitem{Khachatryan:2015hwa}
\hrefCMSnoop {}{{CMS} Collaboration, ``{Performance of electron reconstruction
  and selection with the CMS detector in proton-proton collisions at $\sqrt{s}
  = 8$\TeV}'',} \textit{ JINST} \textbf{ 10} (2015) P06005,
  \href{http://dx.doi.org/10.1088/1748-0221/10/06/P06005}{\doi{10.1088/1748-0221/10/06/P06005}},
\href{http://www.arXiv.org/abs/1502.02701}{\texttt{arXiv:1502.02701}}.

\bibitem{JET}
\hrefCMSnoop {}{{CMS Collaboration}, ``{Jet energy scale and resolution in the
  CMS experiment in pp collisions at 8\,TeV}'',} (2016).
  \href{http://www.arXiv.org/abs/1607.03663}{\texttt{arXiv:1607.03663}}.
Submitted to \textit{JINST}.

\bibitem{BTV}
\href {https://cds.cern.ch/record/2138504}{{CMS Collaboration},
  ``{Identification of b quark jets at the CMS Experiment in the LHC Run 2}'',}
  CMS Physics Analysis Summary CMS-PAS-BTV-15-001, 2016.

\bibitem{Betchart:2013nba}
\hrefCMSnoop {}{B.~A. Betchart, R.~Demina, and A.~Harel, ``Analytic solutions
  for neutrino momenta in decay of top quarks'',} \textit{ Nucl. Instrum. Meth.
  A} \textbf{ 736} (2014) 169,
  \href{http://dx.doi.org/10.1016/j.nima.2013.10.039}{\doi{10.1016/j.nima.2013.10.039}},
\href{http://www.arXiv.org/abs/1305.1878}{\texttt{arXiv:1305.1878}}.

\bibitem{D'Agostini:1994zf}
\hrefCMSnoop {}{G.~D'Agostini, ``{A multidimensional unfolding method based on
  Bayes' theorem}'',} \textit{ Nucl. Instrum. Meth. A} \textbf{ 362} (1995)
  487,
\href{http://dx.doi.org/10.1016/0168-9002(95)00274-X}{\doi{10.1016/0168-9002(95)00274-X}}.

\bibitem{Guzzi:2014wia}
\hrefCMSnoop {}{M.~Guzzi, K.~Lipka, and S.-O. Moch, ``{Top-quark pair
  production at hadron colliders: differential cross section and
  phenomenological applications with DiffTop}'',} \textit{ JHEP} \textbf{ 01}
  (2015) 082,
  \href{http://dx.doi.org/10.1007/JHEP01(2015)082}{\doi{10.1007/JHEP01(2015)082}},
\href{http://www.arXiv.org/abs/1406.0386}{\texttt{arXiv:1406.0386}}.

\bibitem{Dulat:2015mca}
S.~Dulat\hrefCMSnoop {}{ {et~al.}, ``{New parton distribution functions from a
  global analysis of quantum chromodynamics}'',} \textit{ Phys. Rev. D}
  \textbf{ 93} (2016) 033006,
  \href{http://dx.doi.org/10.1103/PhysRevD.93.033006}{\doi{10.1103/PhysRevD.93.033006}},
\href{http://www.arXiv.org/abs/1506.07443}{\texttt{arXiv:1506.07443}}.

\bibitem{ANNNLO}
\hrefCMSnoop {}{N.~Kidonakis, ``{NNNLO soft-gluon corrections for the
  top-antitop pair production cross section}'',} \textit{ Phys. Rev. D}
  \textbf{ 90} (2014) 014006,
  \href{http://dx.doi.org/10.1103/PhysRevD.90.014006}{\doi{10.1103/PhysRevD.90.014006}},
\href{http://www.arXiv.org/abs/1405.7046}{\texttt{arXiv:1405.7046}}.

\bibitem{ANNNLOdiff}
\hrefCMSnoop {}{N.~Kidonakis, ``{NNNLO soft-gluon corrections for the top-quark
  \pt and rapidity distributions}'',} \textit{ Phys. Rev. D} \textbf{ 91}
  (2015) 031501,
  \href{http://dx.doi.org/10.1103/PhysRevD.91.031501}{\doi{10.1103/PhysRevD.91.031501}},
\href{http://www.arXiv.org/abs/1411.2633}{\texttt{arXiv:1411.2633}}.

\bibitem{Martin:2009iq}
\hrefCMSnoop {}{A.~D. Martin, W.~J. Stirling, R.~S. Thorne, and G.~Watt,
  ``{Parton distributions for the LHC}'',} \textit{ Eur. Phys. J. C} \textbf{
  63} (2009) 189,
  \href{http://dx.doi.org/10.1140/epjc/s10052-009-1072-5}{\doi{10.1140/epjc/s10052-009-1072-5}},
\href{http://www.arXiv.org/abs/0901.0002}{\texttt{arXiv:0901.0002}}.

\bibitem{NLONNLL}
\hrefCMSnoop {}{B.~Pecjak, D.~Scott, X.~Wang, and L.~L. Yang, ``{Resummed
  Differential Cross Sections for Top-Quark Pairs at the LHC}'',} \textit{
  Phys. Rev. Lett.} \textbf{ 116} (2016) 202001,
  \href{http://dx.doi.org/10.1103/PhysRevLett.116.202001}{\doi{10.1103/PhysRevLett.116.202001}},
\href{http://www.arXiv.org/abs/1601.07020}{\texttt{arXiv:1601.07020}}.

\bibitem{NNLO}
\hrefCMSnoop {}{M.~Czakon, D.~Heymes, and A.~Mitov, ``{High-Precision
  Differential Predictions for Top-Quark Pairs at the LHC}'',} \textit{ Phys.
  Rev. Lett.} \textbf{ 116} (2016) 082003,
  \href{http://dx.doi.org/10.1103/PhysRevLett.116.082003}{\doi{10.1103/PhysRevLett.116.082003}},
  \href{http://www.arXiv.org/abs/1511.00549}{\texttt{arXiv:1511.00549}}.

\end{thebibliography}\endgroup
\clearpage
\appendix
\begin{appendix}
\section{Tables of parton level cross sections.}
\label{APP1}
\begin{table}[htbp]
\caption{Differential cross section at parton level as a function of $\pt(\tqh)$. The values are shown together with their statistical and systematic uncertainties.}
\centering
\renewcommand{\arraystretch}{1.1}
\begin{scotch}{xr@{$\pm$}c@{$\pm$}l|xr@{$\pm$}c@{$\pm$}l}
\multicolumn{1}{c}{$\pt(\tqh)$} & \multicolumn{3}{c}{$\frac{\rd\sigma}{\rd\pt(\tqh)}$} & \multicolumn{1}{c}{$\pt(\tqh)$} & \multicolumn{3}{c}{$\frac{\rd\sigma}{\rd\pt(\tqh)}$}\\
\multicolumn{1}{c}{[\GeVns{}]} & \multicolumn{3}{c}{[fb\GeV{}$^{-1}$]} & \multicolumn{1}{c}{[\GeVns{}]} & \multicolumn{3}{c}{[fb\GeV{}$^{-1}$]}\\[3pt]
\hline
0,45 & 680&20&180 & 225,270 & 228&6&32\\
45,90 & 1500&20&190 & 270,315 & 119&5&18\\
90,135 & 1290&20&160 & 315,400 & 46&2&7\\
135,180 & 790&10&100 & 400,800 & 5.1&0.3&0.8\\
180,225 & 420&9&59 & \multicolumn{4}{c}{\NA}\\
\end{scotch}
\end{table}
\begin{table}[htbp]
\caption{Differential cross section at parton level as a function of $\abs{y(\tqh)}$. The values are shown together with their statistical and systematic uncertainties.}
\centering
\renewcommand{\arraystretch}{1.1}
\begin{scotch}{xr@{$\pm$}c@{$\pm$}l|xr@{$\pm$}c@{$\pm$}l}
\multicolumn{1}{c}{$\abs{y(\tqh)}$} & \multicolumn{3}{c}{$\frac{\rd\sigma}{\rd \abs{y(\tqh)}}$ [pb]} & \multicolumn{1}{c}{$\abs{y(\tqh)}$} & \multicolumn{3}{c}{$\frac{\rd\sigma}{\rd \abs{y(\tqh)}}$ [pb]}\\[3pt]
\hline
0,0.2 & 142&2&14 & 1,1.3 & 100&2&11\\
0.2,0.4 & 135&2&13 & 1.3,1.6 & 82&2&11\\
0.4,0.7 & 129&2&13 & 1.6,2.5 & 44.0&0.9&6.4\\
0.7,1 & 114&2&12 & \multicolumn{4}{c}{\NA}\\
\end{scotch}
\end{table}
\begin{table}[htbp]
\caption{Differential cross section at parton level as a function of $\pt(\tql)$. The values are shown together with their statistical and systematic uncertainties.}
\centering
\renewcommand{\arraystretch}{1.1}
\begin{scotch}{xr@{$\pm$}c@{$\pm$}l|xr@{$\pm$}c@{$\pm$}l}
\multicolumn{1}{c}{$\pt(\tql)$} & \multicolumn{3}{c}{$\frac{\rd\sigma}{\rd\pt(\tql)}$} & \multicolumn{1}{c}{$\pt(\tql)$} & \multicolumn{3}{c}{$\frac{\rd\sigma}{\rd\pt(\tql)}$}\\
\multicolumn{1}{c}{[\GeVns{}]} & \multicolumn{3}{c}{[fb\GeV{}$^{-1}$]} & \multicolumn{1}{c}{[\GeVns{}]} & \multicolumn{3}{c}{[fb\GeV{}$^{-1}$]}\\[3pt]
\hline
0,45 & 690&10&100 & 225,270 & 218&4&20\\
45,90 & 1470&20&190 & 270,315 & 115&3&14\\
90,135 & 1300&10&150 & 315,400 & 47&1&6\\
135,180 & 810&10&91 & 400,800 & 4.8&0.2&0.5\\
180,225 & 432&7&44 & \multicolumn{4}{c}{\NA}\\
\end{scotch}
\end{table}
\begin{table}[htbp]
\caption{Differential cross section at parton level as a function of $\abs{y(\tql)}$. The values are shown together with their statistical and systematic uncertainties.}
\centering
\renewcommand{\arraystretch}{1.1}
\begin{scotch}{xr@{$\pm$}c@{$\pm$}l|xr@{$\pm$}c@{$\pm$}l}
\multicolumn{1}{c}{$\abs{y(\tql)}$} & \multicolumn{3}{c}{$\frac{\rd\sigma}{\rd \abs{y(\tql)}}$ [pb]} & \multicolumn{1}{c}{$\abs{y(\tql)}$} & \multicolumn{3}{c}{$\frac{\rd\sigma}{\rd \abs{y(\tql)}}$ [pb]}\\[3pt]
\hline
0,0.2 & 135&2&14 & 1,1.3 & 101&1&11\\
0.2,0.4 & 133&1&14 & 1.3,1.6 & 82&1&9\\
0.4,0.7 & 128&1&14 & 1.6,2.5 & 45.5&0.9&5.1\\
0.7,1 & 118&1&13 & \multicolumn{4}{c}{\NA}\\
\end{scotch}
\end{table}
\begin{table}[htbp]
\caption{Differential cross section at parton level as a function of $\pt(\ttbar)$. The values are shown together with their statistical and systematic uncertainties.}
\centering
\renewcommand{\arraystretch}{1.1}
\begin{scotch}{xr@{$\pm$}c@{$\pm$}l|xr@{$\pm$}c@{$\pm$}l}
\multicolumn{1}{c}{$\pt(\ttbar)$} & \multicolumn{3}{c}{$\frac{\rd\sigma}{\rd\pt(\ttbar)}$} & \multicolumn{1}{c}{$\pt(\ttbar)$} & \multicolumn{3}{c}{$\frac{\rd\sigma}{\rd\pt(\ttbar)}$}\\
\multicolumn{1}{c}{[\GeVns{}]} & \multicolumn{3}{c}{[fb\GeV{}$^{-1}$]} & \multicolumn{1}{c}{[\GeVns{}]} & \multicolumn{3}{c}{[fb\GeV{}$^{-1}$]}\\[3pt]
\hline
0,35 & 3050&70&870 & 140,200 & 220&10&30\\
35,80 & 1470&50&370 & 200,500 & 39&1&5\\
80,140 & 570&20&90 & \multicolumn{4}{c}{\NA}\\
\end{scotch}
\end{table}
\begin{table}[htbp]
\caption{Differential cross section at parton level as a function of $M(\ttbar)$. The values are shown together with their statistical and systematic uncertainties.}
\centering
\renewcommand{\arraystretch}{1.1}
\begin{scotch}{xr@{$\pm$}c@{$\pm$}l|xr@{$\pm$}c@{$\pm$}l}
\multicolumn{1}{c}{$M(\ttbar)$} & \multicolumn{3}{c}{$\frac{\rd\sigma}{\rd M(\ttbar)}$} & \multicolumn{1}{c}{$M(\ttbar)$} & \multicolumn{3}{c}{$\frac{\rd\sigma}{\rd M(\ttbar)}$}\\
\multicolumn{1}{c}{[\GeVns{}]} & \multicolumn{3}{c}{[fb\GeV{}$^{-1}$]} & \multicolumn{1}{c}{[\GeVns{}]} & \multicolumn{3}{c}{[fb\GeV{}$^{-1}$]}\\[3pt]
\hline
300,375 & 360&10&160 & 625,740 & 192&6&31\\
375,450 & 990&20&130 & 740,850 & 84&4&10\\
450,530 & 620&10&110 & 850,1100 & 35&2&6\\
530,625 & 373&9&48 & 1100,2000 & 3.6&0.3&0.4\\
\end{scotch}
\end{table}
\begin{table}[htbp]
\caption{Differential cross section at parton level as a function of $\abs{y(\ttbar)}$. The values are shown together with their statistical and systematic uncertainties.}
\centering
\renewcommand{\arraystretch}{1.1}
\begin{scotch}{xr@{$\pm$}c@{$\pm$}l|xr@{$\pm$}c@{$\pm$}l}
\multicolumn{1}{c}{$\abs{y(\ttbar)}$} & \multicolumn{3}{c}{$\frac{\rd\sigma}{\rd \abs{y(\ttbar)}}$ [pb]} & \multicolumn{1}{c}{$\abs{y(\ttbar)}$} & \multicolumn{3}{c}{$\frac{\rd\sigma}{\rd \abs{y(\ttbar)}}$ [pb]}\\[3pt]
\hline
0,0.2 & 166&3&17 & 0.6,0.9 & 137&2&15\\
0.2,0.4 & 157&3&17 & 0.9,1.3 & 103&2&12\\
0.4,0.6 & 149&3&16 & 1.3,2.3 & 48&1&6\\
\end{scotch}
\end{table}
\begin{table}[htbp]
\caption{Cross sections at parton level in bins of the number of additional jets. The values are shown together with their statistical and systematic uncertainties.}
\centering
\renewcommand{\arraystretch}{1.1}
\begin{scotch}{xr@{$\pm$}c@{$\pm$}l|xr@{$\pm$}c@{$\pm$}l}
\multicolumn{1}{c}{Additional jets} & \multicolumn{3}{c}{$\sigma$ [pb]} & \multicolumn{1}{c}{Additional jets} & \multicolumn{3}{c}{$\sigma$ [pb]}\\[3pt]
\hline
\multicolumn{1}{c}{0} & 97&2&7 & \multicolumn{1}{c}{3} & 12.7&0.6&3.1\\
\multicolumn{1}{c}{1} & 77&2&11 & \multicolumn{1}{c}{ $\ge$ 4} & 5.9&0.2&2.1\\
\multicolumn{1}{c}{2} & 36&1&6 & \multicolumn{4}{c}{\NA}\\
\end{scotch}
\end{table}
\begin{table}[htbp]
\caption{Differential cross sections at parton level as a function of $\pt(\tqh)$ in bins of the number of additional jets. The values are shown together with their statistical and systematic uncertainties.}
\centering
\renewcommand{\arraystretch}{1.1}
\begin{scotch}{xr@{$\pm$}c@{$\pm$}l|xr@{$\pm$}c@{$\pm$}l}
\multicolumn{1}{c}{$\pt(\tqh)$} & \multicolumn{3}{c}{$\frac{\rd\sigma}{\rd\pt(\tqh)}$} & \multicolumn{1}{c}{$\pt(\tqh)$} & \multicolumn{3}{c}{$\frac{\rd\sigma}{\rd\pt(\tqh)}$}\\
\multicolumn{1}{c}{[\GeVns{}]} & \multicolumn{3}{c}{[fb\GeV{}$^{-1}$]} & \multicolumn{1}{c}{[\GeVns{}]} & \multicolumn{3}{c}{[fb\GeV{}$^{-1}$]}\\[3pt]
\hline\multicolumn{8}{c}{ Additional jets: 0}\\
0,45 & 340&20&100 & 225,270 & 71&4&9\\
45,90 & 750&20&110 & 270,315 & 29&3&5\\
90,135 & 610&20&70 & 315,400 & 11&1&2\\
135,180 & 310&10&20 & 400,800 & 1.0&0.2&0.1\\
180,225 & 157&7&16 & \multicolumn{4}{c}{\NA}\\
\hline\multicolumn{8}{c}{ Additional jets: 1}\\
0,45 & 206&6&30 & 225,270 & 79&4&11\\
45,90 & 458&9&60 & 270,315 & 42&3&7\\
90,135 & 408&8&69 & 315,400 & 17&1&2\\
135,180 & 267&6&52 & 400,800 & 1.8&0.2&0.4\\
180,225 & 138&5&24 & \multicolumn{4}{c}{\NA}\\
\hline\multicolumn{8}{c}{ Additional jets: 2}\\
0,45 & 92&3&17 & 225,270 & 50&2&9\\
45,90 & 210&5&37 & 270,315 & 29&2&6\\
90,135 & 196&4&35 & 315,400 & 10.6&1.0&1.7\\
135,180 & 136&4&25 & 400,800 & 1.1&0.2&0.2\\
180,225 & 82&3&17 & \multicolumn{4}{c}{\NA}\\
\hline\multicolumn{8}{c}{ Additional jets: $\ge$3}\\
0,45 & 40&2&8 & 225,270 & 28&2&8\\
45,90 & 90&3&20 & 270,315 & 18&1&5\\
90,135 & 94&3&25 & 315,400 & 8.4&0.8&3.1\\
135,180 & 69&2&21 & 400,800 & 1.2&0.2&0.3\\
180,225 & 45&2&14 & \multicolumn{4}{c}{\NA}\\
\end{scotch}
\end{table}
\begin{table}[htbp]
\caption{Differential cross sections at parton level as a function of $\pt(\ttbar)$ in bins of the number of additional jets. The values are shown together with their statistical and systematic uncertainties.}
\centering
\renewcommand{\arraystretch}{1.1}
\begin{scotch}{xr@{$\pm$}c@{$\pm$}l|xr@{$\pm$}c@{$\pm$}l}
\multicolumn{1}{c}{$\pt(\ttbar)$} & \multicolumn{3}{c}{$\frac{\rd\sigma}{\rd\pt(\ttbar)}$} & \multicolumn{1}{c}{$\pt(\ttbar)$} & \multicolumn{3}{c}{$\frac{\rd\sigma}{\rd\pt(\ttbar)}$}\\
\multicolumn{1}{c}{[\GeVns{}]} & \multicolumn{3}{c}{[fb\GeV{}$^{-1}$]} & \multicolumn{1}{c}{[\GeVns{}]} & \multicolumn{3}{c}{[fb\GeV{}$^{-1}$]}\\[3pt]
\hline\multicolumn{8}{c}{ Additional jets: 0}\\
0,35 & 2220&60&530 & 140,200 & 14&5&6\\
35,80 & 420&40&210 & 200,500 & 0.1&0.2&0.1\\
80,140 & 50&10&40 & \multicolumn{4}{c}{\NA}\\
\hline\multicolumn{8}{c}{ Additional jets: 1}\\
0,35 & 610&40&160 & 140,200 & 100&10&20\\
35,80 & 670&30&90 & 200,500 & 9&1&2\\
80,140 & 260&20&40 & \multicolumn{4}{c}{\NA}\\
\hline\multicolumn{8}{c}{ Additional jets: 2}\\
0,35 & 150&10&40 & 140,200 & 68&8&12\\
35,80 & 240&10&60 & 200,500 & 18&1&3\\
80,140 & 180&10&40 & \multicolumn{4}{c}{\NA}\\
\hline\multicolumn{8}{c}{ Additional jets: $\ge$3}\\
0,35 & 42&6&22 & 140,200 & 54&6&13\\
35,80 & 95&8&29 & 200,500 & 14.3&0.8&3.4\\
80,140 & 77&6&23 & \multicolumn{4}{c}{\NA}\\
\end{scotch}
\end{table}
\begin{table}[htbp]
\caption{Double-differential cross section at parton level as a function of $\abs{y(\tqh)}$ \vs $\pt(\tqh)$. The values are shown together with their statistical and systematic uncertainties.}
\centering
\renewcommand{\arraystretch}{1.1}
\begin{scotch}{xr@{$\pm$}c@{$\pm$}l|xr@{$\pm$}c@{$\pm$}l}
\multicolumn{1}{c}{$\pt(\tqh)$} & \multicolumn{3}{c}{$\frac{\rd^2\sigma}{\rd\pt(\tqh) \rd \abs{y(\tqh)}}$} & \multicolumn{1}{c}{$\pt(\tqh)$} & \multicolumn{3}{c}{$\frac{\rd^2\sigma}{\rd\pt(\tqh) \rd \abs{y(\tqh)}}$}\\
\multicolumn{1}{c}{[\GeVns{}]} & \multicolumn{3}{c}{[fb\GeV{}$^{-1}$]} & \multicolumn{1}{c}{[\GeVns{}]} & \multicolumn{3}{c}{[fb\GeV{}$^{-1}$]}\\[3pt]
\hline\multicolumn{8}{c}{$0<\abs{y(\tqh)}<0.5$}\\
0,45 & 370&8&74 & 225,270 & 149&4&19\\
45,90 & 830&10&120 & 270,315 & 81&3&11\\
90,135 & 770&10&80 & 315,400 & 36&2&6\\
135,180 & 493&8&59 & 400,800 & 4.4&0.3&0.6\\
180,225 & 268&6&36 & \multicolumn{4}{c}{\NA}\\
\hline\multicolumn{8}{c}{$0.5<\abs{y(\tqh)}<1$}\\
0,45 & 340&7&56 & 225,270 & 127&4&22\\
45,90 & 730&10&110 & 270,315 & 65&3&11\\
90,135 & 669&10&73 & 315,400 & 26&1&3\\
135,180 & 425&8&49 & 400,800 & 3.3&0.3&0.6\\
180,225 & 238&6&34 & \multicolumn{4}{c}{\NA}\\
\hline\multicolumn{8}{c}{$1<\abs{y(\tqh)}<1.5$}\\
0,45 & 278&7&44 & 225,270 & 88&3&11\\
45,90 & 600&10&70 & 270,315 & 48&2&8\\
90,135 & 528&9&65 & 315,400 & 19&1&3\\
135,180 & 334&7&46 & 400,800 & 1.5&0.2&0.2\\
180,225 & 173&5&25 & \multicolumn{4}{c}{\NA}\\
\hline\multicolumn{8}{c}{$1.5<\abs{y(\tqh)}<2.5$}\\
0,45 & 188&7&24 & 225,270 & 46&2&8\\
45,90 & 385&9&50 & 270,315 & 20&1&4\\
90,135 & 318&7&44 & 315,400 & 6.3&0.6&1.0\\
135,180 & 175&5&22 & 400,800 & 0.50&0.09&0.09\\
180,225 & 91&3&12 & \multicolumn{4}{c}{\NA}\\
\end{scotch}
\end{table}
\begin{table}[htbp]
\caption{Double-differential cross section at parton level as a function of $M(\ttbar)$ \vs $\abs{y(\ttbar)}$. The values are shown together with their statistical and systematic uncertainties.}
\centering
\renewcommand{\arraystretch}{1.1}
\begin{scotch}{xr@{$\pm$}c@{$\pm$}l|xr@{$\pm$}c@{$\pm$}l}
\multicolumn{1}{c}{$\abs{y(\ttbar)}$} & \multicolumn{3}{c}{$\frac{\rd^2\sigma}{\rd M(\ttbar) \rd \abs{y(\ttbar)}}$} & \multicolumn{1}{c}{$\abs{y(\ttbar)}$} & \multicolumn{3}{c}{$\frac{\rd^2\sigma}{\rd M(\ttbar) \rd \abs{y(\ttbar)}}$}\\
\multicolumn{1}{c}{} & \multicolumn{3}{c}{[fb\GeV{}$^{-1}$]} & \multicolumn{1}{c}{} & \multicolumn{3}{c}{[fb\GeV{}$^{-1}$]}\\[3pt]
\hline\multicolumn{8}{c}{$300<M(\ttbar)<450\GeV$}\\
0,0.2 & 418&10&67 & 0.6,0.9 & 374&7&53\\
0.2,0.4 & 418&8&63 & 0.9,1.3 & 307&7&46\\
0.4,0.6 & 409&8&56 & 1.3,2.3 & 162&5&25\\
\hline\multicolumn{8}{c}{$450<M(\ttbar)<625\GeV$}\\
0,0.2 & 359&7&45 & 0.6,0.9 & 303&6&43\\
0.2,0.4 & 343&6&45 & 0.9,1.3 & 224&5&36\\
0.4,0.6 & 331&7&46 & 1.3,2.3 & 99&3&15\\
\hline\multicolumn{8}{c}{$625<M(\ttbar)<850\GeV$}\\
0,0.2 & 123&4&18 & 0.6,0.9 & 87&3&13\\
0.2,0.4 & 108&3&17 & 0.9,1.3 & 62&3&13\\
0.4,0.6 & 92&3&13 & 1.3,2.3 & 24&2&5\\
\hline\multicolumn{8}{c}{$850<M(\ttbar)<2000\GeV$}\\
0,0.2 & 10.0&0.6&1.5 & 0.6,0.9 & 6.9&0.5&0.8\\
0.2,0.4 & 10.1&0.6&1.4 & 0.9,1.3 & 3.7&0.4&0.5\\
0.4,0.6 & 9.1&0.6&1.5 & 1.3,2.3 & 1.0&0.2&0.2\\
\end{scotch}
\end{table}
\begin{table}[htbp]
\caption{Double-differential cross section at parton level as a function of $\pt(\ttbar)$ \vs $M(\ttbar)$. The values are shown together with their statistical and systematic uncertainties.}
\centering
\renewcommand{\arraystretch}{1.1}
\begin{scotch}{xr@{$\pm$}c@{$\pm$}l|xr@{$\pm$}c@{$\pm$}l}
\multicolumn{1}{c}{$M(\ttbar)$} & \multicolumn{3}{c}{$\frac{\rd^2\sigma}{\rd\pt(\ttbar) \rd M(\ttbar)}$} & \multicolumn{1}{c}{$M(\ttbar)$} & \multicolumn{3}{c}{$\frac{\rd^2\sigma}{\rd\pt(\ttbar) \rd M(\ttbar)}$}\\
\multicolumn{1}{c}{[\GeVns{}]} & \multicolumn{3}{c}{[fb\GeV{}$^{-2}$]} & \multicolumn{1}{c}{[\GeVns{}]} & \multicolumn{3}{c}{[fb\GeV{}$^{-2}$]}\\[3pt]
\hline\multicolumn{8}{c}{$0<\pt(\ttbar)<35\GeV$}\\
300,375 & 4.8&0.2&2.0 & 625,740 & 2.18&0.09&0.63\\
375,450 & 13.7&0.3&3.0 & 740,850 & 0.92&0.06&0.18\\
450,530 & 8.5&0.2&3.8 & 850,1100 & 0.36&0.03&0.12\\
530,625 & 4.4&0.1&1.3 & 1100,2000 & 0.039&0.005&0.012\\
\hline\multicolumn{8}{c}{$35<\pt(\ttbar)<80\GeV$}\\
300,375 & 2.25&0.07&1.20 & 625,740 & 1.32&0.04&0.22\\
375,450 & 6.6&0.1&1.6 & 740,850 & 0.60&0.03&0.07\\
450,530 & 4.30&0.08&0.60 & 850,1100 & 0.23&0.01&0.03\\
530,625 & 2.53&0.06&0.29 & 1100,2000 & 0.022&0.002&0.005\\
\hline\multicolumn{8}{c}{$80<\pt(\ttbar)<140\GeV$}\\
300,375 & 0.76&0.03&0.30 & 625,740 & 0.51&0.02&0.07\\
375,450 & 2.24&0.05&0.50 & 740,850 & 0.25&0.01&0.04\\
450,530 & 1.52&0.04&0.19 & 850,1100 & 0.100&0.008&0.026\\
530,625 & 0.96&0.03&0.10 & 1100,2000 & 0.011&0.002&0.002\\
\hline\multicolumn{8}{c}{$140<\pt(\ttbar)<500\GeV$}\\
300,375 & 0.095&0.005&0.025 & 625,740 & 0.068&0.003&0.017\\
375,450 & 0.258&0.008&0.032 & 740,850 & 0.036&0.002&0.004\\
450,530 & 0.185&0.006&0.024 & 850,1100 & 0.016&0.001&0.003\\
530,625 & 0.122&0.005&0.034 & 1100,2000 & 0.0018&0.0003&0.0003\\
\end{scotch}
\end{table}
\clearpage
\section{Tables of particle level cross sections.}
\label{APP2}
\begin{table}[htbp]
\caption{Differential cross section at particle level as a function of $\pt(\tqh)$. The values are shown together with their statistical and systematic uncertainties.}
\centering
\renewcommand{\arraystretch}{1.1}
\begin{scotch}{xr@{$\pm$}c@{$\pm$}l|xr@{$\pm$}c@{$\pm$}l}
\multicolumn{1}{c}{$\pt(\tqh)$} & \multicolumn{3}{c}{$\frac{\rd\sigma}{\rd\pt(\tqh)}$} & \multicolumn{1}{c}{$\pt(\tqh)$} & \multicolumn{3}{c}{$\frac{\rd\sigma}{\rd\pt(\tqh)}$}\\
\multicolumn{1}{c}{[\GeVns{}]} & \multicolumn{3}{c}{[fb\GeV{}$^{-1}$]} & \multicolumn{1}{c}{[\GeVns{}]} & \multicolumn{3}{c}{[fb\GeV{}$^{-1}$]}\\[3pt]
\hline
0,45 & 204&4&18 & 225,270 & 106&2&9\\
45,90 & 461&5&40 & 270,315 & 61&2&6\\
90,135 & 430&5&41 & 315,400 & 27.4&0.9&2.5\\
135,180 & 292&4&27 & 400,800 & 3.2&0.2&0.3\\
180,225 & 179&3&17 & \multicolumn{4}{c}{\NA}\\
\end{scotch}
\end{table}
\begin{table}[htbp]
\caption{Differential cross section at particle level as a function of $\abs{y(\tqh)}$. The values are shown together with their statistical and systematic uncertainties.}
\centering
\renewcommand{\arraystretch}{1.1}
\begin{scotch}{xr@{$\pm$}c@{$\pm$}l|xr@{$\pm$}c@{$\pm$}l}
\multicolumn{1}{c}{$\abs{y(\tqh)}$} & \multicolumn{3}{c}{$\frac{\rd\sigma}{\rd \abs{y(\tqh)}}$ [pb]} & \multicolumn{1}{c}{$\abs{y(\tqh)}$} & \multicolumn{3}{c}{$\frac{\rd\sigma}{\rd \abs{y(\tqh)}}$ [pb]}\\[3pt]
\hline
0,0.2 & 61.3&0.7&5.2 & 1,1.3 & 38.6&0.4&3.7\\
0.2,0.4 & 59.4&0.6&4.9 & 1.3,1.6 & 27.8&0.4&3.1\\
0.4,0.7 & 55.1&0.5&4.7 & 1.6,2.5 & 7.3&0.1&0.8\\
0.7,1 & 47.6&0.5&4.2 & \multicolumn{4}{c}{\NA}\\
\end{scotch}
\end{table}
\begin{table}[htbp]
\caption{Differential cross section at particle level as a function of $\pt(\tql)$. The values are shown together with their statistical and systematic uncertainties.}
\centering
\renewcommand{\arraystretch}{1.1}
\begin{scotch}{xr@{$\pm$}c@{$\pm$}l|xr@{$\pm$}c@{$\pm$}l}
\multicolumn{1}{c}{$\pt(\tql)$} & \multicolumn{3}{c}{$\frac{\rd\sigma}{\rd\pt(\tql)}$} & \multicolumn{1}{c}{$\pt(\tql)$} & \multicolumn{3}{c}{$\frac{\rd\sigma}{\rd\pt(\tql)}$}\\
\multicolumn{1}{c}{[\GeVns{}]} & \multicolumn{3}{c}{[fb\GeV{}$^{-1}$]} & \multicolumn{1}{c}{[\GeVns{}]} & \multicolumn{3}{c}{[fb\GeV{}$^{-1}$]}\\[3pt]
\hline
0,45 & 185&3&17 & 225,270 & 113&2&9\\
45,90 & 425&4&41 & 270,315 & 67&2&5\\
90,135 & 429&4&41 & 315,400 & 30.6&0.9&2.4\\
135,180 & 310&4&28 & 400,800 & 3.7&0.2&0.4\\
180,225 & 194&3&16 & \multicolumn{4}{c}{\NA}\\
\end{scotch}
\end{table}
\begin{table}[htbp]
\caption{Differential cross section at particle level as a function of $\abs{y(\tql)}$. The values are shown together with their statistical and systematic uncertainties.}
\centering
\renewcommand{\arraystretch}{1.1}
\begin{scotch}{xr@{$\pm$}c@{$\pm$}l|xr@{$\pm$}c@{$\pm$}l}
\multicolumn{1}{c}{$\abs{y(\tql)}$} & \multicolumn{3}{c}{$\frac{\rd\sigma}{\rd \abs{y(\tql)}}$ [pb]} & \multicolumn{1}{c}{$\abs{y(\tql)}$} & \multicolumn{3}{c}{$\frac{\rd\sigma}{\rd \abs{y(\tql)}}$ [pb]}\\[3pt]
\hline
0,0.2 & 55.7&0.7&5.0 & 1,1.3 & 38.9&0.5&3.6\\
0.2,0.4 & 54.6&0.6&5.1 & 1.3,1.6 & 29.3&0.4&2.7\\
0.4,0.7 & 52.0&0.5&4.9 & 1.6,2.5 & 10.2&0.2&0.9\\
0.7,1 & 47.2&0.5&4.4 & \multicolumn{4}{c}{\NA}\\
\end{scotch}
\end{table}
\begin{table}[htbp]
\caption{Differential cross section at particle level as a function of $\pt(\ttbar)$. The values are shown together with their statistical and systematic uncertainties.}
\centering
\renewcommand{\arraystretch}{1.1}
\begin{scotch}{xr@{$\pm$}c@{$\pm$}l|xr@{$\pm$}c@{$\pm$}l}
\multicolumn{1}{c}{$\pt(\ttbar)$} & \multicolumn{3}{c}{$\frac{\rd\sigma}{\rd\pt(\ttbar)}$} & \multicolumn{1}{c}{$\pt(\ttbar)$} & \multicolumn{3}{c}{$\frac{\rd\sigma}{\rd\pt(\ttbar)}$}\\
\multicolumn{1}{c}{[\GeVns{}]} & \multicolumn{3}{c}{[fb\GeV{}$^{-1}$]} & \multicolumn{1}{c}{[\GeVns{}]} & \multicolumn{3}{c}{[fb\GeV{}$^{-1}$]}\\[3pt]
\hline
0,35 & 890&10&140 & 140,200 & 92&3&10\\
35,80 & 577&10&62 & 200,500 & 18.4&0.5&1.7\\
80,140 & 219&5&22 & \multicolumn{4}{c}{\NA}\\
\end{scotch}
\end{table}
\begin{table}[htbp]
\caption{Differential cross section at particle level as a function of $M(\ttbar)$. The values are shown together with their statistical and systematic uncertainties.}
\centering
\renewcommand{\arraystretch}{1.1}
\begin{scotch}{xr@{$\pm$}c@{$\pm$}l|xr@{$\pm$}c@{$\pm$}l}
\multicolumn{1}{c}{$M(\ttbar)$} & \multicolumn{3}{c}{$\frac{\rd\sigma}{\rd M(\ttbar)}$} & \multicolumn{1}{c}{$M(\ttbar)$} & \multicolumn{3}{c}{$\frac{\rd\sigma}{\rd M(\ttbar)}$}\\
\multicolumn{1}{c}{[\GeVns{}]} & \multicolumn{3}{c}{[fb\GeV{}$^{-1}$]} & \multicolumn{1}{c}{[\GeVns{}]} & \multicolumn{3}{c}{[fb\GeV{}$^{-1}$]}\\[3pt]
\hline
300,375 & 124&4&14 & 625,740 & 91&2&8\\
375,450 & 247&4&27 & 740,850 & 47&2&4\\
450,530 & 200&4&22 & 850,1100 & 22.3&0.8&2.1\\
530,625 & 144&3&13 & 1100,2000 & 2.7&0.2&0.4\\
\end{scotch}
\end{table}
\begin{table}[htbp]
\caption{Differential cross section at particle level as a function of $\abs{y(\ttbar)}$. The values are shown together with their statistical and systematic uncertainties.}
\centering
\renewcommand{\arraystretch}{1.1}
\begin{scotch}{xr@{$\pm$}c@{$\pm$}l|xr@{$\pm$}c@{$\pm$}l}
\multicolumn{1}{c}{$\abs{y(\ttbar)}$} & \multicolumn{3}{c}{$\frac{\rd\sigma}{\rd \abs{y(\ttbar)}}$ [pb]} & \multicolumn{1}{c}{$\abs{y(\ttbar)}$} & \multicolumn{3}{c}{$\frac{\rd\sigma}{\rd \abs{y(\ttbar)}}$ [pb]}\\[3pt]
\hline
0,0.2 & 76.2&0.9&6.6 & 0.6,0.9 & 55.0&0.6&4.9\\
0.2,0.4 & 71.8&0.7&6.3 & 0.9,1.3 & 35.8&0.5&3.5\\
0.4,0.6 & 66.1&0.7&6.1 & 1.3,2.3 & 7.7&0.2&0.8\\
\end{scotch}
\end{table}
\begin{table}[htbp]
\caption{Cross sections at particle level in bins of the number of additional jets. The values are shown together with their statistical and systematic uncertainties.}
\centering
\renewcommand{\arraystretch}{1.1}
\begin{scotch}{xr@{$\pm$}c@{$\pm$}l|xr@{$\pm$}c@{$\pm$}l}
\multicolumn{1}{c}{Additional jets} & \multicolumn{3}{c}{$\sigma$ [pb]} & \multicolumn{1}{c}{Additional jets} & \multicolumn{3}{c}{$\sigma$ [pb]}\\[3pt]
\hline
\multicolumn{1}{c}{0} & 39.9&0.4&3.0 & \multicolumn{1}{c}{3} & 3.8&0.1&0.6\\
\multicolumn{1}{c}{1} & 25.6&0.3&2.7 & \multicolumn{1}{c}{ $\ge$ 4} & 1.75&0.07&0.36\\
\multicolumn{1}{c}{2} & 10.6&0.2&1.3 & \multicolumn{4}{c}{\NA}\\
\end{scotch}
\end{table}
\begin{table}[htbp]
\caption{Differential cross sections at particle level as a function of $\pt(\tqh)$ in bins of the number of additional jets. The values are shown together with their statistical and systematic uncertainties.}
\centering
\renewcommand{\arraystretch}{1.1}
\begin{scotch}{xr@{$\pm$}c@{$\pm$}l|xr@{$\pm$}c@{$\pm$}l}
\multicolumn{1}{c}{$\pt(\tqh)$} & \multicolumn{3}{c}{$\frac{\rd\sigma}{\rd\pt(\tqh)}$} & \multicolumn{1}{c}{$\pt(\tqh)$} & \multicolumn{3}{c}{$\frac{\rd\sigma}{\rd\pt(\tqh)}$}\\
\multicolumn{1}{c}{[\GeVns{}]} & \multicolumn{3}{c}{[fb\GeV{}$^{-1}$]} & \multicolumn{1}{c}{[\GeVns{}]} & \multicolumn{3}{c}{[fb\GeV{}$^{-1}$]}\\[3pt]
\hline\multicolumn{8}{c}{ Additional jets: 0}\\
0,45 & 108&3&7 & 225,270 & 44&1&4\\
45,90 & 241&4&16 & 270,315 & 22.7&0.9&2.0\\
90,135 & 226&3&16 & 315,400 & 9.7&0.5&1.3\\
135,180 & 146&3&10 & 400,800 & 1.09&0.09&0.15\\
180,225 & 84&2&7 & \multicolumn{4}{c}{\NA}\\
\hline\multicolumn{8}{c}{ Additional jets: 1}\\
0,45 & 60&1&7 & 225,270 & 34.8&0.9&3.6\\
45,90 & 136&2&16 & 270,315 & 20.9&0.7&2.6\\
90,135 & 129&2&13 & 315,400 & 9.4&0.4&0.8\\
135,180 & 92&1&9 & 400,800 & 1.06&0.08&0.14\\
180,225 & 57&1&6 & \multicolumn{4}{c}{\NA}\\
\hline\multicolumn{8}{c}{ Additional jets: 2}\\
0,45 & 24.7&0.5&3.5 & 225,270 & 17.1&0.5&2.1\\
45,90 & 55.8&0.9&7.7 & 270,315 & 10.7&0.4&1.3\\
90,135 & 52.7&0.8&6.9 & 315,400 & 4.9&0.3&0.6\\
135,180 & 38.4&0.7&4.6 & 400,800 & 0.60&0.05&0.08\\
180,225 & 26.0&0.6&3.1 & \multicolumn{4}{c}{\NA}\\
\hline\multicolumn{8}{c}{ Additional jets: $\ge$3}\\
0,45 & 11.6&0.3&2.0 & 225,270 & 9.4&0.4&1.4\\
45,90 & 25.9&0.6&4.4 & 270,315 & 6.5&0.3&1.0\\
90,135 & 26.0&0.6&4.3 & 315,400 & 3.5&0.2&0.6\\
135,180 & 19.2&0.5&2.8 & 400,800 & 0.47&0.05&0.07\\
180,225 & 13.5&0.4&1.8 & \multicolumn{4}{c}{\NA}\\
\end{scotch}
\end{table}
\begin{table}[htbp]
\caption{Differential cross sections at particle level as a function of $\pt(\ttbar)$ in bins of the number of additional jets. The values are shown together with their statistical and systematic uncertainties.}
\centering
\renewcommand{\arraystretch}{1.1}
\begin{scotch}{xr@{$\pm$}c@{$\pm$}l|xr@{$\pm$}c@{$\pm$}l}
\multicolumn{1}{c}{$\pt(\ttbar)$} & \multicolumn{3}{c}{$\frac{\rd\sigma}{\rd\pt(\ttbar)}$} & \multicolumn{1}{c}{$\pt(\ttbar)$} & \multicolumn{3}{c}{$\frac{\rd\sigma}{\rd\pt(\ttbar)}$}\\
\multicolumn{1}{c}{[\GeVns{}]} & \multicolumn{3}{c}{[fb\GeV{}$^{-1}$]} & \multicolumn{1}{c}{[\GeVns{}]} & \multicolumn{3}{c}{[fb\GeV{}$^{-1}$]}\\[3pt]
\hline\multicolumn{8}{c}{ Additional jets: 0}\\
0,35 & 730&10&100 & 140,200 & 7&1&2\\
35,80 & 268&8&31 & 200,500 & 0.19&0.09&0.07\\
80,140 & 33&3&8 & \multicolumn{4}{c}{\NA}\\
\hline\multicolumn{8}{c}{ Additional jets: 1}\\
0,35 & 118&5&19 & 140,200 & 45&3&5\\
35,80 & 222&5&26 & 200,500 & 6.6&0.4&0.7\\
80,140 & 112&4&12 & \multicolumn{4}{c}{\NA}\\
\hline\multicolumn{8}{c}{ Additional jets: 2}\\
0,35 & 25&2&5 & 140,200 & 26&2&3\\
35,80 & 59&3&10 & 200,500 & 6.8&0.4&0.7\\
80,140 & 55&2&8 & \multicolumn{4}{c}{\NA}\\
\hline\multicolumn{8}{c}{ Additional jets: $\ge$3}\\
0,35 & 8.1&1.0&2.0 & 140,200 & 17&1&4\\
35,80 & 23&2&5 & 200,500 & 5.4&0.3&0.8\\
80,140 & 22&1&4 & \multicolumn{4}{c}{\NA}\\
\end{scotch}
\end{table}
\begin{table}[htbp]
\caption{Double-differential cross section at particle level as a function of $\abs{y(\tqh)}$ \vs $\pt(\tqh)$. The values are shown together with their statistical and systematic uncertainties.}
\centering
\renewcommand{\arraystretch}{1.1}
\begin{scotch}{xr@{$\pm$}c@{$\pm$}l|xr@{$\pm$}c@{$\pm$}l}
\multicolumn{1}{c}{$\pt(\tqh)$} & \multicolumn{3}{c}{$\frac{\rd^2\sigma}{\rd\pt(\tqh) \rd \abs{y(\tqh)}}$} & \multicolumn{1}{c}{$\pt(\tqh)$} & \multicolumn{3}{c}{$\frac{\rd^2\sigma}{\rd\pt(\tqh) \rd \abs{y(\tqh)}}$}\\
\multicolumn{1}{c}{[\GeVns{}]} & \multicolumn{3}{c}{[fb\GeV{}$^{-1}$]} & \multicolumn{1}{c}{[\GeVns{}]} & \multicolumn{3}{c}{[fb\GeV{}$^{-1}$]}\\[3pt]
\hline\multicolumn{8}{c}{$0<\abs{y(\tqh)}<0.5$}\\
0,45 & 146&2&12 & 225,270 & 78&2&6\\
45,90 & 330&4&28 & 270,315 & 46&1&4\\
90,135 & 316&4&26 & 315,400 & 21.8&0.8&2.0\\
135,180 & 217&3&18 & 400,800 & 2.7&0.2&0.3\\
180,225 & 129&2&11 & \multicolumn{4}{c}{\NA}\\
\hline\multicolumn{8}{c}{$0.5<\abs{y(\tqh)}<1$}\\
0,45 & 126&2&13 & 225,270 & 63&2&6\\
45,90 & 281&3&25 & 270,315 & 36&1&3\\
90,135 & 267&3&23 & 315,400 & 16.4&0.7&1.4\\
135,180 & 182&3&15 & 400,800 & 2.2&0.1&0.3\\
180,225 & 112&2&10 & \multicolumn{4}{c}{\NA}\\
\hline\multicolumn{8}{c}{$1<\abs{y(\tqh)}<1.5$}\\
0,45 & 88&2&9 & 225,270 & 44&1&4\\
45,90 & 198&3&21 & 270,315 & 25.3&1.0&2.3\\
90,135 & 186&3&18 & 315,400 & 11.1&0.6&1.2\\
135,180 & 130&2&12 & 400,800 & 0.99&0.09&0.11\\
180,225 & 77&2&7 & \multicolumn{4}{c}{\NA}\\
\hline\multicolumn{8}{c}{$1.5<\abs{y(\tqh)}<2.5$}\\
0,45 & 21.9&0.8&3.3 & 225,270 & 12.9&0.5&1.4\\
45,90 & 49&1&6 & 270,315 & 7.0&0.4&0.8\\
90,135 & 48&1&5 & 315,400 & 2.9&0.2&0.3\\
135,180 & 32.2&0.9&3.2 & 400,800 & 0.25&0.03&0.04\\
180,225 & 21.3&0.7&2.0 & \multicolumn{4}{c}{\NA}\\
\end{scotch}
\end{table}
\begin{table}[htbp]
\caption{Double-differential cross section at particle level as a function of $M(\ttbar)$ \vs $\abs{y(\ttbar)}$. The values are shown together with their statistical and systematic uncertainties.}
\centering
\renewcommand{\arraystretch}{1.1}
\begin{scotch}{xr@{$\pm$}c@{$\pm$}l|xr@{$\pm$}c@{$\pm$}l}
\multicolumn{1}{c}{$\abs{y(\ttbar)}$} & \multicolumn{3}{c}{$\frac{\rd^2\sigma}{\rd M(\ttbar) \rd \abs{y(\ttbar)}}$} & \multicolumn{1}{c}{$\abs{y(\ttbar)}$} & \multicolumn{3}{c}{$\frac{\rd^2\sigma}{\rd M(\ttbar) \rd \abs{y(\ttbar)}}$}\\
\multicolumn{1}{c}{} & \multicolumn{3}{c}{[fb\GeV{}$^{-1}$]} & \multicolumn{1}{c}{} & \multicolumn{3}{c}{[fb\GeV{}$^{-1}$]}\\[3pt]
\hline\multicolumn{8}{c}{$300<M(\ttbar)<450\GeV$}\\
0,0.2 & 143&3&12 & 0.6,0.9 & 124&3&11\\
0.2,0.4 & 142&3&12 & 0.9,1.3 & 96&2&9\\
0.4,0.6 & 140&3&12 & 1.3,2.3 & 25.7&0.9&2.5\\
\hline\multicolumn{8}{c}{$450<M(\ttbar)<625\GeV$}\\
0,0.2 & 158&3&15 & 0.6,0.9 & 118&2&12\\
0.2,0.4 & 148&3&15 & 0.9,1.3 & 75&2&9\\
0.4,0.6 & 142&3&14 & 1.3,2.3 & 15.5&0.6&1.7\\
\hline\multicolumn{8}{c}{$625<M(\ttbar)<850\GeV$}\\
0,0.2 & 77&2&6 & 0.6,0.9 & 47&1&4\\
0.2,0.4 & 67&2&6 & 0.9,1.3 & 27&1&3\\
0.4,0.6 & 57&2&5 & 1.3,2.3 & 4.3&0.3&0.4\\
\hline\multicolumn{8}{c}{$850<M(\ttbar)<2000\GeV$}\\
0,0.2 & 8.4&0.4&0.9 & 0.6,0.9 & 4.7&0.3&0.4\\
0.2,0.4 & 8.5&0.4&1.0 & 0.9,1.3 & 1.9&0.1&0.2\\
0.4,0.6 & 6.7&0.3&0.7 & 1.3,2.3 & 0.20&0.03&0.03\\
\end{scotch}
\end{table}
\begin{table}[htbp]
\caption{Double-differential cross section at particle level as a function of $\pt(\ttbar)$ \vs $M(\ttbar)$. The values are shown together with their statistical and systematic uncertainties.}
\centering
\renewcommand{\arraystretch}{1.1}
\begin{scotch}{xr@{$\pm$}c@{$\pm$}l|xr@{$\pm$}c@{$\pm$}l}
\multicolumn{1}{c}{$M(\ttbar)$} & \multicolumn{3}{c}{$\frac{\rd^2\sigma}{\rd\pt(\ttbar) \rd M(\ttbar)}$} & \multicolumn{1}{c}{$M(\ttbar)$} & \multicolumn{3}{c}{$\frac{\rd^2\sigma}{\rd\pt(\ttbar) \rd M(\ttbar)}$}\\
\multicolumn{1}{c}{[\GeVns{}]} & \multicolumn{3}{c}{[fb\GeV{}$^{-2}$]} & \multicolumn{1}{c}{[\GeVns{}]} & \multicolumn{3}{c}{[fb\GeV{}$^{-2}$]}\\[3pt]
\hline\multicolumn{8}{c}{$0<\pt(\ttbar)<35\GeV$}\\
300,375 & 1.44&0.05&0.09 & 625,740 & 0.88&0.02&0.11\\
375,450 & 2.85&0.06&0.41 & 740,850 & 0.48&0.02&0.05\\
450,530 & 2.26&0.05&0.40 & 850,1100 & 0.215&0.010&0.040\\
530,625 & 1.47&0.03&0.22 & 1100,2000 & 0.030&0.003&0.012\\
\hline\multicolumn{8}{c}{$35<\pt(\ttbar)<80\GeV$}\\
300,375 & 0.89&0.02&0.09 & 625,740 & 0.66&0.01&0.06\\
375,450 & 1.76&0.03&0.20 & 740,850 & 0.36&0.01&0.03\\
450,530 & 1.44&0.02&0.16 & 850,1100 & 0.158&0.006&0.020\\
530,625 & 1.03&0.02&0.09 & 1100,2000 & 0.018&0.001&0.004\\
\hline\multicolumn{8}{c}{$80<\pt(\ttbar)<140\GeV$}\\
300,375 & 0.31&0.01&0.03 & 625,740 & 0.249&0.007&0.021\\
375,450 & 0.67&0.02&0.08 & 740,850 & 0.137&0.005&0.016\\
450,530 & 0.55&0.01&0.06 & 850,1100 & 0.059&0.003&0.007\\
530,625 & 0.395&0.010&0.036 & 1100,2000 & 0.0066&0.0007&0.0018\\
\hline\multicolumn{8}{c}{$140<\pt(\ttbar)<500\GeV$}\\
300,375 & 0.035&0.002&0.007 & 625,740 & 0.039&0.001&0.004\\
375,450 & 0.081&0.002&0.009 & 740,850 & 0.022&0.001&0.003\\
450,530 & 0.077&0.002&0.008 & 850,1100 & 0.0107&0.0006&0.0009\\
530,625 & 0.061&0.002&0.008 & 1100,2000 & 0.0016&0.0002&0.0002\\
\end{scotch}
\end{table}
\clearpage
\end{appendix}
\cleardoublepage \section{The CMS Collaboration \label{app:collab}}\begin{sloppypar}\hyphenpenalty=5000\widowpenalty=500\clubpenalty=5000\textbf{Yerevan Physics Institute,  Yerevan,  Armenia}\\*[0pt]
V.~Khachatryan, A.M.~Sirunyan, A.~Tumasyan
\vskip\cmsinstskip
\textbf{Institut f\"{u}r Hochenergiephysik,  Wien,  Austria}\\*[0pt]
W.~Adam, E.~Asilar, T.~Bergauer, J.~Brandstetter, E.~Brondolin, M.~Dragicevic, J.~Er\"{o}, M.~Flechl, M.~Friedl, R.~Fr\"{u}hwirth\cmsAuthorMark{1}, V.M.~Ghete, C.~Hartl, N.~H\"{o}rmann, J.~Hrubec, M.~Jeitler\cmsAuthorMark{1}, A.~K\"{o}nig, I.~Kr\"{a}tschmer, D.~Liko, T.~Matsushita, I.~Mikulec, D.~Rabady, N.~Rad, B.~Rahbaran, H.~Rohringer, J.~Schieck\cmsAuthorMark{1}, J.~Strauss, W.~Waltenberger, C.-E.~Wulz\cmsAuthorMark{1}
% ...
\vskip\cmsinstskip
% ...
1:~~Also at Vienna University of Technology, Vienna, Austria\\
% ...

\end{sloppypar}
\end{document}